%% file: final8.tex
\documentclass[11pt,a4paper]{article}
\usepackage{amsmath}
\usepackage[dvips]{graphicx}

\newcommand{\eps}{\epsilon}
\def\lsim{\mathrel{\rlap{\lower4pt\hbox{\hskip1pt$\sim$}}
    \raise1pt\hbox{$<$}}}                % less than or approx. symbol
\def\gsim{\mathrel{\rlap{\lower4pt\hbox{\hskip1pt$\sim$}}
    \raise1pt\hbox{$>$}}}                % greater than or approx. symbol
\newcommand{\ap}{\ensuremath{\alpha_{I\!\!P}}}

\begin{document}

\thispagestyle{empty}

\begin{center}
\Large{
{\bf Unitarity and the QCD-improved dipole picture}
}
\vspace*{0.5cm}

M. McDermott$^1$, L. Frankfurt$^2$, V. Guzey$^{3,4}$, M. Strikman$^3$ \\

\vspace*{0.5cm}
$^1$Department of Physics and Astronomy,\\
University of Manchester, Manchester M13 9PL England.\\
\vspace*{0.5cm}
$^2$Nuclear Physics Dept., School of Physics and Astronomy,\\
Tel Aviv University, 69978 Tel Aviv, Israel\\
\vspace*{0.5cm}
$^3$Department of Physics, Penn State University,\\
University Park, PA 16802-6300, USA.\\
\vspace*{0.5cm}
$^4$Special Research Centre for the Subatomic Structure of \\
Matter (CSSM), University  of Adelaide,  5005, Australia.\\
\end{center}

\vspace*{0.5cm}

\begin{abstract}
As a consequence of QCD factorization theorems, a wide variety of 
inclusive and exclusive cross sections may be formulated in terms of 
a universal colour dipole cross section at small $x$. 
It is well known that for small transverse size dipoles this cross section is 
related to the leading-log gluon density. Using the measured
pion-proton cross section as a guide, we suggest a reasonable 
extrapolation of the dipole cross section to the large transverse size
region. We point out that the observed magnitude and small $x$ rise of
the gluon density from conventional fits implies that the 
DGLAP approximation has a restricted region of applicability. 
We found that `higher twist' or unitarity 
corrections are required in, or close to, the HERA kinematic region, 
even for small `perturbative' dipoles 
for scattering at central impact parameters. 
This means that the usual perturbative leading
twist description, for moderate virtualities, $1 < Q^2 < 10$ GeV$^2$,
has rather large `higher twist' corrections at small $x$.
In addition, for these virtualities, we also find sizeable contributions 
from large non-perturbative dipoles ($b \gsim 0.4$ fm) to $F_2$, and 
also to $F_L$. This also leads to deviations from the standard 
leading twist DGLAP results, at small $x$ and moderate $Q^2$.
Our model also describes the low $Q^2$ data very well without any 
further tuning.
We generalize the Gribov unitarity limit for the structure functions
of a hadron target to account for the blackening of the interaction at 
central impact parameters and to include scattering at peripheral 
impact parameters which dominate at extremely large energies.
\end{abstract}

\newpage

\section{Introduction}

In the proton rest frame, at small enough $x = Q^2/W^2$ and $Q^2 \gg m_p^2$, 
Deep Inelastic Scattering of a virtual photon from a proton 
may be viewed as being factorized into a three stage process: the 
formation of a state which in general is build of quark, antiquark 
and gluons from the virtual photon, the scattering of this state 
off the static proton and
the subsequent formation of the hadronic final state. 
In QCD, the different quark-antiquark-gluon configurations in the photon
clearly have different interaction strengths with a target.

For states of very small transverse spatial size, $b^2$, the 
dominant scattering state is a quark-antiquark colour dipole state 
(in this simple case $b$ is the transverse diameter of the dipole).
This small dipole has a small scattering probability, which has 
been calculated in perturbative QCD, and at small $x$ 
is related to the gluonic colour field associated with the bound state.  

The fact that such a dipole cross section appears 
in a wide variety of hard small $x$ inclusive, exclusive 
(e.g. heavy vector meson production, DVCS,
etc) and diffractive processes is a consequence of the feasibility of 
separating scales in QCD \cite{cfs}, and may be formulated in terms of
a {\it universal}\footnote{In practice, the kinematical effect of 
skewedness of the amplitude in exclusive processes leads to a partial,
but controlled, breakdown of this universality.} 
dipole cross section at small $x$. So far, general, 
relatively ad hoc, ans\"{a}tze
have been ascribed to this quantity 
\cite{wkgb,kfs} and the phenomenological parameters specified by
successful fits to structure function data. 

States of larger transverse size will in general have much larger
cross sections and will contain many constituents. With the increase of the 
size of the quark-gluon configurations the number of degrees of freedom in the 
photon wavefunction is also increasing (becoming very large
in the non-perturbative QCD regime as a consequence of spontaneously 
broken chiral symmetry in QCD). Nevertheless, the transverse size of the 
scattering state seems to be an appropriate parameter for the smooth 
matching between cross sections in the soft and hard regimes. 
For convenience we will continue to refer to $b^2$ as the {\it dipole}
size, and the cross section as the dipole cross section, but these terms 
should be understood to refer to the transverse size and cross section 
for more general scattering systems for the case of large systems.

The aim of this paper is to exploit the QCD relationship between the 
small dipole cross section (DCS) and the gluon density in the proton 
to build a realistic ansatz, monotonically increasing in $b$, 
for the DCS of all transverse sizes.
In the large $b$ region, we match onto the measured pion-proton 
cross section (at $b_{\pi} = 0.65$ fm), to which we ascribe a 
gentle rise with energy. We suggest a smooth interpolation in $b$ for the DCS 
from small $b$ (where perturbative QCD is valid) up to $b_{\pi}$ 
and a smooth extrapolation for even larger transverse sizes.

Using our ansatz, we then analyze the small $x$ structure functions, 
calculated in $b$-space which we denote 
$F_L^b$ and $F_2^b$. In this picture, they 
are given by convolutions in $b^2$ of the square 
of the light-cone wavefunction of the virtual photons, 
of the appropriate polarization state, with the DCS. 
We observe that for large $Q^2$, $F_L^b$ reproduces approximately 
the same result as the standard leading-log perturbative QCD formula. 
This agreement provides a justification for the 
relationship between four-momentum scales for the gluon density
and transverse dipole sizes which we assume in our ansatz 
($Q^2 = 10/b^2$, cf. \cite{fks1,fks2,fms}).
We then produce values for $F_2$ using our ansatz and, without any fitting, 
find reasonable agreement with the HERA data at small $x$, 
even for the region of low photon virtuality below the input scale for 
QCD evolution, $Q^2 < Q_0^2$, where our ansatz may be expected 
to do less well. 

It is straightforward to calculate how much of the non-perturbative region at 
large $b$ contributes to the structure functions. 
Since $F_2$ is mainly governed
by the transversely polarized photon, the spin structure of the 
$\gamma^{*}_{T} q {\bar q}$ vertex leads to a considerably broader 
integral in $b$ than in longitudinal case. We illustrate this well 
known effect graphically 
(see figs.(\ref{psisqt},\ref{psisql},\ref{psisqtl})). 
We find that for relatively high $Q^2 = 4 - 10 $ GeV$^2$ 
a surprisingly big contribution to the integral (as much as 50\%) 
is coming from this `dangerous' large $b$ region. While much of this 
non-perturbative piece is attributable to the input distributions in
$F_2$, the fact that it is also present in $F_L$ could indicate a
sizeable `higher twist' contribution for these virtualities coming
from the non-perturbative region. This presents a severe challenge to
the use of low, and $x$-independent, input scales in the conventional 
parton density fits (e.g. the recent MRST analysis \cite{mrst} uses 
input scale $Q_0 = 1.0$~GeV). 

We note that the small $x$ rise of the structure function $F_2$ 
observed at HERA, 
which may be translated into a large and steeply-rising gluon density at small
$x$, quickly (after only a few gluon radiations in the ladder) leads to a 
contradiction with unitarity because the DCS for small dipoles 
becomes of the same size as the pion-proton cross section 
(which grows much more slowly with increasing energy). 
To avoid this problem, it is necessary to tame the small $x$
growth of the perturbative DCS. We suggest a way of doing this which 
modifies our ansatz for the DCS at small $x$.
For small enough $x$, this taming
appears to be required within the weak coupling limit
and hence may be related (albeit indirectly and within 
a restricted range of $x$) to the four-gluon to two-gluon calculations
of Bartels and collaborators (for a discussion of perturbative higher
twist effects in QCD, recent developments and references see 
\cite{bartels}).

The unitarity limit for the cross section of a spatially-small colourless 
wave packet with a hadron target (see section(8) of \cite{fks1}), 
and especially the
theoretical analysis of the amount of diffraction in the gluon channel 
predicted by QCD \cite{fs}, show that the restriction to the leading power 
of $1/Q^2$ should break down for small enough $x$, possibly within 
the kinematics of HERA. Moreover, it follows from the
application of Abramovsky-Gribov-Kancheli \cite{agk} cutting rules
that accounting for the next power in $1/Q^2$ (as in \cite{bartels}) 
can lead at most to a $25 \%$ reduction of the leading twist result
without introducing negative cross sections \cite{fks1}.
Thus, it appears that the decomposition into leading and higher twists
becomes ineffective in the kinematical region which can be achieved at
the next generation of proton accelerators (LHC) or maybe even at 
the edge of the kinematics of HERA. 

The use of the optical theorem for the scattering of small size wave packets
off a hadron target makes it possible to deduce a limit
(which is an analog of the Froissart limit for hadron-hadron scattering)
for the amplitude in DIS and to calculate the boundary for
the applicability of perturbative QCD in small $x$ region \cite{fks1}.
It was found that for $x \approx 10^{-4}$, the boundary 
is $Q^2 \lsim 10 $~GeV$^2$.
This estimate suggests a significant contribution from higher twist effects
in the kinematics of HERA for $x \lsim 10^{-3}$.
A more general aim of this paper is to visualize this problem and to evaluate 
structure functions of DIS at very small $x$. 
We show that many features of the very small $x$ behaviour
of structure functions can be understood in terms of the geometry
of the spacetime evolution of high energy QCD processes.

In the black limit approximation, Gribov \cite{gribov} deduced the 
following formulae for the unitarity limit for structure 
functions of DIS\footnote{Gribov considered scattering off a heavy nuclei 
for which the black body limit appears more natural than for a nucleon. 
However, provided one assumes the black limit, 
Gribov's arguments, and hence the formulae, will hold. For a recent discussion 
of the black body limit in QCD for DIS off heavy nuclei 
see \cite{fs,mue} and references.}:
\begin{equation}
F_{T}=\frac{2 \pi r_N^2 Q^2 }{12 \pi^3} \int_{0}^{\delta s}
\frac{M^2 dM^2}
{(M^2+Q^2)^2} \rho(M^2) 
\, . 
\label{black}
\end{equation}
Here $r_{N}$ is the radius of the nucleon and $\rho$ is the 
normalized spectrum of produced hadronic masses
: $\rho(M^2)=\sigma(e^{+} e^{-} \rightarrow \mbox{hadrons})/ 
\sigma( e^{+} e^{-} \rightarrow \mu^{+} \mu^{-})$. 
The upper limit on the $M^2$-integral, which imposes the 
experimentally-observed sharp diffractive peak: 
$-t_{min} B_D \approx \frac{(M^2 + Q^2)^2 m_N^2}{s^2} \frac{r^2_N}{3} 
\ll 1$, leads to a generic logarithmic energy dependence ($B_D $~ 
is the usual diffractive slope parameter, $m_N$ is the nucleon mass).
Strictly speaking this formula is valid for $M^2 \ll s$, or 
 $\delta \ll 1$, satisfying the condition that the interaction 
for a hadronic system of mass $M$ is close to the unitarity limit.
A similar formulae has been obtained for $F_L$ \cite{gribov} 
and for the gluon distribution \cite{fs}. 

It is reasonable to ask if, and if so at which $x$, 
the black limit will begin to be approached in Deep Inelastic Scattering of 
a virtual photon off a proton.
In other words, at which point does the cross section for the scattering 
of a small colour dipole with the proton target move away from 
being transparent 
(due to colour screening) and start to blacken to its geometrical limit 
($\sigma_{tot} = \sigma_{el} + \sigma_{inel} = 2 \sigma_{inel} = 
2 \pi (r_N + b)^2 \approx 40-50 $ mb) ?
This question is especially acute for $F_{L}$ and 
$ \partial F_2 / \partial \ln Q^2$ where the
interaction of spatially small configurations in the wavefunction 
of the photon dominate.
We aim to address this question in a phenomenological fashion
in this paper and to generalize Gribov's unitarity limit to QCD
by accounting for QCD phenomena which are neglected within
the black body limit
(see section(6)).

This subject has a rich and long history. For the 
high energy
scattering of a hadron from a nuclear target many configurations 
in the wave function of the fast hadron 
contribute and it is 
convenient to characterize the interaction by a distribution, 
$P(\sigma)$, of scattering probabilities, $\sigma$, of its constituent states
instead of by the average value of $\sigma$ (this useful realization
pre-dates QCD, see \cite{gw}). The qualitative idea of 
two-gluon exchange as the mediator was suggested by both Low 
and Nussinov \cite{low,nussinov}. Low \cite{low} 
also observed that the dipole cross section should be proportional to the 
transverse area of the object. 
Miettenen and Pumplin \cite{mietp} later suggested that 
scattering eigenstates should be identified with partonic configurations in 
the scattering systems, implying that the scattering cross sections of
particular states should be related to parton densities in the
opposing hadrons. In the modern context, for DIS at small $x$ and for 
sufficiently small $b^2$, 
$\sigma \propto b^2 \alpha_s xg(x,b^2)$ and $P(\sigma)$ follows unambiguously 
from the QCD factorization theorem for hard 
exclusive 
processes \cite{cfs}.
For large $b^2$, our approach is in many respects similar to the 
aligned jet model of \cite{bjk} or QCD-improved aligned jet model of 
\cite{fsaj} where the cross section of small $x$ processes in the 
non-perturbative regime is expressed in terms of universal dipole
cross section at small $x$, which is matched to the soft meson-nucleon
cross section (the similarity holds despite the difference in the 
source of $q\bar q$ pairs).

The paper is organized as follows. 
In section(2) we discuss the structure function $F_L$ in $b$-space, 
introducing a toy model for the DCS for illustrative purposes.
Section(3) sets out our realistic ansatz
for the DCS in detail, section(4) 
compares and contrasts it to other models and ideas in the literature. 
We make some specific, reasonable choices concerning
some of the uncertainties involved in specifying the DCS.
These choices are necessary in order to make quantitative
statements. However, we have also analyzed the precise form of the DCS
in detail numerically and investigated the sensitivity of our results 
to different choices. This analysis will be presented
in a separate paper \cite{fgms}. However, at this point we merely
state that the qualitative statements that we make about unitarity 
and the influence of large dipoles in $F_2$ at small $x$ are robust
with respect to changes in the details. 
We discuss this in more detail in section(6), where we also consider 
the kinematic region in which the Gribov's black limit may be 
reached for scattering at central impact parameters
in DIS for some configurations in the photon wave function.
We point out that certain diffractive processes, for example exclusive 
photoproduction of $J/\psi$, may act as useful precursors 
to the onset of this new QCD regime. We conclude in section(7).

\section{Basic Formulae and a toy ansatz for the dipole cross section}

In this section we examine the cross section $\sigma_{L} (x,Q^2)$ in 
$b$-space using a 
very simple toy model for the extrapolation of the DCS to large $b$.
Our aim is to familiarize the reader with the $b$-space formulation of 
structure functions. For clarity of presentation we employ a very
simple and unphysical ansatz for the DCS at large $b$. 
We will improve on this toy ansatz in the next section.

It is convenient to use the impact factor ($b$-space) representation first 
introduced by Cheng and Wu in considering high energy processes in QED.
In $b$-space the longitudinal structure function, $F_L(x,Q^2)$, may be written
\cite{chengwu,nz} in terms of the DCS convoluted with the light-cone 
wavefunction 
of the virtual photon squared:
\begin{equation}
\sigma_L (x,Q^2) = 2 \, \int_0^{1/2} dz \int d^2 b \, 
{\hat \sigma}(b^2) \, |\psi_{\gamma,L} (z,b)|^2 \, ,
\label{eqsigl}
\end{equation}
\noindent where
\begin{equation}
|\psi_{L}(z,b)|^{2} = \frac{6}{\pi^{2}}\alpha_{e.m.} 
\sum_{q=1}^{n_{f}}e_{q}^
{2}Q^{2}z^{2}(1-z)^{2} K_{0}^{2}(\epsilon b) \, ,
\label{eqpsil}
\end{equation}
\noindent in which $\epsilon^2 = Q^2 (z(1-z)) + m_q^2$ and for now we set 
the light quark mass, $m_q$ to zero. 

For small dipoles, the DCS is governed by perturbative QCD
 \cite{frs,rad} (for an explicit derivation see \cite{rad}):
\begin{equation}
{\hat \sigma}_{pqcd} (b^2,x) = \frac{\pi^2}{3} \, b^2 \, \alpha_s 
({\bar Q^2}) \, x g(x,{\bar Q^2}) \, .
\label{eqshat}
\end{equation}
\noindent We employ a phenomenological scaling ansatz
${\bar Q^2} = \lambda/b^2$ 
to relate transverse sizes to four-momentum scales 
(it is possible to prove that this ansatz is a property of the Fourier 
transform in the LO and NLO approximations but not beyond). 
We also implicitly assume that the DCS
is independent of light-cone momentum sharing variable $z$. 
This is a good approximation for the longitudinal case because the average
$z \sim 1/2$ dominate in the integral and due to the $z \to 1-z $ 
symmetry of the wave function. For the transverse case the end points 
give a larger contribution and hence this assumption is less justified.

The relationship of eqs.(\ref{eqsigl},\ref{eqpsil},\ref{eqshat}) holds to 
leading-log accuracy in $Q^2$ 
(so, for consistency one is forced to use only LO partons and 
$\alpha_s$ at one loop) and involves taking the imaginary part of the 
usual box and crossed box 
graphs. As such this form corresponds only to the dominant inelastic piece of
the DCS. We immediately see a practical problem using
eq.(\ref{eqshat}) in the $b-$integral of eq.(\ref{eqsigl}).
There are always regions in the integral, at large $b$, where the
gluon density is not defined 
and we need to decide what to do. In particular, for fixed $\lambda$, 
the gluon density is not defined for $b^2 > b_{Q0}^2 = \lambda /
Q_0^2$. In the usual treatment this contribution is absorbed into the 
initial condition of the evolution equations.

\begin{figure}[htbp]
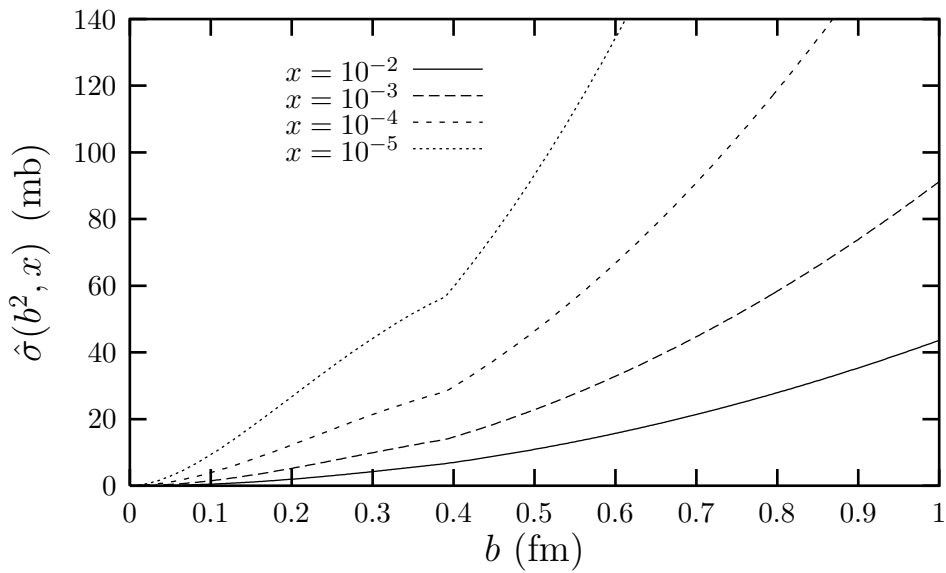

  \begin{center}
     \include{sigl}
    \caption{Dipole cross section in mb for fixed $\lambda = 10$, 
with the toy model ansatz at large $b$. 
For an input scale of $Q_0 = 1.6 $ GeV, $b_{Q0} = 0.39 $ fm marks the 
boundary of the perturbative region.}
\label{fig:1}
\end{center}
\end{figure}

To get started we fix $\lambda = 10$ and simply freeze 
$\alpha_s ({\bar Q^2}) ~x g(x,{\bar Q^2}) $
at its value at $Q_0^2$ for $b^2 > b^2_{Q0}$ in eq.(\ref{eqshat}) 
(we refer to this as ansatz 1). 
This means that the DCS retains the canonical $b^2$-behaviour at large
$b$ in eq.(\ref{eqshat}), and its derivative 
is discontinuous at $b = b_{Q0}$. Fig.(\ref{fig:1}) shows a plot of 
the resultant DCS as a function of $b$ 
for several values of $x$.

\begin{figure}[htbp]
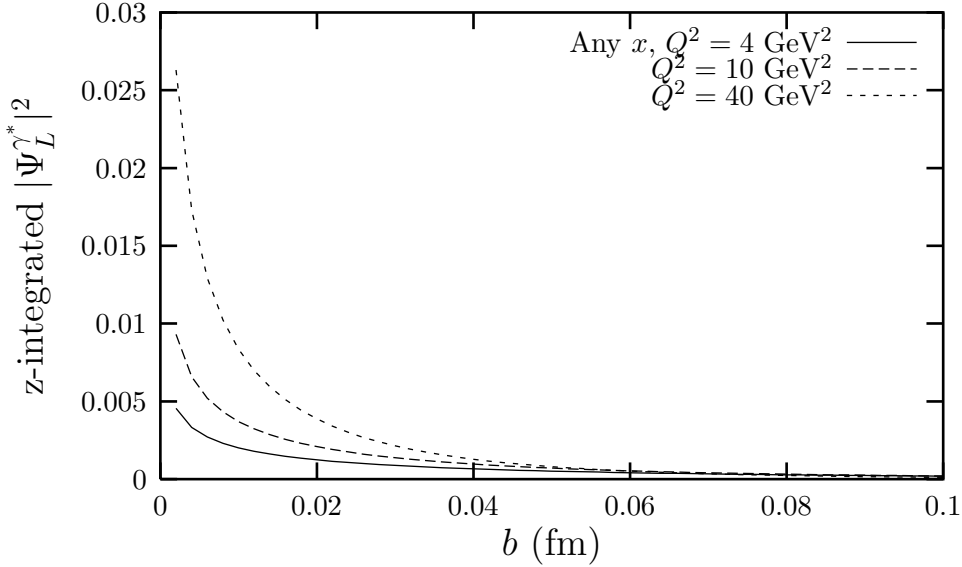

  \begin{center}
     \include{psisq}
    \caption{Longitudinal photon wavefunction squared, integrated over 
z (in units of fm$^{-2}$), for $Q^2=4,10,40$ GeV$^2$.}
    \label{fig:2}
  \end{center}
\end{figure}

We may now examine the dominant regions of the $b$-integral in 
eq.(\ref{eqsigl}):

\begin{equation}
\sigma_L (x,Q^2) = \int_0^{\infty} d b \, I_L (b, x, Q^2) \, ,
\end{equation}
\noindent where,  
\begin{equation}
I_L (b) = 2
 \pi \, b \, {\hat \sigma}(b^2) I_{\gamma, L} \, ,
\end{equation}
\noindent in which $I_{\gamma, L}$ is the integral of 
$|\psi_{L}(z,b)|^{2}$ over $z$. Fig.(\ref{fig:2}) shows 
$I_{\gamma, L}$ as a function of $b$ for three light flavours at
fixed values of $Q^2$, it diverges at small values of $b$ due to the 
logarithmic divergence of the $K_0$ Bessel function at small values of
its argument.

\begin{figure}[htbp]
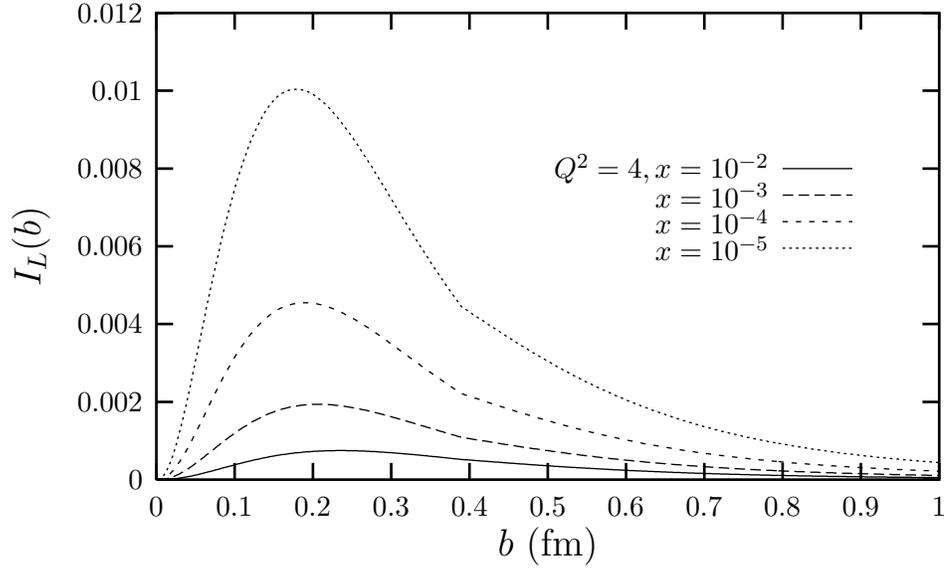

  \begin{center}
     \include{int4l}
    \caption{Integrand of $\sigma_L$, in units of fm,
 for fixed $\lambda = 10$, $Q^2=4$ ~GeV$^2$.}
    \label{fig:3a}
  \end{center}
\end{figure}

\begin{figure}[htbp]
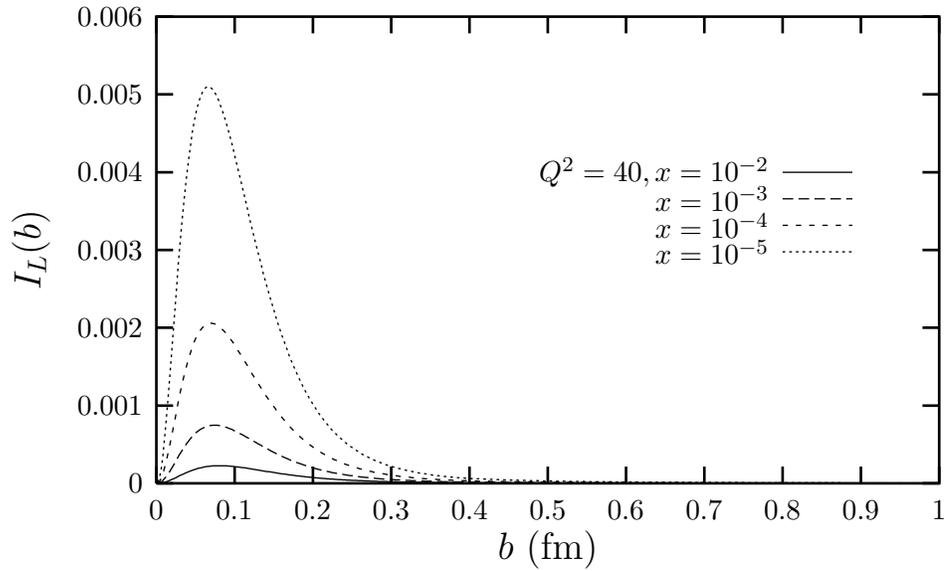

  \begin{center}
     \include{int40l}
    \caption{Integrand of $\sigma_L$, in units of fm, 
for fixed $\lambda = 10$, $Q^2=40$ GeV$^2$.}
    \label{fig:3c}
  \end{center}
\end{figure}

Figs.(\ref{fig:3a},\ref{fig:3c}) show the integrand, $I_L (b)$, for 
two characteristic values of
 $Q^2 = 4, 40$ GeV$^2$ using CTEQ4L gluons \cite{cteq4}. Note in 
fig.(\ref{fig:3a}) one can clearly see that the region 
above $b_{Q0} \approx \frac{\sqrt{10}}{1.6} * 0.197 \approx 0.4$ fm 
(where there is a kink due to the freezing of ansatz 1) contributes 
significantly to the whole integral. 
In contrast, fig.(\ref{fig:3c}) shows that for $Q^2 =40$ GeV$^2$ 
(typical effective scale for $\Upsilon$ photoproduction 
\cite{fms}) this region is completely irrelevant.
This is due to the fact that the photon piece of the integrand 
$I_{\gamma, L}$, multiplying the DCS, 
strongly weights the integrand to progressively smaller $b$ as 
$Q^2$ increases (see fig.(\ref{fig:2})).

As an example, let us focus on the case when $Q^2 = 40$ \,GeV$^2$ and 
$x = 10^{-3}$. 
The $b$-integrand exhibits a strong, slightly asymmetric peak around small 
$ b = b_{\mbox{peak}} \approx 0.08 $ fm, which is slightly skewed to 
larger $b$.
The relationship ${\bar Q^2} = \lambda/b^2$, with $\lambda = 10$, 
implies $Q_{\mbox{peak}}^2 \approx 60 $ \,GeV$^2$, 
and that the typical $b = b_{\mbox{typ}} \approx 0.1$ fm corresponds 
to $Q_{ \mbox{typ}}^2 = 40 $ \,GeV$^2$. 
Clearly, for $b < b_{\mbox{typ}}$ the effective scale will be larger 
than $Q^2$. 
The fact that there is very little contribution from the large
b-region, for large $Q^2$, illustrates 
QCD factorization in $b$-space: the sharply peaked photon piece of the 
integrand ensures 
that only small dipoles contribute significantly in the integral.

We will refer to the structure function $F_L(x,Q^2)$ calculated in 
$b$-space as $F_L^b$. 
It is related to the defined cross section in the following simple way:
\begin{eqnarray}
F_L (x,Q^2) = F_L^b(x,Q^2,\lambda) = 
\frac{Q^2}{4 \pi^2 \alpha_{e.m.}} 
\sigma_L (x,Q^2,\lambda)
\end{eqnarray}
\noindent where we have chosen to write the dependence on $\lambda$ explicitly.
Why have we chosen $\lambda = 10$ ? Lambda should reflect the typical
size of contributing dipoles. 
It may be calculated by defining an average $b$ in the integrand for $F_L$ 
(in \cite{fks1,fks2,fms} a median average of the integral was used). 
Whatever the precise procedure used for defining the average, a value
of $\lambda \approx 10$ comes out for 
large enough photon virtuality $Q^2$. Roughly speaking 
$\lambda = <\!b\!>^2 Q^2$, so that when $b = <\!b\!>$ 
the gluon and $\alpha_s$ are sampled at $Q^2$ in eq.(\ref{eqshat}) as
in the usual leading-log 
in $Q^2$ perturbative QCD formula for $F_{L}$ (see eq.(\ref{eqflq}) later).

\begin{figure}[htbp]
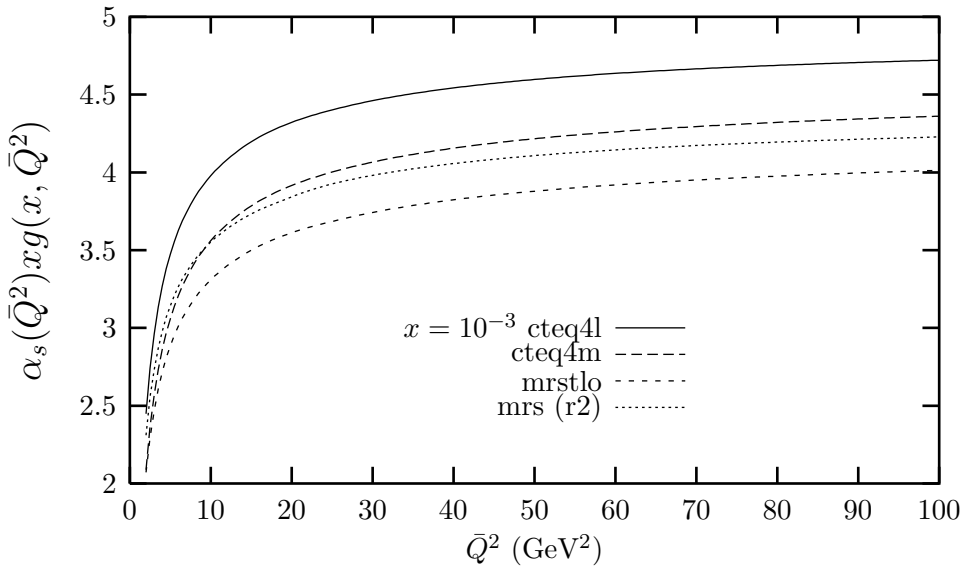

  \begin{center}
  \include{asg3}
    \caption{The function $f({\bar Q^2}) = \alpha_s x g$ at fixed $x$ 
for various parton sets}
    \label{fig:4}
  \end{center}
\end{figure}

In fact $F_L^b$ has a rather weak dependence on $\lambda$ for large
$Q^2$, this fact reflects the 
renormalization group invariance of QCD. We illustrate this in 
fig.(\ref{fig:4}) by plotting $\alpha_s xg$ 
as a function of its argument at fixed $x = 10^{-3}$ for several LO
and NLO parton sets \cite{mrst,cteq4}.
A similar behaviour is observed at other small values of $x<10^{-2}$.
For large ${\bar Q^2}$ it is rather a weak function, and our 
$\lambda$-ansatz translates this into 
a weak dependence on $b$. This in turn implies that the canonical 
${\hat \sigma}_{pqcd} \propto b^2 ~\alpha_s(\lambda^2/b^2) ~xg(x,b^2) 
\propto b^2$ holds approximately for small enough dipoles. For 
$x \lsim 10^{-3}$, $\alpha_s xg$ has positive scaling violations 
implying an effective behaviour slightly softer than $b^2$ from QCD at 
sufficiently small $x$
(the effective ${\bar Q^2}$-power, $\gamma (x,{\bar Q}^2) \lsim 0.1$ 
in fig.(\ref{fig:4}), for large ${\bar Q}^2 \gsim 20$~GeV$^2$).
The peak in $b$ evident in fig.(\ref{fig:3c}) implies that $\alpha_s
xg$ is sampled dominantly only in a small range of ${\bar Q}^2$
in fig.(\ref{fig:4}). For
small dipoles a reasonable change in $\lambda$ corresponds to a shift of this 
dominant region within the fairly flat part of the curve. 
Hence, as we have checked explicitly, for large $Q^2$ $F_L$ is 
insensitive to the precise choice of $\lambda$.

However, for smaller scales $\alpha_s xg$ has a much stronger 
dependence on ${\bar {Q^2}}$
which tames this linear dependence on $b^2$ into something much softer. This 
fact is apparent in the shape of the curves in 
fig.(\ref{fig:1}), for dipole sizes corresponding to the region of 
$Q^2_0 < {\bar Q^2} \leq 10 $ \,GeV$^2$, 
${\hat{\sigma}}_{pqcd}$ deviates considerably from the $b^2$ behaviour
apparent for small ($b \ll b_{Q0}$) and large 
($b > b_{Q0}$) dipoles in our toy ansatz. In \cite{fgms}, 
we investigate this question of interrelation of these scales in more detail.
Finally, we note in passing that $\alpha_s xg$, plotted in fig.(\ref{fig:4}), 
has a comparable numerical value and shape for both LO and NLO gluons. 
Since the dipole cross section is 
proportional to this quantity for small $b$, all statements
that we make about the size of the DCS using LO gluon densities also 
hold at NLO.

\section{A realistic ansatz for the DCS}

At small $x$, because the gluon density dominates over the quark
density, to a good approximation the
LO perturbative QCD formula for $F_{L} (x,Q^2)$ which we denote 
generically as $F_{L}^q$ is:

\begin{equation}
F_{L} (x,Q^2) = \frac{4 \alpha_{s} (Q^2) T_R } {2 \pi} 
\Sigma_{q = 1}^{2 nf} e_q^2 
 \int_x^1 dx' x'g (x',Q^2) \, \frac{x^2}{x^{'3}} (1 - \frac{x}{x'} ) \, .
\label{eqflq}
\end{equation}

\begin{figure}[htbp]
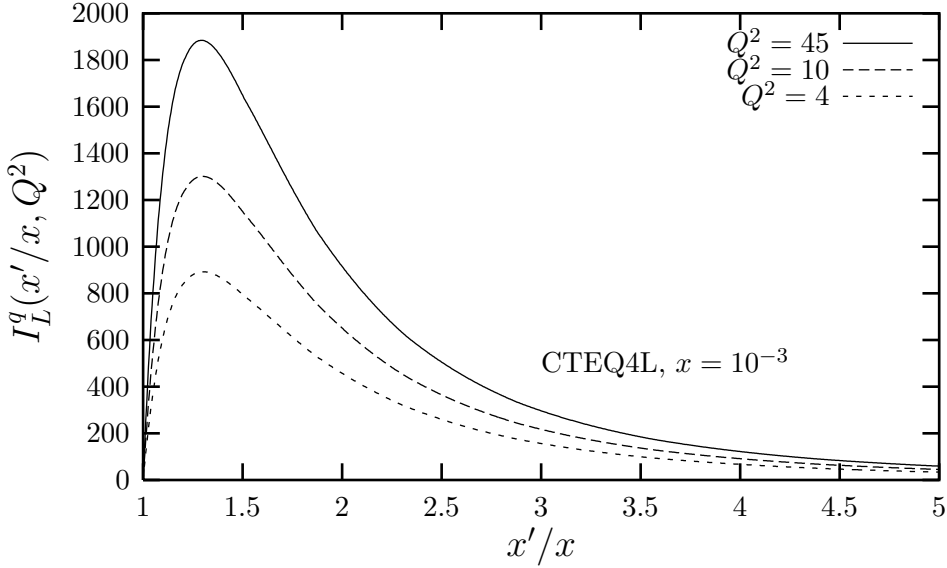

  \begin{center}
     \include{flq}
    \caption{Integrand of $F_L^q$ versus $x^{'}/x$ for various $Q^2$ 
    using CTEQ4L gluons.}
    \label{fig:5}
  \end{center}
\end{figure}

\noindent This conventional expression for $F_L (x,Q^2)$ involves an 
integral over the gluon momentum fraction 
$x^{'}$, of the proton's momentum $p$, which feeds into the quark box. 
In fig.(\ref{fig:5}) we plot the integrand of the above formula
versus $x'/x$, at $x = 10^{-3}$, for various $Q^2$ values. 
The gluon is sampled at a range of values of $x^{'} > x$ with the 
integrand, $I_L^q (x'/x)$, 
peaked around $x'_{peak} \approx 1.3 \, x$, and skewed 
to $x' > x_{peak}$ due to 
the factor multiplying the rising gluon density. We define an average, 
$<\!x\!>$, to be 
that $x'$ up to which one must integrate to obtain half of the full integral: 
it turns out this is always around $x^{'} = <\!x\!> \approx 1.75 ~x$ 
for a wide range of external $x,Q^2$.

Momentum conservation for the fusion of a gluon with a photon, 
of momentum $q$, to produce the quark-antiquark pair, of mass 
$M^2_{q {\bar q}}$, gives
\begin{eqnarray}
(x'p + q)^2 = M^2_{q {\bar q}} & = & \frac{k_t^2 + m_q^2}{z(1-z)} 
\nonumber \\ 
& \ge & 4m_q^2 + \frac{k_t^2}{z(1-z)} \propto \frac{1}{b^2} \, .
\label{eqxmom}
\end{eqnarray}

\noindent The inequality in the second line is satisfied for a non-zero quark 
mass, $m_q$, when $z = 0.5$. In fact the approximation $1/(z(1-z)) \approx 4$ 
holds over a reasonable wide range of $z$ values.
So, taking finite quark masses into account implies a minimum value for 
$x^{'}$ of $x_{min}^{'} = x ~(1 + 4 m_q^2/Q^2) $. 
To account for the important role at large $b$ of confinement and 
spontaneously broken chiral symmetry we choose constituent quark masses. 
With this in mind, for our new ansatz we choose to sample the 
gluon density in eq.(\ref{eqshat}) at

\begin{equation}
x' = x_{min}^{'} (1 + 0.75 \frac{<\!b\!>^2}{b^2}).
\label{eqxprim}
\end{equation}

\noindent This choice guarantees that for the average $b^2$, $x' =<\!x\!>$, 
but allows $x^{'} (b^2)$ to vary according to the inverse of the transverse 
size of the dipoles. Kinematically, a large mass dipole requires a gluon 
carrying a greater than average momentum fraction to produce it
and the ansatz of eq.(\ref{eqxprim}) is designed to reflect this. 
In contrast, for very large, small mass dipoles
$M_{q {\bar q}}^2 \ll Q^2$, and the formula gives $x' \approx x_{min}^{'} $, 
which is approximately $x$ only for light quarks. 
The average $b$ is $<\!b\!>^2 = \lambda / (Q^2 + 4 m_{q}^2)$, in agreement 
with the definition of $\lambda = <\!b\!>^2 Q^2 $ in the DIS region, 
$Q^2 \gg 4 m_q^2$, but is infra-red safe in the photoproduction region 
$Q^2 \lsim 4 m_q^2$.

We also have a constraint on the DCS at large $b$ from the
experimental determination 
of the pion-proton cross section, ${\hat \sigma}_{\pi, N} = 23.78$ mb 
\cite{pipdat}. The DCS should be close to this at 
transverse separations which correspond to the diameter of the pion 
($d_{\pi} \approx 0.65 $ \,fm). For rather large 
$x \approx 10^{-2}$ the magnitude
of the DCS at the interface to the non-perturbative region is
considerably smaller 
(${\hat \sigma}_{pqcd} (x = 10^{-2},b^2_{Q0}) \approx 6 $ mb) than this.

The intermediate region $ b_{Q0} < b < b_{\pi} $ is extremely
interesting and very poorly 
understood, so some modeling is required. 
It is in this region that strong confinement and effects of 
spontaneously broken chiral symmetry
set in to produce the bound state pion. Clearly the dynamics in this 
region will include strong colour 
fields and the creation of light sea pairs from the vacuum. 
As such, it is no longer reasonable to think of $b$ as corresponding 
to the transverse 
size of a dipole. It is better to think of it as corresponding 
to the typical transverse radius of the complicated non-perturbative system, 
which in general will contain many constituents.

The minimum requirement of a interpolating function, ${\hat \sigma}_I$
for the DCS is that it matches appropriately at 
$b=b_{Q0}$ and $b=b_{\pi}$: 
\begin{equation}
{\hat \sigma}_I (x,b^2) = [ \sigma_{\pi p}(x,b^2_{\pi}) - 
{\hat \sigma}_{\mbox{pqcd}} (b^2_{Q0}) ] H(b^2) + 
{\hat \sigma}_{\mbox{pqcd}} (b^2_{Q0}) \, ,
\label{eqbint}
\end{equation}
\noindent with $H(b^2_{Q0}) = 0$ and $H(b^2_{\pi}) = 1$.
On geometrical grounds, we also choose to only consider functions which are 
monotonically increasing as a function of $b$. A very simple function which 
satisfies these requirements is
\begin{equation}
H_1 (b^2) = \frac{ (b^2 - b^2_{Q0}) }{ (b^2_{\pi} - b^2_{Q0})} \, ,
\end{equation}
\noindent which has a linear growth in $b^2$, even for $b \approx
b_{\pi}$. To impose 
a flatter behaviour in this region and a fairly smooth matching close to 
$b \approx b_{Q0}$ we choose the following exponential matching
\begin{equation}
H(b^2) = \frac{e}{(e-1)} [ 1 - \mbox{exp}(-H_1(b^2))] \, ,
\label{eqh}
\end{equation}
\noindent which retains the linear growth in $b^2$ close to $b_{Q0}$.

The pion-proton cross sections is observed to rise slowly as the 
energy increases. 
In order to take this into account we impose a slow growth with 
increasing energy, 
consistent with a Donnachie-Landshoff soft Pomeron \cite{dl1}, in our 
boundary condition at $b=b_{\pi}$:
\begin{equation} 
\sigma_{\pi, N}(b_{\pi},x) = 23.78 \left(\frac{x_0}{x} \right)^{0.08} 
\mbox{mb} \, ,
\label{eqbpi}
\end{equation}

\noindent and choose $x_0 = 0.01$. This behaviour in $x$ is designed to mimic 
the observed $(W^2/W_0^2)^{0.08}$ 
behaviour in energy. The precise value of $23.78$ mb is taken from the 
 Fermilab data \cite{pipdat} from 
pion-proton scattering and corresponds to $W_0^2 = 400 $~GeV$^2$.

As $x$ decreases the magnitude of the DCS for small $b^2$ increases
much more rapidly with energy 
than this soft piece, as a result of the steeply rising gluon density. 
For leading-log gluons at small enough $x$, if unchecked, it can even 
become greater 
than $\sigma_{\pi, N} (b_{\pi}, x) $ for perturbative $b <
b_{Q0}$. This is clearly nonsensical, 
some taming of this rapid growth must occur before this can happen.
Since the gluon density form of eq.(\ref{eqshat}) really represents only the 
inelastic part of the dipole cross section, as it becomes 
comparable to $\sigma_{\pi, N}$ we need 
to include the elastic part of the DCS too, which hitherto was 
implicitly assumed to be negligible. 
In the limit of very large energy the Froissart bound indicates 
that the elastic piece should not exceed the inelastic piece. On this 
basis we really should 
worry about the applicability of our perturbative QCD formula when the
DCS is about $50 \%$ of 
the pion-proton cross section. 
The absolute upper bound for the inelastic small dipole-nucleon 
interaction is $8 \pi B_{2g}/(1 + \eta^2)$ where 
$B_{2g} \sim 4-5 $~GeV$^{-2}$ is the 
slope of the $t$-dependence of the two-gluon form factor of the nucleon, 
as measured in hard exclusive diffractive processes
at HERA, and $\eta$ is the ratio of real and imaginary parts 
of the scattering amplitude \cite{fks1,fs}.
This bound is slightly weaker than the transition point which we assume, but
pushing all the way to the absolute limit appears unrealistic.

In our computer code we test ${\hat \sigma}_{pqcd} ~(x,b^2)$ to see if
the equality is reached in 
the perturbative region $b = b_{crit} < b_{Q0}$, where $ b_{crit}$ is 
defined implicitly by
\begin{eqnarray}
{\hat \sigma} (x,b^2_{crit}) &= \frac{\pi^2 b^2_{crit}}{3} 
\alpha_s (Q^2_{crit}) x' g (x', Q^2_{crit}) 
&= \frac{\sigma(x,b^2_{\pi})}{2} 
\label{eqbcrit}
\end{eqnarray}
\noindent with $Q^2_{crit} = \lambda/b^2_{crit}$ and $x'$ is given 
by eq.(\ref{eqxprim}). If so, we 
use a new interpolation in the region $b_{crit} < b < b_{\pi}$:

\begin{equation}  
{\hat \sigma}_I (b^2,x) = \left( \frac{b^2}{b^2 + a^2} 
\right)^n \sigma_{0} \, .
\label{eqbint2}
\end{equation}  
\noindent Matching at $b=b_{\pi}$ sets the value of 
\begin{equation}  
\sigma_{0} (x) = \sigma_{\pi, N} (b_\pi,x) 
(\frac{b^2_{\pi} + a^2}{b^2_{\pi}})^{ n}.
\label{eqsig0}
\end{equation}  
\noindent The two remaining parameters, $a,n$, are chosen to provide a
 fairly smooth matching at 
$b = b_{crit}$. To achieve this we perform a three parameter fit of 
exactly the same 
form as eq.(\ref{eqbint2}) 
for a given $x$ in the region just below $b_{crit}$, using 
MINUIT \cite{minuit}. 
We then take the effective power, $n_{fit}$, from this fit,
 so that the interpolating ansatz of eq.(\ref{eqbint2}) has approximately 
the correct power in $b^2$ at the boundary.

The last remaining free parameter, the scale $a$, is then specified by
the matching at $b = b_{crit}$:
\begin{equation}
a^2 = \frac{b^2_{crit} ( 1 - (0.5)^{n_{fit}} )}
{ (\frac{b^2_{crit}}{b^2_{\pi}} (0.5)^{n_{fit}} -1 ) } \, .
\label{eqasq}
\end{equation}

\noindent This ensures a fairly smooth behaviour in $b^2$ which takes 
into account the effective behaviour of 
$\alpha_s ({\bar Q^2}) x g(x,{\bar Q^2}) $ close to $b_{crit}$ as
discussed earlier (see fig.(\ref{fig:4})).

For very large dipole sizes, $b>b_{\pi}$, we simply impose a
universal residual slow growth, linked to the value at $b=b_{\pi}$ of the form
\begin{eqnarray}  
{\hat \sigma}_{I} (b^2 > b^2_{\pi}) = \sigma(b^2_{\pi},x) 
\frac{1.5 ~b^2}{(b^2 + b^2_{\pi}/2 )}.
\label{eqbigb}
\end{eqnarray}  

\noindent Numerically this very large $b$ region is totally irrelevant for the 
calculation of DIS structure functions since it is killed by the 
exponential fall-off of the photon part of the integrand due to 
the $K_0, K_1$ Bessel functions (in practice, for moderate $Q^2$ 
we integrate up to $b = 1.0$ fm, for smaller 
$Q^2 < 3.0$ ~GeV$^2$ we extend this out to $2.0$ fm).

Let us briefly summarize our realistic ansatz. Assuming the universal 
scaling relation $\lambda = b^2 {\bar Q^2}$
restricts the region of applicability of any perturbative QCD dipole 
formula to transverse sizes smaller than $b_{Q0}$, 
which corresponds to the input scale of parton densities in $b$-space. 
For small, but not too small $x$, we use the perturbative QCD formula
of eq.(\ref{eqshat}) with the gluon sampled at $x^{'}$ 
(see eq.(\ref{eqxprim})) in the region $0 < b < b_{Q0}$. Between 
$b_{Q0} \approx 0.4$ fm and $b_\pi = 0.65$~fm
we use the interpolating formula of eq.(\ref{eqbint}), employing the 
exponential matching of eq.(\ref{eqh}).
For smaller $x$, we recognize that ${\hat \sigma}_{pqcd}$ gets too
large within the perturbative region $b<b_{Q0}$ 
at some point, $b_{crit}$, defined by eq.(\ref{eqbcrit}). At this
point, we switch from the standard form and use 
the interpolating form of
eqs.(\ref{eqbint2},\ref{eqsig0},\ref{eqasq}), 
which tames the rapid growth and interpolates in the 
region $b_{crit} < b < b_{\pi}$.

\begin{figure}[htbp]
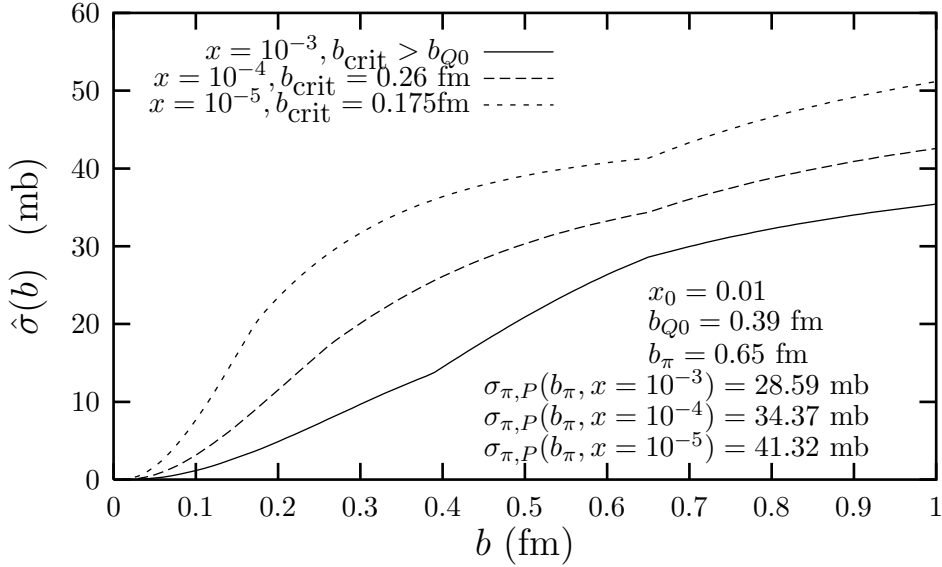

  \begin{center}
     \include{sigx}
    \caption{Dipole cross section in mb for fixed 
$\lambda = 10$, with the realistic ansatz at large $b$. For small 
enough $x$ unitarity corrections are included.}
    \label{fig:6}
  \end{center}
\end{figure}

In both cases, our ansatz matches onto the pion-proton cross section 
of eq.(\ref{eqbpi}), at 
$b = b_{\pi} = 0.65$ ~fm, which is allowed a slow Donnachie-Landshoff 
type energy growth. 
For $b > b_{\pi}$ we use the slow increase given in eq.(\ref{eqbigb}). 
In fig.(\ref{fig:6}) we show our new ansatz for the DCS as a function of $x$. 
Note, we use the ansatz for both heavy and light flavours. 
The NA38 collaboration \cite{psip} recently suggested 
$\sigma _{\psi^{'} N} \approx 24 \pm 5 $ mb, on the basis 
of an observed deficit in the number of $\psi^{'}$ decays to dimuons 
in Sulphur-Uranium collisions relative to well established trends 
in proton-nucleus collisions.
The fact that the large $c{\bar c}(2S)$ bound state 
has a large interaction cross section with the
nucleon in these nuclear collisions, as predicted in \cite{gfssg}, 
is some justification for our flavour blind choice for the DCS in the 
non-perturbative region ($b \sim 0.6 $ fm).

\section{Comparison with other models for the dipole cross section}

\begin{figure}[htbp]
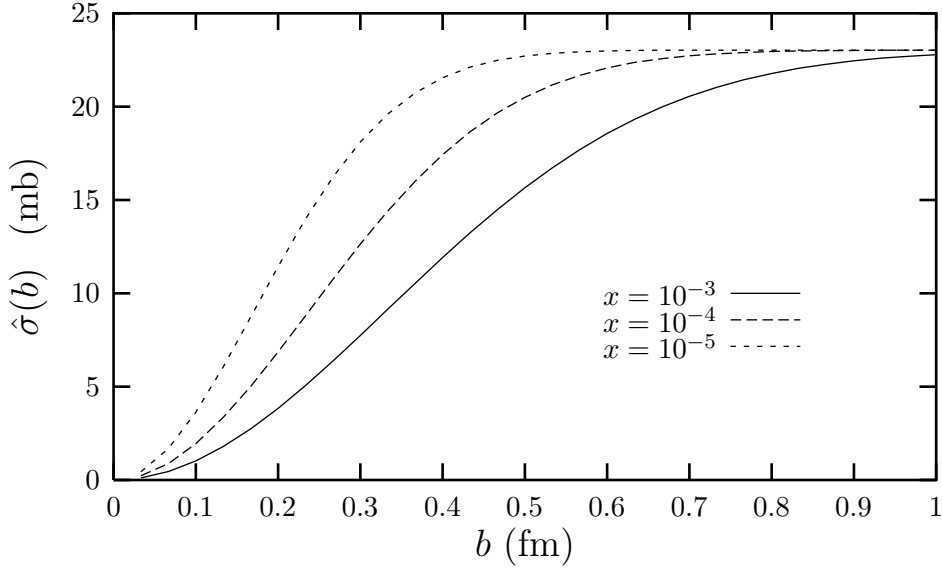

  \begin{center}
     \include{sigwgb}
    \caption{Dipole cross section in mb for 
Wusthoff-Golec-Biernat saturation model \cite{wkgb}.}
    \label{fig:wgb}
  \end{center}
\end{figure}

\begin{figure}[htbp]
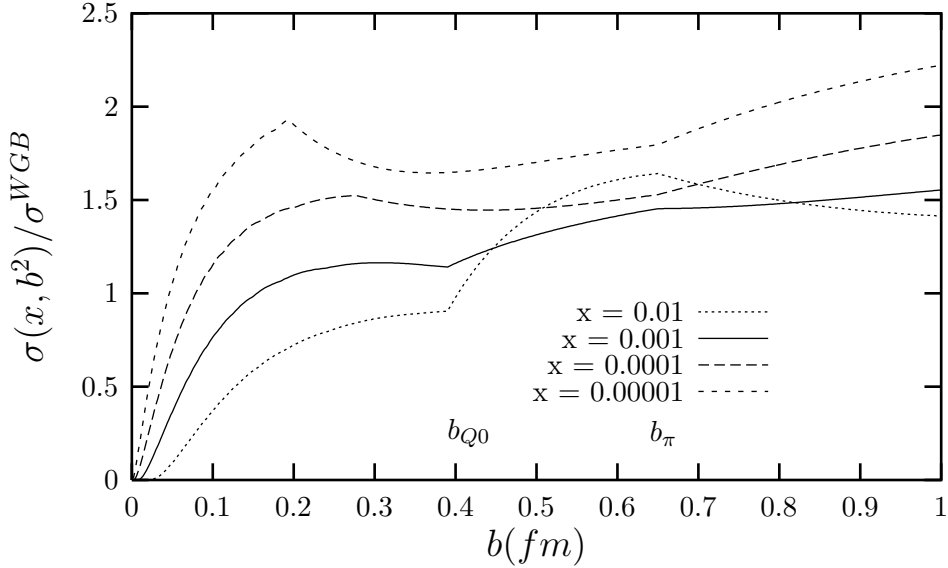

  \begin{center}
     \include{wgbcom}
    \caption{Ratio of our dipole cross section with the 
Wusthoff-Golec-Biernat saturation model \cite{wkgb}.}
    \label{fig:wgbcom}
  \end{center}
\end{figure}

The unitarity correction at small $x$, discussed above, is clearly beyond the 
usual DGLAP leading twist analysis and is similar in spirit to 
the saturating ansatz of W\"{u}sthoff and Golec-Biernat \cite{wkgb}:
\begin{equation}
{\hat \sigma} (x,b^2) = \sigma_0 ( 1 - \exp [-b^2 Q_0^2/ 4 (x/x_0)^\lambda ] ).
\label{eqwgb}
\end{equation}
\noindent A three parameter fit to the HERA data on DIS with $x < 0.01$, 
excluding charm and assuming $Q_0 = 1.0$ GeV, produced the following values 
$\sigma_0 = 23.03 $~mb, $x_0 = 0.0003$, $\lambda = 0.288$ and a 
reasonable $\chi^2$. 
We are encouraged to note that the ``saturation'' cross section coming 
from this fit is considerably
below the lowest value of the black limit 
($\sigma_{tot}^{\mbox{black}} = 2 \pi r_N^2 \approx 40$ mb).

The resultant DCS from eq.(\ref{eqwgb}) is plotted in 
fig.(\ref{fig:wgb}). A comparison of this figure with our model 
for $\sigma$ in fig.(\ref{fig:6}) is shown in fig.(\ref{fig:wgbcom})
which shows the ratio of our $\sigma$ divided by the Wusthoff Golec-Biernat 
form, $\sigma^{WGB}$. It reveals a considerably different $b$-shape, 
normalization and 
energy dependence. Focusing on exclusive processes, which 
are particularly sensitive to the small $b$ region in which the models 
differ most, will help to distinguish between them.
Having emphasized the contrasts, it is worth pointing out that
the two models share some gross features: 
an approximate $b^2$ behaviour and strong rise with $x$ at small $b$, tamed to 
something much softer at large $b$. 
Indeed, the critical point at which we apply our unitarity
corrections, $b_{crit}$, clearly shifts to smaller $b$ as $x$ decreases, as 
a result of the rising gluon density. This is similar to the critical 
line of \cite{wkgb}.

The form of eq.(\ref{eqwgb}) was clearly chosen with simplicity in mind
and is indeed impressively economical in its number of parameters.
We recognize this motivation and so do not wish to criticize it too strongly. 
However, we feel this simple form misses 
several known crucial features which our ansatz includes 
(leading to the differences manifest in fig.(\ref{fig:wgbcom})). 
Firstly, the small dipole form is known from perturbative QCD 
(having specified the relationship between $b$ and $Q^2$, 
see eq.(\ref{eqshat})). 
The fact that the gluon density may be taken from the global fits allows a 
careful study of the deviation from a simple $b^2$ behaviour 
inherent in eq.(\ref{eqwgb}). 
As a result the difference between our $\sigma$ and $\sigma^{WGB}$
for the perturbative region of $b\le 0.4$ ~fm is large and strongly energy 
dependent (cf. fig.(\ref{fig:wgbcom})).
Using the fitted gluon density also allows the precise behaviour in 
$x$ to be incorporated more 
correctly than a single global power implied by eq.(\ref{eqwgb}). 
Secondly, it is well established experimentally 
that soft dipole cross section increases slowly with energy. This may 
be modeled with a small power 
as in eq.(\ref{eqbpi}) or with a logarithm according to theoretical 
prejudice. However, eq.(\ref{eqwgb}) has a flat behaviour 
in energy for large dipoles, leading to a rather large and energy dependent 
difference for large b.
Thirdly, as discussed above, the region where unitarity has to be
included is known from the requirement of smallness of the elastic cross
section as compared to the total cross section.

The eikonal form assumed in \cite{wkgb}, i.e. eq.(\ref{eqwgb}), was inspired 
by an earlier work by Gotsman, Levin and Maor \cite{glm1} (see 
also \cite{glm2}) in which the hard cross section 
$\sigma_{pqcd}^{inel} \propto \alpha_s xg$ \cite{frs} 
explicitly appears in the eikonal
\footnote{However, we remind the reader that 
if $\sigma^{inel}_{pqcd}$ is large enough to require eikonalization 
then $\sigma^{el}_{pqcd}$ will also be so large that 
in the black limit one has
$\sigma^{tot} = \int d^2 \rho (2 - 2 \exp(-\sigma^{tot}_{pqcd} T(\rho)/2)) $ 
within the eikonal approximation.}.
At the same time the eikonal approximation has problems in accounting for
the generic properties of QCD. In particular, the eikonal approximation 
assumes conservation of bare particles 
in the wave function of the photon despite the fact that operators of 
Lorentz boosts do not commute with the operator of the number of
bare particles. 
A related problem is that the eikonal 
approximation neglects the energy lost 
by an energetic particle in the inelastic collisions.
Hence, the taming of the increase of parton distribution by this 
method strongly overestimates 
%underestimates
the energy released in inelastic collisions.

The recent paper of these authors with Naftali \cite{glmn}, appears to
be somewhat orthogonal to their earlier works \cite{glm1,glm2} in that
the problem of $\sigma^{inel}_{pqcd}$ getting too large is not addressed.
They discuss different regions in the mass of the scattering state, 
in contrast to dipole sizes (cf. eq.(\ref{black})).
Roughly speaking, large masses correspond to small dipoles 
(at least for the lowest $q \bar{q}$ Fock state) and are related to 
the unintegrated gluon density in this hard contribution. 
The small mass region is modeled using a Regge analysis. As we have discussed 
in the current paper (see also \cite{abram}), they also stressed 
the interplay of short and long distance physics to all processes at small $x$.
This correspondence between masses and regions in transverse size is known to 
break down for higher order Fock states 
due to the possibility of large size and large mass aligned jet type 
configurations (see e.g. \cite{bhm}). 

An important point of our analysis, which has been stressed 
elsewhere including in \cite{fks1,glm1,glm2}, is that it is 
the magnitude and increase of the gluon density within the 
DGLAP approximation as $x$ decreases that leads to conflict with unitarity. 
However, in contrast to \cite{wkgb,glm1,glm2} we predict that structure 
functions continue to increase significantly with energy above the unitarity 
limit as a result of important role of peripheral collisions. 
This is because unitarity only restricts the contribution to 
structure functions related to the collisions 
at central impact parameters (see section(6)).

Forshaw, Kerley and Shaw \cite{kfs} propose a general, rather 
\textit{ad hoc}, ansatz for the DCS, 
which they stress is modeled as a function of $b^2$ and $W^2$ (rather 
than $x$), with a soft Pomeron and a hard Pomeron piece:
\begin{equation}
{\hat \sigma} (W^2,b) = a\frac{P^{2}_{s}(b)}{1 + 
P^{2}_{s}(b)}(b^{2} W^2)^{\lambda_{s}} + b^2 P^{2}_{h}(b) 
\exp(-\nu_{h}^{2}b) (b^{2} W^2)^{\lambda_{h}} \, , 
\label{eqkfs}
\end{equation}
\noindent where $P_{s}(b)$ and $P_{h}(b) $ are polynomials in b. 
They successfully fit this general form to the data.
{}From the point of view of very large dipoles (of the order of a meson
size) one may think that the DCS 
should be a function of $W^2$ rather than $x=Q^2/W^2$, since it should
just depend on the energy of the collision 
(here there is no hard scale with which to specify an `$x$'). 
However the presence of a finite $x$ imposes off-forward, or skewed, 
kinematics which cannot be ignored even in the limit of soft interactions. 
One can see this for example in the Aligned Jet Model
where the propagator for a transition from negative mass-squared ($-Q^2$)
to positive mass-squared of the $q\bar q$ pair is present for DIS.
Hence, in order to achieve the observed approximate Bjorken scaling
of $F_2$ it was necessary to model the DCS in eq.(\ref{eqkfs}) 
as a function of $b^2 W^2$, rather than $W^2$ alone. 
It is interesting to note that with our ansatz for the relationship 
between small dipole sizes and 
hard four-momentum scales, this reduces to an $x^{'}$-dependence for 
small dipoles 
$(b^2 W^2)^n \rightarrow (\lambda W^2 /Q^2)^n \propto x^{-n}$. For 
small dipoles, in our approach, there is an 
identifiable set of diagrams which contain the fusion of a gluon with 
the virtual photon, 
hence the DCS can and must depend on the momentum fraction $x^{'}$ of 
the incoming gluon.
How, and whether, this $x-$dependence becomes a $W^2$ dependence for 
large dipoles, and whether it 
is possible to define the DCS as a unique function over the whole
range in $b$, 
are interesting and open questions which deserve further consideration. 
We also note that the DCS which results from the fit of
eq.(\ref{eqkfs}), has the 
rather unattractive feature at large (fixed) energies that it is not 
a monotonically increasing function of $b$, it contains a minimum in 
the region 
$b \approx 0.6$~fm for $s = 10^5 $~GeV$^2$ (see fig.(6,7) of \cite{kfs}). 
By design, our ansatz specifically avoids this, for fixed $x$.

{}From a radically different point of view, we have arrived at a broadly
similar picture to Donnachie and Landshoff's two Pomeron model 
\cite{dl2}, with the soft 
Pomeron becoming of increasing less importance as the hardness of the 
process increases (commonly called `higher twist'). 
The approximate scaling observed in $\alpha_s xg$ in 
fig.(\ref{fig:4}) explains 
why a single power in energy will work fairly well for the hard piece.
At the same time there are important differences, in our picture
for intermediate $b$ the power is different from either of the Pomeron powers 
of \cite{dl2} and description of it by the sum of two powers is only
approximate and looks rather artificial from the point of view of the 
dipole picture. Even more importantly, due to unitarity effects 
and the important role of peripheral collisions
we expect a change 
of the power at higher energies making it closer to the soft Pomeron case. 
For example, in exclusive production of $J/\psi$ the logic of our ansatz 
suggests that the large power observed at HERA will become tamed to a 
smaller power at even higher energies due to unitarity corrections. 
In contrast, 
in the two Pomeron picture of \cite{dl2} the harder Pomeron would 
completely dominate at higher energies.

\section{Testing our ansatz: comparison with structure functions}

In section(2), we argued that for very large $Q^2$ the integrand in 
$\sigma_L$ is strongly peaked in the perturbative region and has 
very little influence from non-perturbative effects in the large $b$ region. 
If this is really the case, and our ansatz is reasonable, we should 
be able to reproduce values for the structure function $F_L (x,Q^2)$
which are in close 
agreement with the standard leading-log perturbative QCD formula of 
eq.(\ref{eqflq}). 
As $Q^2$ decreases towards the input scale $Q_0^2$ we might expect the 
two formula to deviate since the large $b$ region is implicitly 
excluded from the leading twist perturbative QCD
formula. For consistency we use the same parton set in each and to 
avoid the complexities of 
treating massive charm, in this theoretical cross check, we run with
only three light flavours of quarks 
(we also set the three light quark masses to zero).

Table.(\ref{tabfl}) reveals the excellent agreement of the $b$-space 
formula with perturbative QCD at large $Q^2$ 
and also displays the deviation of the two formula for low $Q^2$.
Also shown, in the last column is the percentage, $R_L(\%)$, 
of the b-integral coming from the non-perturbative region above
$b>b_{Q0}$. As expected this decreases with increasing $Q^2$. 
We used CTEQ4L parton distributions which have an input scale of 
$Q_0^2 = 2.56$~GeV$^2$.
\begin{table}[htbp]
\begin{center}
\begin{tabular}{|l|c|c|c|c|} \hline
$x$    & $F_L^q$ & $F_L^b $ & $\%$ diff. = 100 $\times$ & $R_L (\%)$ \\ 
CTEQ4L &         &          & $(F_L^q - F_L^b)/F_L^q$ & \\ \hline
\, \, \, \, \, \, $Q^2 = 45$  &    &  &  & \\ \hline
$ 10^{-2}$  & 0.0638 & 0.0716 & -12.3  &5.1  \\ \hline
$ 10^{-3}$  & 0.224  & 0.225  & -0.3   &2.4  \\ \hline
$ 10^{-4}$  & 0.620  & 0.594  &  4.1   &1.3  \\ \hline
$ 10^{-5}$  & 1.53   & 1.40   &  8.2   &0.7  \\ \hline
 \, \, \, \, \, \, $Q^2 = 10$  & &  & & \\ 
$ 10^{-2}$  & 0.0704  & 0.0750 & -6.53 & 26.2  \\ \hline
$ 10^{-3}$  & 0.204  & 0.187   & 8.46  & 15.2  \\ \hline
$ 10^{-4}$  & 0.493  & 0.427   & 13.4  & 9.87  \\ \hline
$ 10^{-5}$  & 1.10  & 0.869    & 21.1  & 6.38  \\ \hline
 \, \, \, \, \, \, $Q^2 = 4$   & &  & & \\ 
$ 10^{-2}$  & 0.0714 & 0.0759  & -6.32 & 54.3  \\ \hline
$ 10^{-3}$  & 0.174  & 0.153   & 12.0  & 37.5  \\ \hline
$ 10^{-4}$  & 0.378  & 0.307   & 18.6  & 26.9  \\ \hline
$ 10^{-5}$  & 0.787  & 0.558   & 29.1  & 19.4  \\ \hline
%MRSTL & & & &  \\ \hline
% \, \, \, \, \, \, $Q^2 = 45$  & &  & & \\ 
%$ 10^{-2}$  & 0.0568 & 0.0640   & -12.8 & 1.33  \\ \hline
%$ 10^{-3}$  & 0.191  & 0.190    & 0.395 & 0.539 \\ \hline
%$ 10^{-4}$  & 0.483  & 0.456    & 5.50  & 0.271 \\ \hline
%$ 10^{-5}$  & 1.12   & 1.04     & 7.78 & 0.143 \\ \hline
% \, \, \, \, \, \, $Q^2 = 10$ &  &  & & \\ 
%$ 10^{-2}$  & 0.0620 & 0.0682  & -10.2 & 6.56  \\ \hline
%$ 10^{-3}$  & 0.173  & 0.159   &  7.59 & 3.38  \\ \hline
%$ 10^{-4}$  & 0.369  & 0.323   &  12.6 & 2.01  \\ \hline
%$ 10^{-5}$  & &  & &  &  \\ \hline
% \, \, \, \, \, \, $Q^2 = 4$   & &  & & \\ 
%$ 10^{-2}$  & 0.0628 & 0.0711  & -13.2 & 17.5  \\ \hline
%$ 10^{-3}$  & 0.147  & 0.136   &  7.6  & 11.1  \\ \hline
%$ 10^{-4}$  & 0.268  & 0.239   & 10.7  & 7.54  \\ \hline
%$ 10^{-5}$  & 0.461  & 0.429   & 6.93  & 5.06  \\ \hline
\end{tabular}
\end{center}
\caption{A comparison of the $b$-space formula, $F_L^b$, using our 
ansatz for the DCS, 
with the standard perturbative QCD result, 
$F_L^q$, for $F_L^{n_{f} = 3} (x,Q^2)$
as a function of $x$ for $Q^2 = 4,10,45$~GeV$^2$. We used CTEQ4L gluons.}
\label{tabfl}
\end{table}

\noindent Having found reasonable agreement with this theoretical 
cross check of our ansatz for the DCS, we proceed to 
calculate its predictions for $F_2 (x,Q^2)$, which we denote, $F_2^b$. 
In order to calculate this we need to 
know the wavefunction squared for transverse photons:
\begin{equation}
  |\psi_{T}(z,b)|^{2} = \frac{3}{2 \pi^{2}} \alpha_{e.m.} 
\sum_{q=1}^{n_{f}}e_{q}^
{2} \left[ (z^2 + (1-z)^2) \eps^2 K_{1}^{2}(\eps b) + m_q^2 K_0^2 (\eps b) 
\right] \, ,
\label{eqpsit}
\end{equation}
\noindent where $\eps^2 = Q^2 (z(1-z)) + m_q^2$. 
From now on we will use $m_q = 300$~MeV, for
$u,d,s$ and $m_c = 1.5$ GeV, for both the longitudinal and transversely 
polarized photon wavefunctions. The small light quark constituent mass 
only affects the structure function seriously in the case of very small 
$Q^2 < Q_0^2$. It acts as a regulator for the divergence of the Bessel 
function in the photoproduction limit $Q^2 \rightarrow 0$.

We are interested in how $|\psi_{T}|^2$, integrated over z, weights the 
dipole cross section in 
\begin{equation}
\sigma_T (x,Q^2) = 2 \pi \,  2 \, \int_0^{1/2} dz \int_0^{\infty} b db \, 
{\hat \sigma}(b^2) \, |\psi_{\gamma,T} (z,b)|^2 \,  = 
\int_0^{\infty} db I_T (b,x,Q^2) \, ,
\label{eqsigt}
\end{equation}
\noindent where $I_T (b) = 2 \pi b \, {\hat \sigma} \, I_{\gamma,T}$ and
\begin{equation}
I_{\gamma,T} (b) = 2 \int_0^{1/2} dz \frac{3 \alpha_{e.m} 
\sum_{q=1}^{n_{f}} e_{q}^2}{2 \pi^2} 
\left[\left(z^{2} + (1-z)^{2} \right) \epsilon^2 
K^2_1 (\epsilon b) + m_q^2 K_0^2 (\eps b) \right] \, . 
\end{equation}

\begin{figure}[htbp]
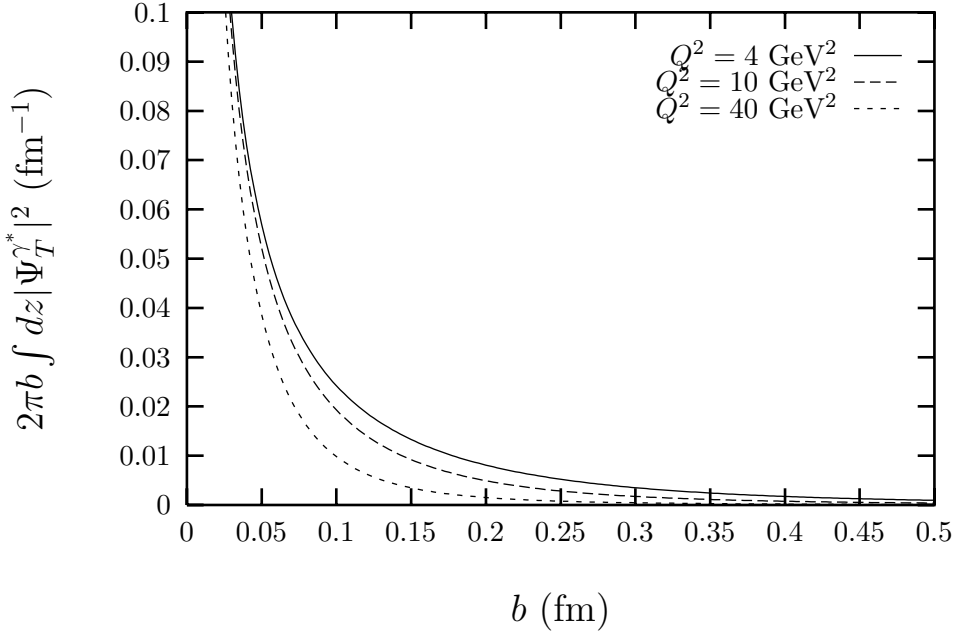

  \begin{center}
     \include{psisqt}
    \caption{The weight given to dipole cross section by the 
transversely polarized photon as a function of transverse size.}
    \label{psisqt}
  \end{center}
\end{figure}

\begin{figure}[htbp]
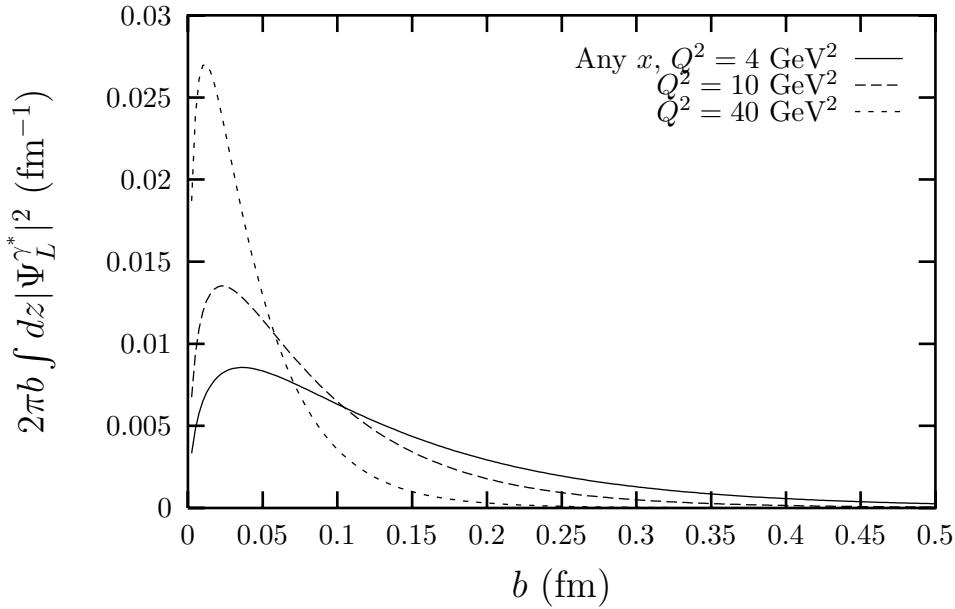

  \begin{center}
     \include{psisql}
    \caption{The weight given to dipole cross section by 
the longitudinally polarized photon as a function of transverse size.}
    \label{psisql}
  \end{center}
\end{figure}

This integrand, multiplied by the Jacobian factor $2 \pi b$, 
is shown in fig.(\ref{psisqt}) as a function of $b$ for three 
light flavours. 
At small $b$, $K_1(a) \propto 1/a$ so $I_{\gamma,T}$ is approximately 
independent of $Q^2$ and the residual 
$1/b^2$, apparent in fig.(\ref{psisqt}) cancels the $b^2$ in 
eq.(\ref{eqshat}). 
As a result, in the perturbative region the 
transverse cross section, which dominates $F_2$, is particularly 
sensitive to the effective behaviour of $\alpha_s xg$ in $b$.
For comparison, fig.(\ref{psisql}) shows $2 \pi b ~I_{\gamma,L}$.
In contrast, in the longitudinal case $K_0 (a) \propto - \ln (a)$ 
at small $a$, leaving almost the full power of $Q^2$ evident 
in eq.(\ref{eqpsil}), and effects the approximate $b^2$-behaviour 
of ${\hat \sigma}$ very little.

\begin{figure}[htbp]
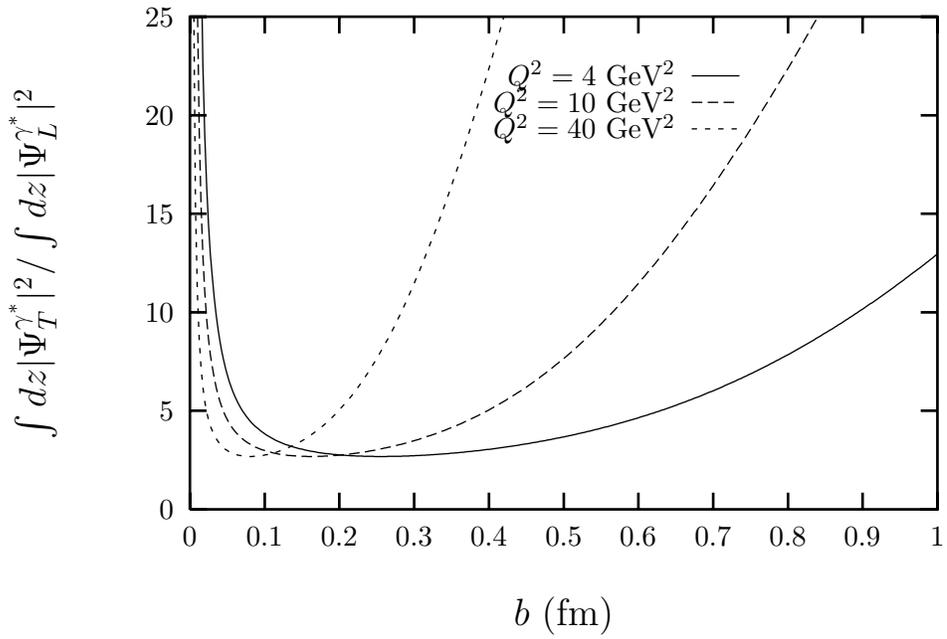

  \begin{center}
     \include{psisqtl}
   \caption{The ratio of weights, transverse divided by longitudinal, given to 
the dipole cross section for the two different photon polarizations. 
Three light flavours are included.}
    \label{psisqtl}
  \end{center}
\end{figure}

\begin{figure}[htbp]
  \begin{center}
     \include{itb40}
     \include{ilb40}
\caption{Integrands in $b$-space (units mb fm$^{-1}$) for $Q^2=40 $ GeV$^2$.}
    \label{itlb40}
  \end{center}
\end{figure}

\begin{figure}[htbp]
  \begin{center}
     \include{itb4}
     \include{ilb4}
\caption{Integrands in $b$-space (units mb fm$^{-1}$) for $Q^2=4 $ GeV$^2$.}
    \label{itlb4}
  \end{center}
\end{figure}

\begin{figure}[htbp]
  \begin{center}
     \includegraphics[height=18cm,width=13cm]{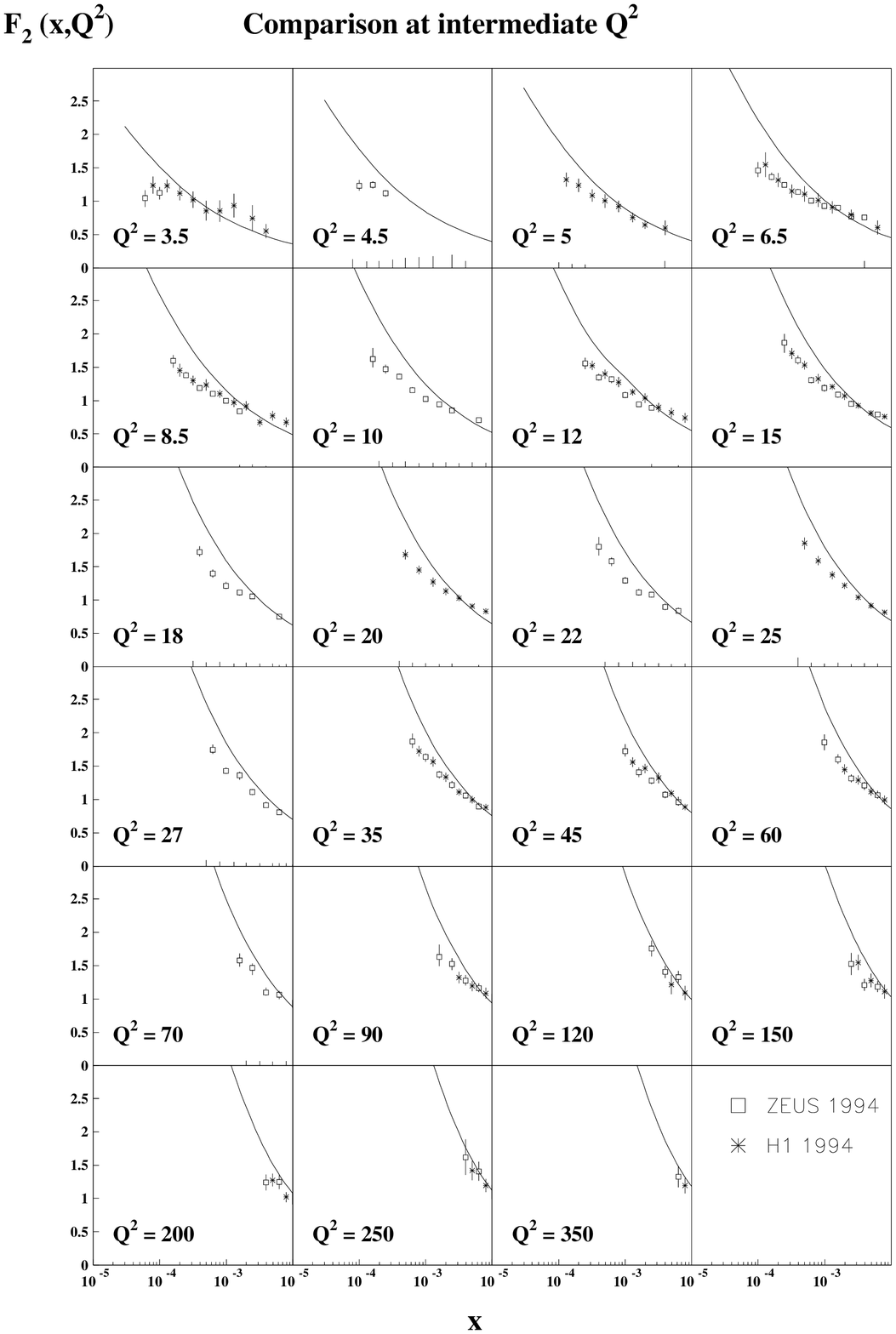}
    \caption{A comparison of the results with the 1994 data \cite{ZEUS94}, 
\cite{H194}, without any fitting procedures.} 
    \label{f2hq}
  \end{center}
\end{figure}

\begin{figure}[htbp]
  \begin{center}
     \includegraphics[height=18cm,width=13cm]{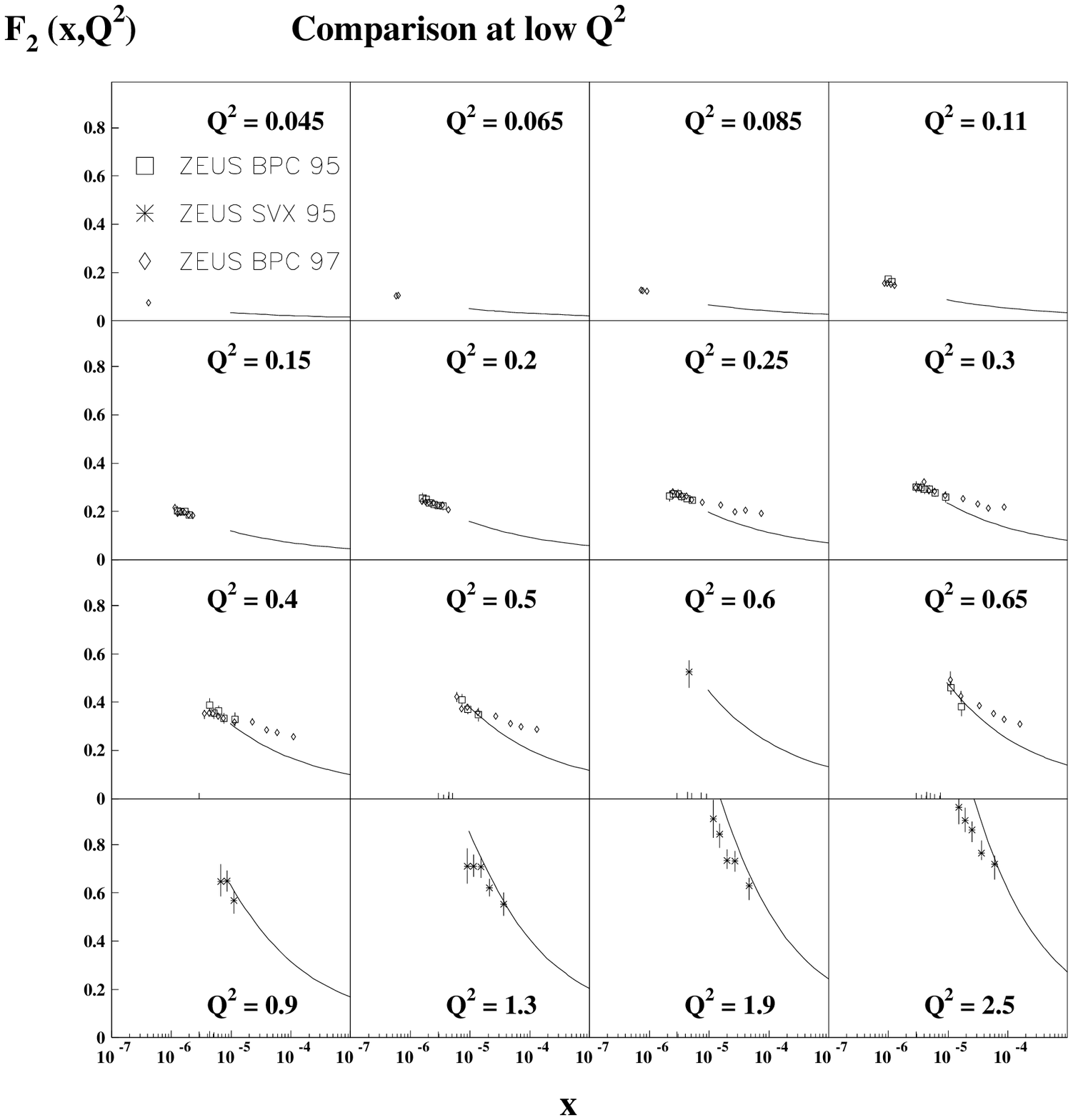}
    \caption{A comparison of the results with the ZEUS data at 
low $Q^2$, without any fitting procedures.} 
    \label{f2lq}
  \end{center}
\end{figure}

In order to make a comparison of the relative weight that the 
transverse photon gives to the DCS, we plot 
the ratio of $I_{\gamma,T}$ and $I_{\gamma,L}$ in fig(\ref{psisqtl}). 
This plot clearly shows that the transverse photon provides support
over a much broader range in $b$,
i.e. both at smaller and larger $b$, than the longitudinal photon. This
is especially true at high $Q^2$ and leads to a very 
broadly peaked integrand $I_T$. 
The peak corresponds approximately to $Q^2$ and
smaller dipoles $b < b_{peak}$ strictly speaking lie outside of the usual 
leading-log approximation (which result from logarithmic 
integrations in the perturbative QCD ladder in $k_t$ up to $Q^2$; 
the variable $b$ is conjugate to the $k_t$ in the upper quark loop of
the ladder in this formulation).

The fact that the transverse photon wavefunction squared is so broad
in $b$-space reflects Gribov's paradox (or Bjorken's aligned jet 
model). At large $Q^2$ the large dipoles are being produced 
by asymmetric splittings ($z \ll 1$). While such splittings are unlikely, 
the large dipoles they produce interact with a large hadronic cross section.
The detailed link between large dipoles and asymmetric splittings, and
the possibility of a $z$-dependence in ${\hat \sigma}$, 
requires further study. 
Figs.(\ref{itlb40},\ref{itlb4}) 
compare the resultant integrands for the transverse and 
longitudinal photons for large ($Q^2 = 40$~GeV$^2$) and moderate 
($Q^2 = 4$~GeV$^2$) values of the photon virtuality, respectively. 
For the latter case, one can clearly seen the unitarity corrections starting 
to affect the integrands at the smallest values of $x$. 
The relative contributions of perturbative and non-perturbative 
regions are very clear from these figures.

To calculate $F_2^{b (n_f =4)}$ we use
\begin{equation}
F_2 (x,Q^2) = F_T + F_L = \frac{Q^2}{4 \pi^2 \alpha_{e.m}} 
(\sigma_T + \sigma_L) 
\end{equation}
\noindent with $\sigma_T$ defined precisely analogously to
$\sigma_L$. 
We also attempt to include threshold effects, albeit in a rather crude way, 
by imposing that the momentum fraction of the gluon must be 
sufficient to generate the charm quark pair, using a theta function 
$\Theta ( x' - (Q^2 + M^2_{cc})/W^2 ) $. This condition is included in our 
ansatz of eq.(\ref{eqxprim}) and ensures that 
$x' > x_{min}^{'} = x ~(1 + 4 m_c^2 /Q^2) $ for any value of $b$. 
This procedure leads to approximately the correct ratio of charm in $F_2$. 

In fig.(\ref{f2hq}) we compare our results (solid curves) with the 1994 
HERA data \cite{ZEUS94}, \cite{H194} for the larger 
$Q^2 > Q_0^2 = 2.56 $~GeV$^2$ data. 
Our physical ansatz is in reasonable agreement with the data at small $x$ 
for moderate values of $Q^2$. This gives us faith in some of the choices 
we made in specifying our ansatz for the dipole cross section. 
Recall that no fitting or minimization procedure has been applied 
to tune the available parameters to give an excellent fit, 
although of course it would be possible to do so.

In addition, in fig.(\ref{f2lq}) for completeness we show an explicit 
comparison of our model with the lower 
$Q^2 < Q_0^2$ ZEUS BPC \cite{ZEUS95BPC,ZEUS97}
and a selection of the ZEUS SVX \cite{ZEUS95SVX} data.
We could equally have chosen to compare to the low $Q^2$
H1 data from 1995 \cite{H195}, which is binned differently in $Q^2$.
It is interesting to note that our model does a reasonably good 
job in this low virtuality region too, which is most sensitive to 
our ansatz for the DCS at large $b$. 
Again recall that no fitting procedure has been applied. 

More recent H1 data, over a wider kinematic range than in \cite{H195} 
was recently presented \cite{KLEIN}. Unfortunately the data tables were 
not publicly available to include in our plots. 

Note that in this region of small $Q^2$ we clearly cannot use DGLAP evolution.
At the same time, due to the possibility of separating contributions of small
transverse distances in the wave function of the virtual photon 
with small virtuality we can estimate with rather small
uncertainties the short-distance contribution which leads to a fast
increase of the cross section with $W^2$ at fixed $Q^2$.

\section{The onset of the new QCD regime}

The small $x$ dipole formulation of high energy processes is
attractive in that the contributions from perturbative and 
non-perturbative regions become very clear. We have extrapolated the 
perturbative QCD formula for small dipoles into the non-perturbative regime,
using the pion-proton cross section as a guide. 
For very small $x$ it was necessary to tame the growth of the small
dipoles to avoid conflict with unitarity.

In our ansatz, small dipoles are governed by the steeply 
rising gluon density. In principle all high energy processes contain 
perturbative and non-perturbative contributions, which rise quickly 
and slowly with increasing energy, respectively. Having specified the
dipole cross section, the light-cone wavefunctions of the external particles 
pick out a given region in $b$ according to their hardness. A 
reanalysis of fairly soft exclusive processes,
(e.g. electroproduction of $\rho$-mesons, or DVCS) will be crucial 
in constraining the large dipole region further.

So, where does this leave the conventional DGLAP analyses of 
inclusive structure functions which always produce qualitative good 
fits to the data? In our opinion, the linear nature of the 
evolution equations mean that the global fits are almost guaranteed to work. 
When new data comes out it is always possible to fine tune the 
functional form of the input, and the many input parameters at the 
starting scale, to reproduce the slow logarithmic changes in $Q^2$ in $F_2$. 
However, when the resulting quark and gluon densities produce 
apparent contradictions it is surely time to extend the conventional 
picture to include other (higher twist) contributions. This is of
course very far from straightforward in practice although some
interesting attempts have
been made \cite{bartels,mclerran}.
A good way to establish experimentally the important role of higher 
twist effects is to measure diffraction in the 
gluon channel and/or diffractive charm and beauty production in DIS 
(see \cite{fs} for discussion).

As well as discussing the transverse size of the scattering system 
it is also useful, and usual, to discuss the typical 
\textit{impact parameter}, $\rho$, of a given configuration on the
target (i.e. the conjugate variable to the transverse momentum 
transfer $q_t: t = - q^2$). Although it is standard to discuss the
impact parameter representation of scattering process (see e.g. the 
textbook by Eden \cite{eden}), we have noticed that there is often a 
confusion between $\rho$ and transverse size $b$ in recent literature.
Some useful explicit formulae in this regard are given in the recent paper 
by Gieseke and Qiao \cite{gq}. For particular processes 
the typical impact parameters are usually expressed in terms of the 
slope parameter $B$ which specifies the $t$-dependence (assumed to 
be exponential). 

In hard exclusive diffractive processes in DIS, within the kinematics of
HERA, the interaction is typically dominated by scattering at 
central impact parameters: $\sigma(\rho^2) \propto \exp(-\rho^2/2B)$, 
where $B\approx 4.5$\, GeV$^{-2}$. Thus, very approximately, for these 
processes $\rho \approx \sqrt{B} \sim 0.6$ fm which 
is significantly less than the electromagnetic quadratic radius of a nucleon 
$r^{trans.}_{N} \approx \sqrt{2 r^2_{N}/3} \approx 0.84$ fm. In
contrast, in hadron-hadron collisions peripheral (large $\rho$) interactions 
play an important, and perhaps dominant, role. For example, an 
analysis of elastic 
proton-proton collisions at Fermilab collider energy range shows that 
$B\approx 17$\, GeV$^{-2}$ and therefore peripheral 
$\rho \approx \sqrt{2B} \sim 1.2$ fm are essential in the total cross section. 
On this basis we can conclude that the spacetime development initiated 
by a spatially small wave packet of quarks and gluons in DIS is different at 
the achieved energy ranges from the pattern known from proton-proton 
collisions, where soft Pomeron physics dominates. 

{}From the analysis of partial waves in high energy 
($\sqrt{s} \ge 50 $~GeV) elastic $p {\bar p}$ collisions we know that
non-perturbative interactions at central impact parameters are close 
to the black limit and that peripheral collisions play a crucial role.
The distinctive feature of DIS, in the kinematics of HERA, which 
we understand from the above analysis of slopes, is that the 
scattering at central impact parameters dominates and peripheral 
scattering, where the interaction is far from black in a wide range 
of small $x$, is a correction.
The advantage of considering the black regime in DIS is that it is 
possible to make a complete evaluation of structure 
functions of a proton, 
(at least at not too large $Q^2\approx 1 $~GeV$^2$, i.e. in the regime where 
nonperturbative QCD physics dominates in the structure functions) 
because in this limit all configurations interact
with the same cross section.
Since geometrically, large transverse size dipoles are more likely to 
correspond to large impact parameters, peripheral collisions should 
play a much bigger role in $\sigma_T$ than in $\sigma_L$ for DIS 
since the large $b$ plays a much more significant 
role (cf. fig.(\ref{psisqtl})).

To demonstrate that our analysis is rather general and almost model 
independent it is useful to rewrite the equations of the dipole model 
in a form which accounts for the blackness of interaction for
$b^2 > b_{0}^2$, i.e. to evaluate the black limit of cross section:

\begin{eqnarray}  
\sigma_{tot}(\gamma^{*}_{L,T} + p\rightarrow X) & = & \int
dz d^2b ~{\hat \sigma} (z,x,b^2) |\psi^{L,T}_{\gamma}(z,b,Q^2)|^2 \nonumber \\
& = & 2 \pi r_{N}^2 \int dz d^2b |\psi^{L,T}_{\gamma}(z,b,Q^2)|^2 \nonumber \\
& - & \! \! \int dz d^2b [2\pi r_{N}^2 - {\hat \sigma} (z,x,b^2)] 
|\psi_{\gamma}^{L,T} (z,b,Q)|^2 \theta(b_{0}^2-b^2) \nonumber \\ 
 & + & \int dz d^2b ~{\hat \sigma}_{peripheral}(z,x,b^2) 
|\psi^{L,T}_{\gamma}(z,b,Q^2)|^2 \, . 
\label{ul}
\end{eqnarray}  

\noindent The first term in the last expression is relevant for 
central collisions and assumes blackness for all dipole 
sizes\footnote{It is unclear whether the hard amplitude actually achieves 
the black limit or if the increase of the perturbative QCD amplitude with 
energy is tamed earlier. Either case leads to the unitarity limit for 
collisions at central impact parameters.} 
: $\sigma(z,x,b^2)= 2 \pi r_{N}^2$ 
(including inelastic and elastic contributions). 
Alternatively, this black limit of Gribov \cite{gribov} 
may be expressed in terms of the structure functions
\begin{eqnarray} 
F_{T} & = & \frac{2 \pi r_{N}^2}{12 \pi^3} 
\int^{\delta s}_{0} \frac{dM^2 \rho(M^2) M^2 Q^2}{(M^2+Q^2)^2} \nonumber \\
F_{L} & = & \frac{2 \pi r_{N}^2}{12 \pi^3} 
\int^{\delta s}_{0} \frac {dM^2 Q^4 \rho(M^2)}{(M^2+Q^2)^2} \, .
\label{u2l}
\end{eqnarray}  
\noindent As we mentioned in the introduction (cf. eq.(\ref{black})) 
the integral over the mass, $M^2$, of the intermediate state 
leads to a logarithm in energy, or equivalently in $\delta/x$:
\begin{equation} 
F_{2} = \frac{2 \pi r_{N}^2 Q^2 \rho}{12 \pi^3} \ln (\delta/x) \, .
\end{equation}  
This estimate is rather close to that obtained within the BFKL
approximation (\cite{mue} and references therein) which we 
don't explore in this paper. 
Based on the equations for the black limit suggested in this paper,
which account for the characteristic suppression of the 
interaction for sufficiently small configurations
we may identify $\delta$ as the critical scale 
$x_{crit}=M^2/s$ at which unitarity corrections become necessary
for a given $s$ and $M^2$. 
An analysis of the black limit in QCD, to be published in a separate paper,
shows that for fixed $Q^2$ $\delta$ decreases as $x$ increases, 
leading to a reduction of the coefficient of $\ln (1/x)$ by a factor 
of $2-4$ from one for a fixed $\delta$ (depending on $Q^2$).
The size of this reduction depends on the rate of increase of $xg(x,Q^2)$ with 
decreasing $x$ and increasing $Q^2$.

For $Q^2 $ of a few GeV$^2$, $x\sim 10^{-5}$ using our estimates 
of the kinematic range in which unitarity effects may become important 
we find for $\delta$:
$10^{-3} < \delta \sim x_{crit} < 10^{-4}$. 
This allows us to estimate the numerical values of $F_2$ 
for which black limit corrections will be important. For $Q^2 = 1 \, $GeV$^2$
$x \approx 10^{-5}$, and taking a reasonable constant 
$\rho \approx 2.5$, one obtains $F_2^{\mbox{black}} \approx 1.5 - 3$. This 
estimate is only a factor of about three bigger than the current 
HERA measurements.

The second term in eq.(\ref{ul}) corrects for the fact that for 
small $b < b_0$ this black limit will not yet have been reached, 
at the $x$ value concerned. This is completely general and does not 
require a detailed knowledge of the wavefunction of the photon in the 
large $b^2$ region. 
A hypothesis on the blackening of the interaction permits a 
calculation of the structure functions at very small $x$ which 
otherwise can not be evaluated within the existing methods of QCD. 
The second term accounts for the details of QCD phenomena: because 
of colour screening phenomenon and asymptotic freedom, which
are built into the QCD expression for small dipoles 
(cf. eq.(\ref{eqshat})), the Gribov unitarity limit can not be 
achieved for all configurations in the wave function of a photon. 
At any particular $x$, there will exist configurations whose interaction
is far from the black limit. 
So, to evaluate this term it is necessary to make some additional assumptions
concerning how the black limit is reached in practice. This introduces 
some model dependence. For example, one may choose to use our ansatz for 
$\sigma (z,x,b^2)$ with its specific assumption of a 
smooth interpolation of it to unitarity limit.

The third term of eq.(\ref{ul}) concerns peripheral collisions which 
will dominate for any initial configuration in the photon wavefunction 
at extremely small $x$, due to Gribov diffusion in the parton ladder. 
The reason for this asymptotic dominance is that for a 
particular transverse size 
central collisions will freeze at their black limit, whereas peripheral 
collisions can continue to grow (albeit slowly) with energy. For example, for 
a Donnachie-Landshoff soft Pomeron parameterization \cite{dl1}
they would continue to grow like 
$\sigma_{\mbox{peripheral}} \propto (s/s_{\mbox{black}})^{\epsilon}$.

In this paper we have not tried to elaborate in detail the behaviour of
the structure functions in the kinematics where the increase with energy 
of the perturbative QCD amplitude is slowed down
(i.e. where the dipole-nucleon cross section is not far from unitarity limit).
In such a kinematic region a dipole of given transverse size $b^2$ expands 
with decrease of $x$ to a soft hadronic scale
leading to a switch from the perturbative QCD unitarity regime to the 
soft QCD regime.
A distinctive feature of such a scenario will be a fast increase with 
energy of the slope of the $t$-dependence of hard exclusive processes.

A similar analysis of the black limit also applies to exclusive process 
such as diffractive photoproduction and electroproduction of $J/\psi$. 
In this case, the black limit applies to the forward scattering amplitude:
\begin{equation}
\frac{A (\gamma^{*} + P \rightarrow J/\psi + P, t=0 )}{i s} = 
2 \pi r_{N}^2 \int dz d^2b \, \psi_{\gamma} (z,b^2,Q^2) \, \psi_{J/\psi} (z,b^2) 
\, . \label{psidif}
\end{equation}
\noindent One can then further refine this, at a given $x$, 
by breaking the amplitude down into black, not-yet black and 
peripheral contributions in direct analogy 
with the decomposition for total cross sections in eq.(\ref{ul}). 
It follows from this formulae that within the black limit approximation the 
cross sections of exclusive processes at central impact parameters 
should not depend on energy. The difference from the structure functions
case is because the wavefunction of vector meson suppresses the 
contribution of small transverse sizes. However, a residual, relatively slow, 
increase of amplitude with energy is expected due to contributions 
from peripheral collisions neglected in the above formulae. 
Since the analysis of quark Fermi motion effects \cite {fks1,fks2}
revealed that $J/\psi$ photoproduction is dominated by 
relatively large $b$, where the unitarity corrections set in early, 
this is an excellent process in which to search for the onset of 
the black limit.
   
Finally, proximity to Gribov's black body limit has 
important implications for diffraction, which in this limit 
is equivalent to elastic scattering of quark-gluon configurations 
in the photon wave function and contributes half of the total cross section 
(recall that in the black 
limit $\sigma_{el} = \sigma_{inel} = \sigma_{tot}/2$). 
Since all component states in the photon have the same cross section 
non-diagonal transitions in $M^2$ are absent. 
We can therefore unfold the $M^2$-integral in eq.(\ref{u2l}) 
and divide by two to get to the diffractive mass spectrum:

\begin{equation}
\frac{dF_T^{D(2)} (x,Q^2)}{dM^2} = \frac{\pi r^2_{target}}{12 \pi^3} 
\frac{Q^2 M^2 \rho(M^2)}{(M^2 + Q^2)^2}
\label{diffbl}
\end{equation}

\noindent Hence for $M^2 \gg Q^2$ one obtains an expression similar 
to the triple Pomeron limit with $\ap (0)=1$. The 
contribution from scattering at peripheral impact parameters,
neglected in the black body scenario, would correspond to $\ap (0)\ge 1$.
The distinctive feature of the final state is that the average 
transverse momenta are $\propto M/2$.
Let us assume that the black body limit is reached in nature, for example 
in DIS off heavy nuclei. 
Since it should be reached at lower energies for large dipoles, 
the diffractive production of 
small masses ($M \approx $ few GeV) may serve as an \textit{early signal} for 
the onset of the Gribov regime. Thus, for a given $Q^2$ eq.(\ref{diffbl}) will
be valid up to $M^2 \sim Q^2_{sat}$ so that decreasing $x$ expands the 
region of applicability of this equation.

Note also, as we have argued in \cite{fs}, that the current 
diffractive parton density analyses 
of the HERA diffractive data (see e.g \cite{actw}) indicate 
that gluon induced diffraction at HERA energies maybe already 
be close to this saturation limit.
Unfortunately the lack of measurements of the $t$-slope of the 
diffractive amplitude around $Q^2 \sim 4-6$ GeV$^2$ in gluon induced 
channels precludes distinguishing between two competing scenarios. Either 
small sized colour octet dipoles with $\sigma_{eff} \sim$ 30-40 mb 
dominate (expected $B \sim B_{J/\psi} \approx 4.5 $ GeV$^{-2}$, 
$\alpha^{'} \ll \alpha^{'}_{soft}, \ap -1 \approx 0.2$) or 
large colour-triplet dipole with with $\sigma_{eff} \sim$ 50-60 mb
dominate (expected $B \sim B_{inclusive} \approx 7 $ GeV$^{-2}$, 
$\alpha^{'} \sim \alpha^{'}_{soft}, \ap - 1 \approx \alpha_{soft}$ ).
The first scenario naturally leads to a behaviour analogous to 
eq.(\ref{diffbl}) for the gluon channel for scattering off nuclei.
We will consider these phenomena elsewhere.

\section{Conclusions}

We have proposed a physically motivated ansatz for the dipole cross 
section (DCS) relevant to a wide range of small $x$ scattering processes. 
The small dipole cross section is governed by 
the leading log gluon density at small $x$. 
Using this and the measured pion-proton cross section as a guide, we 
construct an ansatz for it in the 
non-perturbative region, below the input scale at which the input 
density for the gluon is defined. 
At very small values of $x$, as a result of the large and steeply 
rising gluon density, the DCS 
threatens to become larger at small perturbative $b$ than the
pion-proton cross section, in conflict with unitarity. 
To prevent this from happening we tame the rapid growth using a smooth
ansatz that ensures a monotonically 
increasing function of $b$ at fixed $x$.

The resultant DCS produces values for $F_L(x,Q^2)$ which are in good 
agreement with those from perturbative QCD in the 
large $Q^2$ and the high end of the small $x$ region, where it would
be expected to.
Our DCS compares reasonably well with all available small $x$ data on $F_2$ 
from HERA, without any further tuning of parameters.
Interestingly, in the moderate $Q^2$ region a significant fraction of 
the cross section appears to be coming 
from the region of large non-perturbative dipoles.
If the perturbative unitarity corrections are neglected our model 
would continue to grow very steeply in the very small $x$ region. 
More detailed studies of the choices that we make for the precise form
of the ansatz are being carried out 
and will be reported in a separate paper \cite{fgms}. Although these 
clearly affect the results quantitatively to a certain extent, 
we are confident of the qualitative conclusion of the paper that 
unitarity corrections must set in within or near to 
the HERA kinematic region of small $x$ and moderate $Q^2$ (we estimate
that the $Q^2 < 10$ GeV$^2$ will be affected 
at the smallest attainable values of $x$). 
This calls into question the use of the low, and $x$-independent, input scales
used in the standard DGLAP fits \cite{mrst,cteq4,grv98,cteq5} 
($Q_0^2 \approx 0.8 - 2.6 $~GeV$^2$).

In section(6) we generalized Gribov's black limit to the perturbative QCD
analysis of DIS to make it compatible with the applicability of the 
DGLAP approximation at sufficiently large $Q^2$ but fixed $x$ and 
with the dominance of peripheral collisions at fixed $Q^2$ 
at extremely high energies.
We estimate that this black limit is close to the lower 
edge of the HERA kinematic region in $x$ for $Q^2= 1 $ GeV$^2$. 
We suggest that certain diffractive 
processes may acts as earlier indicators of the onset of this new regime.

\end{document}

%% file: sigl.tex
% GNUPLOT: LaTeX picture with Postscript
\begingroup%
  \makeatletter%
  \newcommand{\GNUPLOTspecial}{%
    \@sanitize\catcode`\%=14\relax\special}%
  \setlength{\unitlength}{0.1bp}%
\begin{picture}(3600,2160)(0,0)%
\special{psfile=sigl llx=0 lly=0 urx=720 ury=504 rwi=7200}
\put(1417,1571){\makebox(0,0)[r]{$x=10^{-5}$}}%
\put(1417,1671){\makebox(0,0)[r]{$x=10^{-4}$}}%
\put(1417,1771){\makebox(0,0)[r]{$x=10^{-3}$}}%
\put(1417,1871){\makebox(0,0)[r]{$x=10^{-2}$}}%
\put(1925,50){\makebox(0,0){\Large $b$ (fm)}}%
\put(100,1180){%
\special{ps: gsave currentpoint currentpoint translate
270 rotate neg exch neg exch translate}%
\makebox(0,0)[b]{\shortstack{\Large $\hat{\sigma} (b^2, x) \,  $ (mb)}}%
\special{ps: currentpoint grestore moveto}%
}%
\put(3450,200){\makebox(0,0){1}}%
\put(3145,200){\makebox(0,0){0.9}}%
\put(2840,200){\makebox(0,0){0.8}}%
\put(2535,200){\makebox(0,0){0.7}}%
\put(2230,200){\makebox(0,0){0.6}}%
\put(1925,200){\makebox(0,0){0.5}}%
\put(1620,200){\makebox(0,0){0.4}}%
\put(1315,200){\makebox(0,0){0.3}}%
\put(1010,200){\makebox(0,0){0.2}}%
\put(705,200){\makebox(0,0){0.1}}%
\put(400,200){\makebox(0,0){0}}%
\put(350,2060){\makebox(0,0)[r]{140}}%
\put(350,1809){\makebox(0,0)[r]{120}}%
\put(350,1557){\makebox(0,0)[r]{100}}%
\put(350,1306){\makebox(0,0)[r]{80}}%
\put(350,1054){\makebox(0,0)[r]{60}}%
\put(350,803){\makebox(0,0)[r]{40}}%
\put(350,551){\makebox(0,0)[r]{20}}%
\put(350,300){\makebox(0,0)[r]{0}}%
\end{picture}%
\endgroup
 

%% file: psisq.tex
% GNUPLOT: LaTeX picture with Postscript
\begingroup%
  \makeatletter%
  \newcommand{\GNUPLOTspecial}{%
    \@sanitize\catcode`\%=14\relax\special}%
  \setlength{\unitlength}{0.1bp}%
\begin{picture}(3600,2160)(0,0)%
\special{psfile=psisq llx=0 lly=0 urx=720 ury=504 rwi=7200}
\put(3037,1747){\makebox(0,0)[r]{$Q^2=40$ GeV$^2$}}%
\put(3037,1847){\makebox(0,0)[r]{$Q^2=10$ GeV$^2$}}%
\put(3037,1947){\makebox(0,0)[r]{Any $x$, $Q^2=4$ GeV$^2$}}%
\put(1975,50){\makebox(0,0){\Large $b$ (fm)}}%
\put(100,1180){%
\special{ps: gsave currentpoint currentpoint translate
270 rotate neg exch neg exch translate}%
\makebox(0,0)[b]{\shortstack{\Large  z-integrated $|\Psi^{\gamma^*}_{L} |^2 $}}%
\special{ps: currentpoint grestore moveto}%
}%
\put(3450,200){\makebox(0,0){0.1}}%
\put(2860,200){\makebox(0,0){0.08}}%
\put(2270,200){\makebox(0,0){0.06}}%
\put(1680,200){\makebox(0,0){0.04}}%
\put(1090,200){\makebox(0,0){0.02}}%
\put(500,200){\makebox(0,0){0}}%
\put(450,2060){\makebox(0,0)[r]{0.03}}%
\put(450,1767){\makebox(0,0)[r]{0.025}}%
\put(450,1473){\makebox(0,0)[r]{0.02}}%
\put(450,1180){\makebox(0,0)[r]{0.015}}%
\put(450,887){\makebox(0,0)[r]{0.01}}%
\put(450,593){\makebox(0,0)[r]{0.005}}%
\put(450,300){\makebox(0,0)[r]{0}}%
\end{picture}%
\endgroup
 

%% file: int4l.tex
% GNUPLOT: LaTeX picture with Postscript
\begingroup%
  \makeatletter%
  \newcommand{\GNUPLOTspecial}{%
    \@sanitize\catcode`\%=14\relax\special}%
  \setlength{\unitlength}{0.1bp}%
\begin{picture}(3600,2160)(0,0)%
\special{psfile=int4l llx=0 lly=0 urx=720 ury=504 rwi=7200}
\put(2810,1173){\makebox(0,0)[r]{$x=10^{-5}$}}%
\put(2810,1273){\makebox(0,0)[r]{$x=10^{-4}$}}%
\put(2810,1373){\makebox(0,0)[r]{$x=10^{-3}$}}%
\put(2810,1473){\makebox(0,0)[r]{$Q^2 =4, x=10^{-2}$}}%
\put(1975,50){\makebox(0,0){\Large $b$ (fm)}}%
\put(100,1180){%
\special{ps: gsave currentpoint currentpoint translate
270 rotate neg exch neg exch translate}%
\makebox(0,0)[b]{\shortstack{\Large  $I_{L} (b) $ }}%
\special{ps: currentpoint grestore moveto}%
}%
\put(3450,200){\makebox(0,0){1}}%
\put(3155,200){\makebox(0,0){0.9}}%
\put(2860,200){\makebox(0,0){0.8}}%
\put(2565,200){\makebox(0,0){0.7}}%
\put(2270,200){\makebox(0,0){0.6}}%
\put(1975,200){\makebox(0,0){0.5}}%
\put(1680,200){\makebox(0,0){0.4}}%
\put(1385,200){\makebox(0,0){0.3}}%
\put(1090,200){\makebox(0,0){0.2}}%
\put(795,200){\makebox(0,0){0.1}}%
\put(500,200){\makebox(0,0){0}}%
\put(450,2060){\makebox(0,0)[r]{0.012}}%
\put(450,1767){\makebox(0,0)[r]{0.01}}%
\put(450,1473){\makebox(0,0)[r]{0.008}}%
\put(450,1180){\makebox(0,0)[r]{0.006}}%
\put(450,887){\makebox(0,0)[r]{0.004}}%
\put(450,593){\makebox(0,0)[r]{0.002}}%
\put(450,300){\makebox(0,0)[r]{0}}%
\end{picture}%
\endgroup
 

%% file: int40l.tex
% GNUPLOT: LaTeX picture with Postscript
\begingroup%
  \makeatletter%
  \newcommand{\GNUPLOTspecial}{%
    \@sanitize\catcode`\%=14\relax\special}%
  \setlength{\unitlength}{0.1bp}%
\begin{picture}(3600,2160)(0,0)%
\special{psfile=int40l llx=0 lly=0 urx=720 ury=504 rwi=7200}
\put(2810,1173){\makebox(0,0)[r]{$x=10^{-5}$}}%
\put(2810,1273){\makebox(0,0)[r]{$x=10^{-4}$}}%
\put(2810,1373){\makebox(0,0)[r]{$x=10^{-3}$}}%
\put(2810,1473){\makebox(0,0)[r]{$Q^2 =40, x=10^{-2}$}}%
\put(1975,50){\makebox(0,0){\Large $b$ (fm)}}%
\put(100,1180){%
\special{ps: gsave currentpoint currentpoint translate
270 rotate neg exch neg exch translate}%
\makebox(0,0)[b]{\shortstack{\Large  $I_{L} (b) $ }}%
\special{ps: currentpoint grestore moveto}%
}%
\put(3450,200){\makebox(0,0){1}}%
\put(3155,200){\makebox(0,0){0.9}}%
\put(2860,200){\makebox(0,0){0.8}}%
\put(2565,200){\makebox(0,0){0.7}}%
\put(2270,200){\makebox(0,0){0.6}}%
\put(1975,200){\makebox(0,0){0.5}}%
\put(1680,200){\makebox(0,0){0.4}}%
\put(1385,200){\makebox(0,0){0.3}}%
\put(1090,200){\makebox(0,0){0.2}}%
\put(795,200){\makebox(0,0){0.1}}%
\put(500,200){\makebox(0,0){0}}%
\put(450,2060){\makebox(0,0)[r]{0.006}}%
\put(450,1767){\makebox(0,0)[r]{0.005}}%
\put(450,1473){\makebox(0,0)[r]{0.004}}%
\put(450,1180){\makebox(0,0)[r]{0.003}}%
\put(450,887){\makebox(0,0)[r]{0.002}}%
\put(450,593){\makebox(0,0)[r]{0.001}}%
\put(450,300){\makebox(0,0)[r]{0}}%
\end{picture}%
\endgroup
 

%% file: asg3.tex
% GNUPLOT: LaTeX picture with Postscript
\begingroup%
  \makeatletter%
  \newcommand{\GNUPLOTspecial}{%
    \@sanitize\catcode`\%=14\relax\special}%
  \setlength{\unitlength}{0.1bp}%
\begin{picture}(3600,2160)(0,0)%
\special{psfile=asg3 llx=0 lly=0 urx=720 ury=504 rwi=7200}
\put(2180,587){\makebox(0,0)[r]{mrs (r2)}}%
\put(2180,687){\makebox(0,0)[r]{mrstlo}}%
\put(2180,787){\makebox(0,0)[r]{ cteq4m}}%
\put(2180,887){\makebox(0,0)[r]{$x = 10^{-3}$  cteq4l}}%
\put(1925,50){\makebox(0,0){${\bar Q^2}$ (GeV$^2$)}}%
\put(100,1180){%
\special{ps: gsave currentpoint currentpoint translate
270 rotate neg exch neg exch translate}%
\makebox(0,0)[b]{\shortstack{\Large  $\alpha_s ({\bar Q^2}) x g (x,{\bar Q^2})  $}}%
\special{ps: currentpoint grestore moveto}%
}%
\put(3450,200){\makebox(0,0){100}}%
\put(3145,200){\makebox(0,0){90}}%
\put(2840,200){\makebox(0,0){80}}%
\put(2535,200){\makebox(0,0){70}}%
\put(2230,200){\makebox(0,0){60}}%
\put(1925,200){\makebox(0,0){50}}%
\put(1620,200){\makebox(0,0){40}}%
\put(1315,200){\makebox(0,0){30}}%
\put(1010,200){\makebox(0,0){20}}%
\put(705,200){\makebox(0,0){10}}%
\put(400,200){\makebox(0,0){0}}%
\put(350,2060){\makebox(0,0)[r]{5}}%
\put(350,1767){\makebox(0,0)[r]{4.5}}%
\put(350,1473){\makebox(0,0)[r]{4}}%
\put(350,1180){\makebox(0,0)[r]{3.5}}%
\put(350,887){\makebox(0,0)[r]{3}}%
\put(350,593){\makebox(0,0)[r]{2.5}}%
\put(350,300){\makebox(0,0)[r]{2}}%
\end{picture}%
\endgroup
 

%% file: flq.tex
% GNUPLOT: LaTeX picture with Postscript
\begingroup%
  \makeatletter%
  \newcommand{\GNUPLOTspecial}{%
    \@sanitize\catcode`\%=14\relax\special}%
  \setlength{\unitlength}{0.1bp}%
\begin{picture}(3600,2160)(0,0)%
\special{psfile=flq llx=0 lly=0 urx=720 ury=504 rwi=7200}
\put(3037,1747){\makebox(0,0)[r]{$Q^2 = 4$}}%
\put(3037,1847){\makebox(0,0)[r]{$Q^2 = 10$}}%
\put(3037,1947){\makebox(0,0)[r]{$Q^2 = 45$}}%
\put(1950,740){\makebox(0,0)[l]{CTEQ4L, $x = 10^{-3}$}}%
\put(1950,50){\makebox(0,0){\Large $x'/x$}}%
\put(100,1180){%
\special{ps: gsave currentpoint currentpoint translate
270 rotate neg exch neg exch translate}%
\makebox(0,0)[b]{\shortstack{\Large $I_L^{q} (x'/x,Q^2) $}}%
\special{ps: currentpoint grestore moveto}%
}%
\put(3450,200){\makebox(0,0){5}}%
\put(3075,200){\makebox(0,0){4.5}}%
\put(2700,200){\makebox(0,0){4}}%
\put(2325,200){\makebox(0,0){3.5}}%
\put(1950,200){\makebox(0,0){3}}%
\put(1575,200){\makebox(0,0){2.5}}%
\put(1200,200){\makebox(0,0){2}}%
\put(825,200){\makebox(0,0){1.5}}%
\put(450,200){\makebox(0,0){1}}%
\put(400,2060){\makebox(0,0)[r]{2000}}%
\put(400,1884){\makebox(0,0)[r]{1800}}%
\put(400,1708){\makebox(0,0)[r]{1600}}%
\put(400,1532){\makebox(0,0)[r]{1400}}%
\put(400,1356){\makebox(0,0)[r]{1200}}%
\put(400,1180){\makebox(0,0)[r]{1000}}%
\put(400,1004){\makebox(0,0)[r]{800}}%
\put(400,828){\makebox(0,0)[r]{600}}%
\put(400,652){\makebox(0,0)[r]{400}}%
\put(400,476){\makebox(0,0)[r]{200}}%
\put(400,300){\makebox(0,0)[r]{0}}%
\end{picture}%
\endgroup
 

%% file: sigx.tex
% GNUPLOT: LaTeX picture with Postscript
\begingroup%
  \makeatletter%
  \newcommand{\GNUPLOTspecial}{%
    \@sanitize\catcode`\%=14\relax\special}%
  \setlength{\unitlength}{0.1bp}%
\begin{picture}(3600,2160)(0,0)%
\special{psfile=sigx llx=0 lly=0 urx=720 ury=504 rwi=7200}
\put(1695,1713){\makebox(0,0)[r]{$x = 10^{-5}, b_{\mbox{crit}} = 0.175$fm}}%
\put(1695,1813){\makebox(0,0)[r]{$x = 10^{-4}, b_{\mbox{crit}} = 0.26$ fm}}%
\put(1695,1913){\makebox(0,0)[r]{$x = 10^{-3}, b_{\mbox{crit}} > b_{Q0}$}}%
\put(1745,417){\makebox(0,0)[l]{$\sigma_{\pi,P} (b_{\pi},x =  10^{-5}) = 41.32 $ mb}}%
\put(1745,535){\makebox(0,0)[l]{$\sigma_{\pi,P} (b_{\pi},x =  10^{-4}) = 34.37 $ mb}}%
\put(1745,652){\makebox(0,0)[l]{$\sigma_{\pi,P} (b_{\pi},x =  10^{-3}) = 28.59 $ mb}}%
\put(2365,1004){\makebox(0,0)[l]{$x_0 = 0.01$}}%
\put(2365,887){\makebox(0,0)[l]{$b_{Q0} = 0.39$ fm}}%
\put(2365,769){\makebox(0,0)[l]{$b_{\pi} = 0.65$ fm}}%
\put(1900,50){\makebox(0,0){\Large $b$ (fm)}}%
\put(100,1180){%
\special{ps: gsave currentpoint currentpoint translate
270 rotate neg exch neg exch translate}%
\makebox(0,0)[b]{\shortstack{\Large $\hat{\sigma} (b) $ \, (mb)}}%
\special{ps: currentpoint grestore moveto}%
}%
\put(3450,200){\makebox(0,0){1}}%
\put(3140,200){\makebox(0,0){0.9}}%
\put(2830,200){\makebox(0,0){0.8}}%
\put(2520,200){\makebox(0,0){0.7}}%
\put(2210,200){\makebox(0,0){0.6}}%
\put(1900,200){\makebox(0,0){0.5}}%
\put(1590,200){\makebox(0,0){0.4}}%
\put(1280,200){\makebox(0,0){0.3}}%
\put(970,200){\makebox(0,0){0.2}}%
\put(660,200){\makebox(0,0){0.1}}%
\put(350,200){\makebox(0,0){0}}%
\put(300,2060){\makebox(0,0)[r]{60}}%
\put(300,1767){\makebox(0,0)[r]{50}}%
\put(300,1473){\makebox(0,0)[r]{40}}%
\put(300,1180){\makebox(0,0)[r]{30}}%
\put(300,887){\makebox(0,0)[r]{20}}%
\put(300,593){\makebox(0,0)[r]{10}}%
\put(300,300){\makebox(0,0)[r]{0}}%
\end{picture}%
\endgroup
 

%% file: sigwgb.tex
% GNUPLOT: LaTeX picture with Postscript
\begingroup%
  \makeatletter%
  \newcommand{\GNUPLOTspecial}{%
    \@sanitize\catcode`\%=14\relax\special}%
  \setlength{\unitlength}{0.1bp}%
\begin{picture}(3600,2160)(0,0)%
\special{psfile=sigwgb llx=0 lly=0 urx=720 ury=504 rwi=7200}
\put(2625,804){\makebox(0,0)[r]{$x = 10^{-5}$}}%
\put(2625,904){\makebox(0,0)[r]{$x = 10^{-4}$}}%
\put(2625,1004){\makebox(0,0)[r]{$x = 10^{-3}$}}%
\put(1900,50){\makebox(0,0){\Large $b$ (fm)}}%
\put(100,1180){%
\special{ps: gsave currentpoint currentpoint translate
270 rotate neg exch neg exch translate}%
\makebox(0,0)[b]{\shortstack{\Large $\hat{\sigma} (b) $ \, (mb)}}%
\special{ps: currentpoint grestore moveto}%
}%
\put(3450,200){\makebox(0,0){1}}%
\put(3140,200){\makebox(0,0){0.9}}%
\put(2830,200){\makebox(0,0){0.8}}%
\put(2520,200){\makebox(0,0){0.7}}%
\put(2210,200){\makebox(0,0){0.6}}%
\put(1900,200){\makebox(0,0){0.5}}%
\put(1590,200){\makebox(0,0){0.4}}%
\put(1280,200){\makebox(0,0){0.3}}%
\put(970,200){\makebox(0,0){0.2}}%
\put(660,200){\makebox(0,0){0.1}}%
\put(350,200){\makebox(0,0){0}}%
\put(300,2060){\makebox(0,0)[r]{25}}%
\put(300,1708){\makebox(0,0)[r]{20}}%
\put(300,1356){\makebox(0,0)[r]{15}}%
\put(300,1004){\makebox(0,0)[r]{10}}%
\put(300,652){\makebox(0,0)[r]{5}}%
\put(300,300){\makebox(0,0)[r]{0}}%
\end{picture}%
\endgroup
 

%% file: wgbcom.tex
% GNUPLOT: LaTeX picture with Postscript
\begingroup%
  \makeatletter%
  \newcommand{\GNUPLOTspecial}{%
    \@sanitize\catcode`\%=14\relax\special}%
  \setlength{\unitlength}{0.1bp}%
\begin{picture}(3600,2160)(0,0)%
\special{psfile=wgbcom llx=0 lly=0 urx=720 ury=504 rwi=7200}
\put(2485,634){\makebox(0,0)[r]{x = 0.00001}}%
\put(2485,734){\makebox(0,0)[r]{x = 0.0001}}%
\put(2485,834){\makebox(0,0)[r]{x = 0.001}}%
\put(2485,934){\makebox(0,0)[r]{x = 0.01}}%
\put(2352,476){\makebox(0,0)[l]{$b_{\pi}$}}%
\put(1590,476){\makebox(0,0)[l]{$b_{Q0}$}}%
\put(1925,50){\makebox(0,0){\Large $b (fm)$}}%
\put(100,1180){%
\special{ps: gsave currentpoint currentpoint translate
270 rotate neg exch neg exch translate}%
\makebox(0,0)[b]{\shortstack{\Large $\sigma(x,b^2) / \sigma^{WGB}$}}%
\special{ps: currentpoint grestore moveto}%
}%
\put(3450,200){\makebox(0,0){1}}%
\put(3145,200){\makebox(0,0){0.9}}%
\put(2840,200){\makebox(0,0){0.8}}%
\put(2535,200){\makebox(0,0){0.7}}%
\put(2230,200){\makebox(0,0){0.6}}%
\put(1925,200){\makebox(0,0){0.5}}%
\put(1620,200){\makebox(0,0){0.4}}%
\put(1315,200){\makebox(0,0){0.3}}%
\put(1010,200){\makebox(0,0){0.2}}%
\put(705,200){\makebox(0,0){0.1}}%
\put(400,200){\makebox(0,0){0}}%
\put(350,2060){\makebox(0,0)[r]{2.5}}%
\put(350,1708){\makebox(0,0)[r]{2}}%
\put(350,1356){\makebox(0,0)[r]{1.5}}%
\put(350,1004){\makebox(0,0)[r]{1}}%
\put(350,652){\makebox(0,0)[r]{0.5}}%
\put(350,300){\makebox(0,0)[r]{0}}%
\end{picture}%
\endgroup
 

%% file: psisqt.tex
% GNUPLOT: LaTeX picture with Postscript
\setlength{\unitlength}{0.1bp}
\special{!
%!PS-Adobe-2.0
%%Creator: gnuplot
%%DocumentFonts: Helvetica
%%BoundingBox: 50 50 770 554
%%Pages: (atend)
%%EndComments
/gnudict 40 dict def
gnudict begin
/Color false def
/Solid false def
/gnulinewidth 5.000 def
/vshift -33 def
/dl {10 mul} def
/hpt 31.5 def
/vpt 31.5 def
/M {moveto} bind def
/L {lineto} bind def
/R {rmoveto} bind def
/V {rlineto} bind def
/vpt2 vpt 2 mul def
/hpt2 hpt 2 mul def
/Lshow { currentpoint stroke M
  0 vshift R show } def
/Rshow { currentpoint stroke M
  dup stringwidth pop neg vshift R show } def
/Cshow { currentpoint stroke M
  dup stringwidth pop -2 div vshift R show } def
/DL { Color {setrgbcolor Solid {pop []} if 0 setdash }
 {pop pop pop Solid {pop []} if 0 setdash} ifelse } def
/BL { stroke gnulinewidth 2 mul setlinewidth } def
/AL { stroke gnulinewidth 2 div setlinewidth } def
/PL { stroke gnulinewidth setlinewidth } def
/LTb { BL [] 0 0 0 DL } def
/LTa { AL [1 dl 2 dl] 0 setdash 0 0 0 setrgbcolor } def
/LT0 { PL [] 0 1 0 DL } def
/LT1 { PL [4 dl 2 dl] 0 0 1 DL } def
/LT2 { PL [2 dl 3 dl] 1 0 0 DL } def
/LT3 { PL [1 dl 1.5 dl] 1 0 1 DL } def
/LT4 { PL [5 dl 2 dl 1 dl 2 dl] 0 1 1 DL } def
/LT5 { PL [4 dl 3 dl 1 dl 3 dl] 1 1 0 DL } def
/LT6 { PL [2 dl 2 dl 2 dl 4 dl] 0 0 0 DL } def
/LT7 { PL [2 dl 2 dl 2 dl 2 dl 2 dl 4 dl] 1 0.3 0 DL } def
/LT8 { PL [2 dl 2 dl 2 dl 2 dl 2 dl 2 dl 2 dl 4 dl] 0.5 0.5 0.5 DL } def
/P { stroke [] 0 setdash
  currentlinewidth 2 div sub M
  0 currentlinewidth V stroke } def
/D { stroke [] 0 setdash 2 copy vpt add M
  hpt neg vpt neg V hpt vpt neg V
  hpt vpt V hpt neg vpt V closepath stroke
  P } def
/A { stroke [] 0 setdash vpt sub M 0 vpt2 V
  currentpoint stroke M
  hpt neg vpt neg R hpt2 0 V stroke
  } def
/B { stroke [] 0 setdash 2 copy exch hpt sub exch vpt add M
  0 vpt2 neg V hpt2 0 V 0 vpt2 V
  hpt2 neg 0 V closepath stroke
  P } def
/C { stroke [] 0 setdash exch hpt sub exch vpt add M
  hpt2 vpt2 neg V currentpoint stroke M
  hpt2 neg 0 R hpt2 vpt2 V stroke } def
/T { stroke [] 0 setdash 2 copy vpt 1.12 mul add M
  hpt neg vpt -1.62 mul V
  hpt 2 mul 0 V
  hpt neg vpt 1.62 mul V closepath stroke
  P  } def
/S { 2 copy A C} def
end
}
\begin{picture}(3600,2160)(0,0)
\special{"
gnudict begin
gsave
50 50 translate
0.100 0.100 scale
0 setgray
/Helvetica findfont 100 scalefont setfont
newpath
-500.000000 -500.000000 translate
LTa
600 251 M
2817 0 V
600 251 M
0 1858 V
LTb
600 251 M
63 0 V
2754 0 R
-63 0 V
600 437 M
63 0 V
2754 0 R
-63 0 V
600 623 M
63 0 V
2754 0 R
-63 0 V
600 808 M
63 0 V
2754 0 R
-63 0 V
600 994 M
63 0 V
2754 0 R
-63 0 V
600 1180 M
63 0 V
2754 0 R
-63 0 V
600 1366 M
63 0 V
2754 0 R
-63 0 V
600 1552 M
63 0 V
2754 0 R
-63 0 V
600 1737 M
63 0 V
2754 0 R
-63 0 V
600 1923 M
63 0 V
2754 0 R
-63 0 V
600 2109 M
63 0 V
2754 0 R
-63 0 V
600 251 M
0 63 V
0 1795 R
0 -63 V
882 251 M
0 63 V
0 1795 R
0 -63 V
1163 251 M
0 63 V
0 1795 R
0 -63 V
1445 251 M
0 63 V
0 1795 R
0 -63 V
1727 251 M
0 63 V
0 1795 R
0 -63 V
2009 251 M
0 63 V
0 1795 R
0 -63 V
2290 251 M
0 63 V
0 1795 R
0 -63 V
2572 251 M
0 63 V
0 1795 R
0 -63 V
2854 251 M
0 63 V
0 1795 R
0 -63 V
3135 251 M
0 63 V
0 1795 R
0 -63 V
3417 251 M
0 63 V
0 1795 R
0 -63 V
600 251 M
2817 0 V
0 1858 V
-2817 0 V
600 251 L
LT0
3114 1946 M
180 0 V
768 2109 M
1 -10 V
14 -151 V
14 -130 V
14 -113 V
14 -99 V
14 -87 V
15 -78 V
14 -70 V
14 -63 V
14 -57 V
14 -53 V
14 -47 V
14 -44 V
14 -40 V
14 -38 V
14 -34 V
14 -32 V
14 -30 V
15 -28 V
14 -27 V
14 -24 V
14 -23 V
14 -22 V
14 -21 V
14 -19 V
14 -19 V
14 -17 V
14 -17 V
14 -15 V
14 -15 V
15 -14 V
14 -14 V
14 -13 V
14 -12 V
14 -12 V
14 -11 V
14 -11 V
14 -11 V
14 -9 V
14 -10 V
14 -9 V
15 -9 V
14 -8 V
14 -9 V
14 -7 V
14 -8 V
14 -7 V
14 -7 V
14 -7 V
14 -6 V
14 -7 V
14 -6 V
14 -6 V
15 -5 V
14 -6 V
14 -5 V
14 -5 V
14 -5 V
14 -5 V
14 -5 V
14 -4 V
14 -5 V
14 -4 V
14 -4 V
14 -4 V
15 -4 V
14 -3 V
14 -4 V
14 -4 V
14 -3 V
14 -3 V
14 -4 V
14 -3 V
14 -3 V
14 -3 V
14 -3 V
14 -3 V
15 -2 V
14 -3 V
14 -3 V
14 -2 V
14 -2 V
14 -3 V
14 -2 V
14 -2 V
14 -3 V
14 -2 V
14 -2 V
15 -2 V
14 -2 V
14 -2 V
14 -2 V
14 -2 V
14 -1 V
14 -2 V
14 -2 V
14 -2 V
14 -1 V
14 -2 V
14 -1 V
15 -2 V
14 -1 V
14 -2 V
14 -1 V
14 -2 V
14 -1 V
14 -1 V
14 -2 V
14 -1 V
14 -1 V
14 -1 V
14 -1 V
15 -2 V
14 -1 V
14 -1 V
14 -1 V
14 -1 V
14 -1 V
14 -1 V
14 -1 V
14 -1 V
14 -1 V
14 -1 V
14 -1 V
15 0 V
14 -1 V
14 -1 V
14 -1 V
14 -1 V
14 -1 V
14 0 V
14 -1 V
14 -1 V
14 -1 V
14 0 V
14 -1 V
15 -1 V
14 0 V
14 -1 V
14 -1 V
14 0 V
14 -1 V
14 0 V
14 -1 V
14 -1 V
14 0 V
14 -1 V
15 0 V
14 -1 V
14 0 V
14 -1 V
14 0 V
14 -1 V
14 0 V
14 -1 V
14 0 V
14 -1 V
14 0 V
14 -1 V
15 0 V
14 0 V
14 -1 V
14 0 V
14 -1 V
14 0 V
14 0 V
14 -1 V
14 0 V
14 0 V
14 -1 V
14 0 V
15 0 V
14 -1 V
14 0 V
14 0 V
14 -1 V
14 0 V
14 0 V
14 -1 V
14 0 V
14 0 V
14 -1 V
14 0 V
15 0 V
14 0 V
14 -1 V
14 0 V
14 0 V
14 0 V
LT1
3114 1846 M
180 0 V
762 2109 M
7 -85 V
14 -153 V
14 -133 V
14 -114 V
14 -101 V
14 -90 V
15 -79 V
14 -71 V
14 -64 V
14 -59 V
14 -53 V
14 -48 V
14 -44 V
14 -41 V
14 -38 V
14 -35 V
14 -32 V
14 -30 V
15 -28 V
14 -27 V
14 -24 V
14 -23 V
14 -22 V
14 -20 V
14 -19 V
14 -18 V
14 -17 V
14 -16 V
14 -15 V
14 -15 V
15 -13 V
14 -13 V
14 -13 V
14 -11 V
14 -12 V
14 -10 V
14 -10 V
14 -10 V
14 -9 V
14 -9 V
14 -8 V
15 -8 V
14 -8 V
14 -7 V
14 -7 V
14 -7 V
14 -6 V
14 -6 V
14 -6 V
14 -6 V
14 -5 V
14 -5 V
14 -5 V
15 -5 V
14 -5 V
14 -4 V
14 -4 V
14 -5 V
14 -3 V
14 -4 V
14 -4 V
14 -3 V
14 -4 V
14 -3 V
14 -3 V
15 -3 V
14 -3 V
14 -3 V
14 -3 V
14 -2 V
14 -3 V
14 -2 V
14 -3 V
14 -2 V
14 -2 V
14 -2 V
14 -2 V
15 -2 V
14 -2 V
14 -2 V
14 -2 V
14 -2 V
14 -1 V
14 -2 V
14 -2 V
14 -1 V
14 -2 V
14 -1 V
15 -1 V
14 -2 V
14 -1 V
14 -1 V
14 -1 V
14 -2 V
14 -1 V
14 -1 V
14 -1 V
14 -1 V
14 -1 V
14 -1 V
15 -1 V
14 -1 V
14 -1 V
14 -1 V
14 -1 V
14 0 V
14 -1 V
14 -1 V
14 -1 V
14 0 V
14 -1 V
14 -1 V
15 -1 V
14 0 V
14 -1 V
14 0 V
14 -1 V
14 -1 V
14 0 V
14 -1 V
14 0 V
14 -1 V
14 0 V
14 -1 V
15 0 V
14 -1 V
14 0 V
14 -1 V
14 0 V
14 -1 V
14 0 V
14 0 V
14 -1 V
14 0 V
14 -1 V
14 0 V
15 0 V
14 -1 V
14 0 V
14 0 V
14 -1 V
14 0 V
14 0 V
14 -1 V
14 0 V
14 0 V
14 0 V
15 -1 V
14 0 V
14 0 V
14 0 V
14 -1 V
14 0 V
14 0 V
14 0 V
14 -1 V
14 0 V
14 0 V
14 0 V
15 0 V
14 -1 V
14 0 V
14 0 V
14 0 V
14 0 V
14 -1 V
14 0 V
14 0 V
14 0 V
14 0 V
14 0 V
15 -1 V
14 0 V
14 0 V
14 0 V
14 0 V
14 0 V
14 0 V
14 -1 V
14 0 V
14 0 V
14 0 V
14 0 V
15 0 V
14 0 V
14 0 V
14 -1 V
14 0 V
14 0 V
LT2
3114 1746 M
180 0 V
746 2109 M
9 -145 V
14 -186 V
14 -157 V
14 -135 V
14 -116 V
14 -101 V
14 -90 V
15 -78 V
14 -70 V
14 -63 V
14 -56 V
14 -50 V
14 -46 V
14 -41 V
14 -38 V
14 -34 V
14 -31 V
14 -29 V
14 -26 V
15 -24 V
14 -22 V
14 -21 V
14 -18 V
14 -18 V
14 -16 V
14 -15 V
14 -14 V
14 -12 V
14 -12 V
14 -11 V
14 -11 V
15 -9 V
14 -9 V
14 -9 V
14 -8 V
14 -7 V
14 -7 V
14 -6 V
14 -6 V
14 -6 V
14 -5 V
14 -5 V
15 -5 V
14 -4 V
14 -4 V
14 -4 V
14 -4 V
14 -3 V
14 -3 V
14 -3 V
14 -3 V
14 -3 V
14 -3 V
14 -2 V
15 -2 V
14 -3 V
14 -2 V
14 -2 V
14 -2 V
14 -1 V
14 -2 V
14 -2 V
14 -1 V
14 -2 V
14 -1 V
14 -1 V
15 -1 V
14 -2 V
14 -1 V
14 -1 V
14 -1 V
14 -1 V
14 -1 V
14 -1 V
14 0 V
14 -1 V
14 -1 V
14 -1 V
15 -1 V
14 0 V
14 -1 V
14 0 V
14 -1 V
14 -1 V
14 0 V
14 -1 V
14 0 V
14 -1 V
14 0 V
15 -1 V
14 0 V
14 0 V
14 -1 V
14 0 V
14 -1 V
14 0 V
14 0 V
14 -1 V
14 0 V
14 0 V
14 0 V
15 -1 V
14 0 V
14 0 V
14 -1 V
14 0 V
14 0 V
14 0 V
14 0 V
14 -1 V
14 0 V
14 0 V
14 0 V
15 0 V
14 -1 V
14 0 V
14 0 V
14 0 V
14 0 V
14 0 V
14 -1 V
14 0 V
14 0 V
14 0 V
14 0 V
15 0 V
14 0 V
14 -1 V
14 0 V
14 0 V
14 0 V
14 0 V
14 0 V
14 0 V
14 0 V
14 0 V
14 -1 V
15 0 V
14 0 V
14 0 V
14 0 V
14 0 V
14 0 V
14 0 V
14 0 V
14 0 V
14 0 V
14 0 V
15 0 V
14 -1 V
14 0 V
14 0 V
14 0 V
14 0 V
14 0 V
14 0 V
14 0 V
14 0 V
14 0 V
14 0 V
15 0 V
14 0 V
14 0 V
14 0 V
14 0 V
14 0 V
14 0 V
14 0 V
14 -1 V
14 0 V
14 0 V
14 0 V
15 0 V
14 0 V
14 0 V
14 0 V
14 0 V
14 0 V
14 0 V
14 0 V
14 0 V
14 0 V
14 0 V
14 0 V
15 0 V
14 0 V
14 0 V
14 0 V
14 0 V
14 0 V
stroke
grestore
end
showpage
}
\put(3054,1746){\makebox(0,0)[r]{$Q^2=40$ GeV$^2$}}
\put(3054,1846){\makebox(0,0)[r]{$Q^2=10$ GeV$^2$}}
\put(3054,1946){\makebox(0,0)[r]{$Q^2= 4$ GeV$^2$}}
\put(2008,-149){\makebox(0,0){\Large $b$ (fm)}}
\put(100,1180){%
\special{ps: gsave currentpoint currentpoint translate
270 rotate neg exch neg exch translate}%
\makebox(0,0)[b]{\shortstack{\Large $2 \pi b \int dz | \Psi^{\gamma^*}_{T} |^2$ (fm$^{-1}$) }}%
\special{ps: currentpoint grestore moveto}%
}
\put(3417,151){\makebox(0,0){0.5}}
\put(3135,151){\makebox(0,0){0.45}}
\put(2854,151){\makebox(0,0){0.4}}
\put(2572,151){\makebox(0,0){0.35}}
\put(2290,151){\makebox(0,0){0.3}}
\put(2009,151){\makebox(0,0){0.25}}
\put(1727,151){\makebox(0,0){0.2}}
\put(1445,151){\makebox(0,0){0.15}}
\put(1163,151){\makebox(0,0){0.1}}
\put(882,151){\makebox(0,0){0.05}}
\put(600,151){\makebox(0,0){0}}
\put(540,2109){\makebox(0,0)[r]{0.1}}
\put(540,1923){\makebox(0,0)[r]{0.09}}
\put(540,1737){\makebox(0,0)[r]{0.08}}
\put(540,1552){\makebox(0,0)[r]{0.07}}
\put(540,1366){\makebox(0,0)[r]{0.06}}
\put(540,1180){\makebox(0,0)[r]{0.05}}
\put(540,994){\makebox(0,0)[r]{0.04}}
\put(540,808){\makebox(0,0)[r]{0.03}}
\put(540,623){\makebox(0,0)[r]{0.02}}
\put(540,437){\makebox(0,0)[r]{0.01}}
\put(540,251){\makebox(0,0)[r]{0}}
\end{picture}

%% file: psisql.tex
% GNUPLOT: LaTeX picture with Postscript
\setlength{\unitlength}{0.1bp}
\special{!
%!PS-Adobe-2.0
%%Creator: gnuplot
%%DocumentFonts: Helvetica
%%BoundingBox: 50 50 770 554
%%Pages: (atend)
%%EndComments
/gnudict 40 dict def
gnudict begin
/Color false def
/Solid false def
/gnulinewidth 5.000 def
/vshift -33 def
/dl {10 mul} def
/hpt 31.5 def
/vpt 31.5 def
/M {moveto} bind def
/L {lineto} bind def
/R {rmoveto} bind def
/V {rlineto} bind def
/vpt2 vpt 2 mul def
/hpt2 hpt 2 mul def
/Lshow { currentpoint stroke M
  0 vshift R show } def
/Rshow { currentpoint stroke M
  dup stringwidth pop neg vshift R show } def
/Cshow { currentpoint stroke M
  dup stringwidth pop -2 div vshift R show } def
/DL { Color {setrgbcolor Solid {pop []} if 0 setdash }
 {pop pop pop Solid {pop []} if 0 setdash} ifelse } def
/BL { stroke gnulinewidth 2 mul setlinewidth } def
/AL { stroke gnulinewidth 2 div setlinewidth } def
/PL { stroke gnulinewidth setlinewidth } def
/LTb { BL [] 0 0 0 DL } def
/LTa { AL [1 dl 2 dl] 0 setdash 0 0 0 setrgbcolor } def
/LT0 { PL [] 0 1 0 DL } def
/LT1 { PL [4 dl 2 dl] 0 0 1 DL } def
/LT2 { PL [2 dl 3 dl] 1 0 0 DL } def
/LT3 { PL [1 dl 1.5 dl] 1 0 1 DL } def
/LT4 { PL [5 dl 2 dl 1 dl 2 dl] 0 1 1 DL } def
/LT5 { PL [4 dl 3 dl 1 dl 3 dl] 1 1 0 DL } def
/LT6 { PL [2 dl 2 dl 2 dl 4 dl] 0 0 0 DL } def
/LT7 { PL [2 dl 2 dl 2 dl 2 dl 2 dl 4 dl] 1 0.3 0 DL } def
/LT8 { PL [2 dl 2 dl 2 dl 2 dl 2 dl 2 dl 2 dl 4 dl] 0.5 0.5 0.5 DL } def
/P { stroke [] 0 setdash
  currentlinewidth 2 div sub M
  0 currentlinewidth V stroke } def
/D { stroke [] 0 setdash 2 copy vpt add M
  hpt neg vpt neg V hpt vpt neg V
  hpt vpt V hpt neg vpt V closepath stroke
  P } def
/A { stroke [] 0 setdash vpt sub M 0 vpt2 V
  currentpoint stroke M
  hpt neg vpt neg R hpt2 0 V stroke
  } def
/B { stroke [] 0 setdash 2 copy exch hpt sub exch vpt add M
  0 vpt2 neg V hpt2 0 V 0 vpt2 V
  hpt2 neg 0 V closepath stroke
  P } def
/C { stroke [] 0 setdash exch hpt sub exch vpt add M
  hpt2 vpt2 neg V currentpoint stroke M
  hpt2 neg 0 R hpt2 vpt2 V stroke } def
/T { stroke [] 0 setdash 2 copy vpt 1.12 mul add M
  hpt neg vpt -1.62 mul V
  hpt 2 mul 0 V
  hpt neg vpt 1.62 mul V closepath stroke
  P  } def
/S { 2 copy A C} def
end
}
\begin{picture}(3600,2160)(0,0)
\special{"
gnudict begin
gsave
50 50 translate
0.100 0.100 scale
0 setgray
/Helvetica findfont 100 scalefont setfont
newpath
-500.000000 -500.000000 translate
LTa
600 251 M
2817 0 V
600 251 M
0 1858 V
LTb
600 251 M
63 0 V
2754 0 R
-63 0 V
600 561 M
63 0 V
2754 0 R
-63 0 V
600 870 M
63 0 V
2754 0 R
-63 0 V
600 1180 M
63 0 V
2754 0 R
-63 0 V
600 1490 M
63 0 V
2754 0 R
-63 0 V
600 1799 M
63 0 V
2754 0 R
-63 0 V
600 2109 M
63 0 V
2754 0 R
-63 0 V
600 251 M
0 63 V
0 1795 R
0 -63 V
882 251 M
0 63 V
0 1795 R
0 -63 V
1163 251 M
0 63 V
0 1795 R
0 -63 V
1445 251 M
0 63 V
0 1795 R
0 -63 V
1727 251 M
0 63 V
0 1795 R
0 -63 V
2009 251 M
0 63 V
0 1795 R
0 -63 V
2290 251 M
0 63 V
0 1795 R
0 -63 V
2572 251 M
0 63 V
0 1795 R
0 -63 V
2854 251 M
0 63 V
0 1795 R
0 -63 V
3135 251 M
0 63 V
0 1795 R
0 -63 V
3417 251 M
0 63 V
0 1795 R
0 -63 V
600 251 M
2817 0 V
0 1858 V
-2817 0 V
600 251 L
LT0
3114 1946 M
180 0 V
614 457 M
14 91 V
14 61 V
14 44 V
14 33 V
15 25 V
14 20 V
14 15 V
14 12 V
14 8 V
14 7 V
14 4 V
14 2 V
14 2 V
14 0 V
14 -1 V
14 -2 V
15 -3 V
14 -4 V
14 -3 V
14 -5 V
14 -5 V
14 -5 V
14 -5 V
14 -6 V
14 -6 V
14 -6 V
14 -7 V
14 -6 V
15 -7 V
14 -7 V
14 -6 V
14 -7 V
14 -7 V
14 -7 V
14 -6 V
14 -7 V
14 -7 V
14 -7 V
14 -7 V
14 -6 V
15 -7 V
14 -7 V
14 -6 V
14 -7 V
14 -6 V
14 -6 V
14 -7 V
14 -6 V
14 -6 V
14 -6 V
14 -6 V
15 -6 V
14 -6 V
14 -5 V
14 -6 V
14 -6 V
14 -5 V
14 -6 V
14 -5 V
14 -5 V
14 -5 V
14 -5 V
14 -5 V
15 -5 V
14 -5 V
14 -5 V
14 -4 V
14 -5 V
14 -4 V
14 -5 V
14 -4 V
14 -4 V
14 -4 V
14 -4 V
14 -4 V
15 -4 V
14 -4 V
14 -4 V
14 -4 V
14 -3 V
14 -4 V
14 -3 V
14 -4 V
14 -3 V
14 -3 V
14 -4 V
14 -3 V
15 -3 V
14 -3 V
14 -3 V
14 -3 V
14 -3 V
14 -3 V
14 -2 V
14 -3 V
14 -3 V
14 -2 V
14 -3 V
15 -2 V
14 -3 V
14 -2 V
14 -3 V
14 -2 V
14 -2 V
14 -2 V
14 -3 V
14 -2 V
14 -2 V
14 -2 V
14 -2 V
15 -2 V
14 -2 V
14 -2 V
14 -1 V
14 -2 V
14 -2 V
14 -2 V
14 -1 V
14 -2 V
14 -2 V
14 -1 V
14 -2 V
15 -1 V
14 -2 V
14 -1 V
14 -2 V
14 -1 V
14 -2 V
14 -1 V
14 -1 V
14 -2 V
14 -1 V
14 -1 V
14 -1 V
15 -1 V
14 -2 V
14 -1 V
14 -1 V
14 -1 V
14 -1 V
14 -1 V
14 -1 V
14 -1 V
14 -1 V
14 -1 V
14 -1 V
15 -1 V
14 -1 V
14 -1 V
14 -1 V
14 -1 V
14 0 V
14 -1 V
14 -1 V
14 -1 V
14 0 V
14 -1 V
15 -1 V
14 -1 V
14 0 V
14 -1 V
14 -1 V
14 0 V
14 -1 V
14 -1 V
14 0 V
14 -1 V
14 -1 V
14 0 V
15 -1 V
14 0 V
14 -1 V
14 0 V
14 -1 V
14 0 V
14 -1 V
14 0 V
14 -1 V
14 0 V
14 -1 V
14 0 V
15 -1 V
14 0 V
14 -1 V
14 0 V
14 0 V
14 -1 V
14 0 V
14 -1 V
14 0 V
14 0 V
14 -1 V
14 0 V
15 0 V
14 -1 V
14 0 V
14 0 V
14 -1 V
14 0 V
LT1
3114 1846 M
180 0 V
614 669 M
14 161 V
14 97 V
14 63 V
14 42 V
15 27 V
14 17 V
14 9 V
14 4 V
14 -2 V
14 -5 V
14 -8 V
14 -10 V
14 -12 V
14 -13 V
14 -15 V
14 -15 V
15 -16 V
14 -16 V
14 -17 V
14 -17 V
14 -17 V
14 -17 V
14 -17 V
14 -17 V
14 -17 V
14 -17 V
14 -16 V
14 -16 V
15 -16 V
14 -15 V
14 -16 V
14 -15 V
14 -14 V
14 -15 V
14 -14 V
14 -13 V
14 -13 V
14 -13 V
14 -13 V
14 -12 V
15 -12 V
14 -12 V
14 -11 V
14 -11 V
14 -11 V
14 -10 V
14 -10 V
14 -10 V
14 -10 V
14 -9 V
14 -9 V
15 -9 V
14 -8 V
14 -8 V
14 -8 V
14 -8 V
14 -7 V
14 -7 V
14 -7 V
14 -7 V
14 -7 V
14 -6 V
14 -6 V
15 -6 V
14 -6 V
14 -5 V
14 -6 V
14 -5 V
14 -5 V
14 -5 V
14 -5 V
14 -4 V
14 -5 V
14 -4 V
14 -4 V
15 -4 V
14 -4 V
14 -4 V
14 -4 V
14 -3 V
14 -3 V
14 -4 V
14 -3 V
14 -3 V
14 -3 V
14 -3 V
14 -3 V
15 -2 V
14 -3 V
14 -3 V
14 -2 V
14 -2 V
14 -3 V
14 -2 V
14 -2 V
14 -2 V
14 -2 V
14 -2 V
15 -2 V
14 -2 V
14 -2 V
14 -1 V
14 -2 V
14 -2 V
14 -1 V
14 -2 V
14 -1 V
14 -1 V
14 -2 V
14 -1 V
15 -1 V
14 -2 V
14 -1 V
14 -1 V
14 -1 V
14 -1 V
14 -1 V
14 -1 V
14 -1 V
14 -1 V
14 -1 V
14 -1 V
15 -1 V
14 -1 V
14 0 V
14 -1 V
14 -1 V
14 -1 V
14 0 V
14 -1 V
14 -1 V
14 0 V
14 -1 V
14 -1 V
15 0 V
14 -1 V
14 0 V
14 -1 V
14 0 V
14 -1 V
14 0 V
14 -1 V
14 0 V
14 -1 V
14 0 V
14 0 V
15 -1 V
14 0 V
14 -1 V
14 0 V
14 0 V
14 -1 V
14 0 V
14 0 V
14 -1 V
14 0 V
14 0 V
15 0 V
14 -1 V
14 0 V
14 0 V
14 0 V
14 -1 V
14 0 V
14 0 V
14 0 V
14 -1 V
14 0 V
14 0 V
15 0 V
14 0 V
14 -1 V
14 0 V
14 0 V
14 0 V
14 0 V
14 0 V
14 -1 V
14 0 V
14 0 V
14 0 V
15 0 V
14 0 V
14 0 V
14 0 V
14 -1 V
14 0 V
14 0 V
14 0 V
14 0 V
14 0 V
14 0 V
14 0 V
15 0 V
14 -1 V
14 0 V
14 0 V
14 0 V
14 0 V
LT2
3114 1746 M
180 0 V
614 1409 M
14 320 V
14 139 V
14 52 V
14 3 V
15 -26 V
14 -44 V
14 -56 V
14 -62 V
14 -66 V
14 -68 V
14 -68 V
14 -68 V
14 -66 V
14 -64 V
14 -61 V
14 -60 V
15 -56 V
14 -54 V
14 -51 V
14 -49 V
14 -46 V
14 -43 V
14 -41 V
14 -38 V
14 -37 V
14 -34 V
14 -32 V
14 -30 V
15 -29 V
14 -26 V
14 -25 V
14 -24 V
14 -21 V
14 -21 V
14 -19 V
14 -18 V
14 -17 V
14 -16 V
14 -15 V
14 -14 V
15 -13 V
14 -12 V
14 -11 V
14 -11 V
14 -10 V
14 -9 V
14 -9 V
14 -8 V
14 -8 V
14 -7 V
14 -7 V
15 -6 V
14 -6 V
14 -6 V
14 -5 V
14 -5 V
14 -4 V
14 -4 V
14 -4 V
14 -4 V
14 -3 V
14 -4 V
14 -3 V
15 -3 V
14 -2 V
14 -3 V
14 -2 V
14 -2 V
14 -3 V
14 -1 V
14 -2 V
14 -2 V
14 -2 V
14 -1 V
14 -2 V
15 -1 V
14 -1 V
14 -1 V
14 -1 V
14 -1 V
14 -1 V
14 -1 V
14 -1 V
14 -1 V
14 -1 V
14 0 V
14 -1 V
15 -1 V
14 0 V
14 -1 V
14 0 V
14 -1 V
14 0 V
14 -1 V
14 0 V
14 -1 V
14 0 V
14 0 V
15 -1 V
14 0 V
14 0 V
14 -1 V
14 0 V
14 0 V
14 0 V
14 -1 V
14 0 V
14 0 V
14 0 V
14 0 V
15 -1 V
14 0 V
14 0 V
14 0 V
14 0 V
14 0 V
14 0 V
14 0 V
14 -1 V
14 0 V
14 0 V
14 0 V
15 0 V
14 0 V
14 0 V
14 0 V
14 0 V
14 0 V
14 0 V
14 -1 V
14 0 V
14 0 V
14 0 V
14 0 V
15 0 V
14 0 V
14 0 V
14 0 V
14 0 V
14 0 V
14 0 V
14 0 V
14 0 V
14 0 V
14 0 V
14 0 V
15 0 V
14 0 V
14 0 V
14 0 V
14 0 V
14 0 V
14 0 V
14 0 V
14 0 V
14 0 V
14 0 V
15 0 V
14 -1 V
14 0 V
14 0 V
14 0 V
14 0 V
14 0 V
14 0 V
14 0 V
14 0 V
14 0 V
14 0 V
15 0 V
14 0 V
14 0 V
14 0 V
14 0 V
14 0 V
14 0 V
14 0 V
14 0 V
14 0 V
14 0 V
14 0 V
15 0 V
14 0 V
14 0 V
14 0 V
14 0 V
14 0 V
14 0 V
14 0 V
14 0 V
14 0 V
14 0 V
14 0 V
15 0 V
14 0 V
14 0 V
14 0 V
14 0 V
14 0 V
stroke
grestore
end
showpage
}
\put(3054,1746){\makebox(0,0)[r]{$Q^2=40$ GeV$^2$}}
\put(3054,1846){\makebox(0,0)[r]{$Q^2=10$ GeV$^2$}}
\put(3054,1946){\makebox(0,0)[r]{Any $x$, $Q^2=4$ GeV$^2$}}
\put(2008,-49){\makebox(0,0){\Large $b$ (fm)}}
\put(100,1180){%
\special{ps: gsave currentpoint currentpoint translate
270 rotate neg exch neg exch translate}%
\makebox(0,0)[b]{\shortstack{\Large $2 \pi b  \int dz | \Psi^{\gamma^*}_{L} |^2 $ (fm$^{-1}$)}}%
\special{ps: currentpoint grestore moveto}%
}
\put(3417,151){\makebox(0,0){0.5}}
\put(3135,151){\makebox(0,0){0.45}}
\put(2854,151){\makebox(0,0){0.4}}
\put(2572,151){\makebox(0,0){0.35}}
\put(2290,151){\makebox(0,0){0.3}}
\put(2009,151){\makebox(0,0){0.25}}
\put(1727,151){\makebox(0,0){0.2}}
\put(1445,151){\makebox(0,0){0.15}}
\put(1163,151){\makebox(0,0){0.1}}
\put(882,151){\makebox(0,0){0.05}}
\put(600,151){\makebox(0,0){0}}
\put(540,2109){\makebox(0,0)[r]{0.03}}
\put(540,1799){\makebox(0,0)[r]{0.025}}
\put(540,1490){\makebox(0,0)[r]{0.02}}
\put(540,1180){\makebox(0,0)[r]{0.015}}
\put(540,870){\makebox(0,0)[r]{0.01}}
\put(540,561){\makebox(0,0)[r]{0.005}}
\put(540,251){\makebox(0,0)[r]{0}}
\end{picture}

%% file: psisqtl.tex
% GNUPLOT: LaTeX picture with Postscript
\setlength{\unitlength}{0.1bp}
\special{!
%!PS-Adobe-2.0
%%Creator: gnuplot
%%DocumentFonts: Helvetica
%%BoundingBox: 50 50 770 554
%%Pages: (atend)
%%EndComments
/gnudict 40 dict def
gnudict begin
/Color false def
/Solid false def
/gnulinewidth 5.000 def
/vshift -33 def
/dl {10 mul} def
/hpt 31.5 def
/vpt 31.5 def
/M {moveto} bind def
/L {lineto} bind def
/R {rmoveto} bind def
/V {rlineto} bind def
/vpt2 vpt 2 mul def
/hpt2 hpt 2 mul def
/Lshow { currentpoint stroke M
  0 vshift R show } def
/Rshow { currentpoint stroke M
  dup stringwidth pop neg vshift R show } def
/Cshow { currentpoint stroke M
  dup stringwidth pop -2 div vshift R show } def
/DL { Color {setrgbcolor Solid {pop []} if 0 setdash }
 {pop pop pop Solid {pop []} if 0 setdash} ifelse } def
/BL { stroke gnulinewidth 2 mul setlinewidth } def
/AL { stroke gnulinewidth 2 div setlinewidth } def
/PL { stroke gnulinewidth setlinewidth } def
/LTb { BL [] 0 0 0 DL } def
/LTa { AL [1 dl 2 dl] 0 setdash 0 0 0 setrgbcolor } def
/LT0 { PL [] 0 1 0 DL } def
/LT1 { PL [4 dl 2 dl] 0 0 1 DL } def
/LT2 { PL [2 dl 3 dl] 1 0 0 DL } def
/LT3 { PL [1 dl 1.5 dl] 1 0 1 DL } def
/LT4 { PL [5 dl 2 dl 1 dl 2 dl] 0 1 1 DL } def
/LT5 { PL [4 dl 3 dl 1 dl 3 dl] 1 1 0 DL } def
/LT6 { PL [2 dl 2 dl 2 dl 4 dl] 0 0 0 DL } def
/LT7 { PL [2 dl 2 dl 2 dl 2 dl 2 dl 4 dl] 1 0.3 0 DL } def
/LT8 { PL [2 dl 2 dl 2 dl 2 dl 2 dl 2 dl 2 dl 4 dl] 0.5 0.5 0.5 DL } def
/P { stroke [] 0 setdash
  currentlinewidth 2 div sub M
  0 currentlinewidth V stroke } def
/D { stroke [] 0 setdash 2 copy vpt add M
  hpt neg vpt neg V hpt vpt neg V
  hpt vpt V hpt neg vpt V closepath stroke
  P } def
/A { stroke [] 0 setdash vpt sub M 0 vpt2 V
  currentpoint stroke M
  hpt neg vpt neg R hpt2 0 V stroke
  } def
/B { stroke [] 0 setdash 2 copy exch hpt sub exch vpt add M
  0 vpt2 neg V hpt2 0 V 0 vpt2 V
  hpt2 neg 0 V closepath stroke
  P } def
/C { stroke [] 0 setdash exch hpt sub exch vpt add M
  hpt2 vpt2 neg V currentpoint stroke M
  hpt2 neg 0 R hpt2 vpt2 V stroke } def
/T { stroke [] 0 setdash 2 copy vpt 1.12 mul add M
  hpt neg vpt -1.62 mul V
  hpt 2 mul 0 V
  hpt neg vpt 1.62 mul V closepath stroke
  P  } def
/S { 2 copy A C} def
end
}
\begin{picture}(3600,2160)(0,0)
\special{"
gnudict begin
gsave
50 50 translate
0.100 0.100 scale
0 setgray
/Helvetica findfont 100 scalefont setfont
newpath
-500.000000 -500.000000 translate
LTa
600 251 M
2817 0 V
600 251 M
0 1858 V
LTb
600 251 M
63 0 V
2754 0 R
-63 0 V
600 623 M
63 0 V
2754 0 R
-63 0 V
600 994 M
63 0 V
2754 0 R
-63 0 V
600 1366 M
63 0 V
2754 0 R
-63 0 V
600 1737 M
63 0 V
2754 0 R
-63 0 V
600 2109 M
63 0 V
2754 0 R
-63 0 V
600 251 M
0 63 V
0 1795 R
0 -63 V
882 251 M
0 63 V
0 1795 R
0 -63 V
1163 251 M
0 63 V
0 1795 R
0 -63 V
1445 251 M
0 63 V
0 1795 R
0 -63 V
1727 251 M
0 63 V
0 1795 R
0 -63 V
2009 251 M
0 63 V
0 1795 R
0 -63 V
2290 251 M
0 63 V
0 1795 R
0 -63 V
2572 251 M
0 63 V
0 1795 R
0 -63 V
2854 251 M
0 63 V
0 1795 R
0 -63 V
3135 251 M
0 63 V
0 1795 R
0 -63 V
3417 251 M
0 63 V
0 1795 R
0 -63 V
600 251 M
2817 0 V
0 1858 V
-2817 0 V
600 251 L
LT0
2491 1886 M
180 0 V
646 2109 M
3 -187 V
7 -259 V
7 -190 V
7 -146 V
7 -114 V
8 -92 V
7 -75 V
7 -62 V
7 -53 V
7 -45 V
7 -39 V
7 -33 V
7 -30 V
7 -26 V
7 -24 V
7 -20 V
7 -19 V
7 -17 V
7 -15 V
7 -14 V
7 -13 V
7 -12 V
7 -10 V
7 -10 V
7 -9 V
7 -9 V
7 -8 V
7 -7 V
7 -7 V
8 -6 V
7 -6 V
7 -6 V
7 -5 V
7 -5 V
7 -5 V
7 -4 V
7 -4 V
7 -4 V
7 -4 V
7 -4 V
7 -3 V
7 -3 V
7 -3 V
7 -3 V
7 -3 V
7 -2 V
7 -2 V
7 -3 V
7 -2 V
7 -2 V
7 -2 V
7 -2 V
8 -2 V
7 -1 V
7 -2 V
7 -2 V
7 -1 V
7 -2 V
7 -1 V
7 -1 V
7 -1 V
7 -2 V
7 -1 V
7 -1 V
7 -1 V
7 -1 V
7 -1 V
7 0 V
7 -1 V
7 -1 V
7 -1 V
7 0 V
7 -1 V
7 -1 V
7 0 V
7 -1 V
8 0 V
7 -1 V
7 0 V
7 -1 V
7 0 V
7 0 V
7 -1 V
7 0 V
7 0 V
7 0 V
7 -1 V
7 0 V
7 0 V
7 0 V
7 0 V
7 0 V
7 -1 V
7 0 V
7 0 V
7 0 V
7 0 V
7 0 V
7 0 V
8 1 V
7 0 V
7 0 V
7 0 V
7 0 V
7 0 V
7 0 V
7 0 V
7 1 V
7 0 V
7 0 V
7 0 V
7 1 V
7 0 V
7 0 V
7 1 V
7 0 V
7 0 V
7 1 V
7 0 V
7 0 V
7 1 V
7 0 V
7 1 V
8 0 V
7 1 V
7 0 V
7 1 V
7 0 V
7 1 V
7 0 V
7 1 V
7 0 V
7 1 V
7 1 V
7 0 V
7 1 V
7 0 V
7 1 V
7 1 V
7 0 V
7 1 V
7 1 V
7 1 V
7 0 V
7 1 V
7 1 V
8 1 V
7 0 V
7 1 V
7 1 V
7 1 V
7 1 V
7 1 V
7 0 V
7 1 V
7 1 V
7 1 V
7 1 V
7 1 V
7 1 V
7 1 V
7 1 V
7 1 V
7 1 V
7 1 V
7 1 V
7 1 V
7 1 V
7 1 V
7 1 V
8 2 V
7 1 V
7 1 V
7 1 V
7 1 V
7 1 V
7 2 V
7 1 V
7 1 V
7 1 V
7 2 V
7 1 V
7 1 V
7 1 V
7 2 V
7 1 V
7 1 V
7 2 V
7 1 V
7 2 V
7 1 V
7 1 V
7 2 V
8 1 V
7 2 V
7 1 V
7 2 V
7 1 V
7 2 V
7 1 V
7 2 V
7 2 V
7 1 V
7 2 V
7 1 V
7 2 V
7 2 V
7 1 V
7 2 V
7 2 V
7 2 V
7 1 V
7 2 V
7 2 V
7 2 V
7 2 V
7 1 V
8 2 V
7 2 V
7 2 V
7 2 V
7 2 V
7 2 V
7 2 V
7 2 V
7 2 V
7 2 V
7 2 V
7 2 V
7 2 V
7 2 V
7 2 V
7 2 V
7 2 V
7 3 V
7 2 V
7 2 V
7 2 V
7 2 V
7 3 V
7 2 V
8 2 V
7 3 V
7 2 V
7 2 V
7 3 V
7 2 V
7 2 V
7 3 V
7 2 V
7 3 V
7 2 V
7 3 V
7 2 V
7 3 V
7 3 V
7 2 V
7 3 V
7 2 V
7 3 V
7 3 V
7 2 V
7 3 V
7 3 V
8 3 V
7 2 V
7 3 V
7 3 V
7 3 V
7 3 V
7 3 V
7 3 V
7 2 V
7 3 V
7 3 V
7 3 V
7 3 V
7 3 V
7 3 V
7 4 V
7 3 V
7 3 V
7 3 V
7 3 V
7 3 V
7 3 V
7 4 V
7 3 V
8 3 V
7 4 V
7 3 V
7 3 V
7 4 V
7 3 V
7 3 V
7 4 V
7 3 V
7 4 V
7 3 V
7 4 V
7 3 V
7 4 V
7 3 V
7 4 V
7 4 V
7 3 V
7 4 V
7 4 V
7 3 V
7 4 V
7 4 V
8 4 V
7 4 V
7 3 V
7 4 V
7 4 V
7 4 V
7 4 V
7 4 V
7 4 V
7 4 V
7 4 V
7 4 V
7 4 V
7 4 V
7 4 V
7 4 V
7 4 V
7 5 V
7 4 V
7 4 V
7 4 V
7 5 V
7 4 V
7 4 V
8 5 V
7 4 V
7 4 V
7 5 V
7 4 V
7 5 V
7 4 V
7 4 V
7 5 V
7 5 V
7 4 V
7 5 V
7 4 V
7 5 V
7 5 V
7 4 V
7 5 V
7 5 V
7 4 V
7 5 V
7 5 V
7 5 V
7 5 V
8 4 V
7 5 V
7 5 V
7 5 V
7 5 V
7 5 V
7 5 V
7 5 V
7 5 V
7 5 V
7 5 V
7 5 V
7 5 V
7 6 V
7 5 V
7 5 V
7 5 V
7 5 V
7 6 V
7 5 V
7 5 V
7 6 V
7 5 V
7 5 V
8 6 V
7 5 V
7 5 V
7 6 V
7 5 V
7 6 V
7 5 V
7 6 V
7 5 V
7 6 V
7 6 V
7 5 V
LT1
2491 1786 M
180 0 V
629 2109 M
6 -425 V
7 -287 V
7 -191 V
7 -135 V
7 -100 V
7 -77 V
7 -60 V
8 -49 V
7 -40 V
7 -34 V
7 -28 V
7 -25 V
7 -20 V
7 -19 V
7 -16 V
7 -14 V
7 -12 V
7 -12 V
7 -10 V
7 -9 V
7 -8 V
7 -7 V
7 -7 V
7 -6 V
7 -6 V
7 -5 V
7 -5 V
7 -4 V
7 -5 V
7 -3 V
7 -4 V
8 -3 V
7 -3 V
7 -3 V
7 -2 V
7 -2 V
7 -3 V
7 -2 V
7 -1 V
7 -2 V
7 -2 V
7 -1 V
7 -2 V
7 -1 V
7 -1 V
7 -1 V
7 -1 V
7 -1 V
7 -1 V
7 0 V
7 -1 V
7 0 V
7 -1 V
7 0 V
8 -1 V
7 0 V
7 0 V
7 0 V
7 -1 V
7 0 V
7 0 V
7 0 V
7 1 V
7 0 V
7 0 V
7 0 V
7 0 V
7 1 V
7 0 V
7 0 V
7 1 V
7 0 V
7 1 V
7 0 V
7 1 V
7 1 V
7 0 V
7 1 V
8 1 V
7 1 V
7 0 V
7 1 V
7 1 V
7 1 V
7 1 V
7 1 V
7 1 V
7 1 V
7 1 V
7 1 V
7 1 V
7 2 V
7 1 V
7 1 V
7 1 V
7 2 V
7 1 V
7 2 V
7 1 V
7 2 V
7 1 V
8 2 V
7 1 V
7 2 V
7 1 V
7 2 V
7 2 V
7 2 V
7 1 V
7 2 V
7 2 V
7 2 V
7 2 V
7 2 V
7 2 V
7 2 V
7 2 V
7 2 V
7 2 V
7 3 V
7 2 V
7 2 V
7 3 V
7 2 V
7 2 V
8 3 V
7 2 V
7 3 V
7 2 V
7 3 V
7 3 V
7 2 V
7 3 V
7 3 V
7 3 V
7 2 V
7 3 V
7 3 V
7 3 V
7 3 V
7 3 V
7 3 V
7 4 V
7 3 V
7 3 V
7 3 V
7 4 V
7 3 V
8 4 V
7 3 V
7 4 V
7 3 V
7 4 V
7 4 V
7 3 V
7 4 V
7 4 V
7 4 V
7 4 V
7 4 V
7 4 V
7 4 V
7 4 V
7 4 V
7 4 V
7 5 V
7 4 V
7 4 V
7 5 V
7 4 V
7 5 V
7 4 V
8 5 V
7 5 V
7 4 V
7 5 V
7 5 V
7 5 V
7 5 V
7 5 V
7 5 V
7 5 V
7 5 V
7 6 V
7 5 V
7 5 V
7 6 V
7 5 V
7 6 V
7 5 V
7 6 V
7 6 V
7 5 V
7 6 V
7 6 V
8 6 V
7 6 V
7 6 V
7 6 V
7 6 V
7 6 V
7 7 V
7 6 V
7 6 V
7 7 V
7 6 V
7 7 V
7 6 V
7 7 V
7 7 V
7 7 V
7 6 V
7 7 V
7 7 V
7 7 V
7 7 V
7 7 V
7 8 V
7 7 V
8 7 V
7 7 V
7 8 V
7 7 V
7 8 V
7 7 V
7 8 V
7 8 V
7 7 V
7 8 V
7 8 V
7 8 V
7 8 V
7 8 V
7 8 V
7 8 V
7 8 V
7 8 V
7 9 V
7 8 V
7 8 V
7 9 V
7 8 V
7 9 V
8 9 V
7 8 V
7 9 V
7 9 V
7 8 V
7 9 V
7 9 V
7 9 V
7 9 V
7 9 V
7 9 V
7 10 V
7 9 V
7 9 V
7 9 V
7 10 V
7 9 V
7 10 V
7 9 V
7 10 V
7 9 V
7 10 V
7 10 V
8 9 V
7 10 V
7 10 V
7 10 V
7 10 V
7 10 V
7 10 V
7 10 V
7 10 V
7 10 V
7 11 V
7 10 V
7 10 V
7 10 V
7 11 V
7 10 V
7 11 V
7 10 V
7 11 V
7 11 V
7 10 V
7 11 V
7 11 V
7 10 V
8 11 V
7 11 V
7 11 V
7 11 V
7 11 V
7 11 V
7 11 V
7 11 V
7 11 V
7 12 V
7 11 V
7 11 V
7 12 V
7 11 V
7 11 V
7 12 V
7 11 V
7 12 V
7 11 V
7 12 V
7 12 V
7 11 V
7 12 V
8 12 V
7 12 V
7 11 V
7 12 V
7 12 V
7 12 V
7 12 V
7 12 V
7 12 V
7 12 V
7 12 V
7 13 V
7 12 V
7 12 V
7 12 V
7 13 V
7 12 V
7 12 V
7 13 V
LT2
2491 1686 M
180 0 V
615 2109 M
6 -712 V
7 -326 V
7 -177 V
7 -109 V
7 -74 V
7 -53 V
7 -39 V
7 -30 V
7 -24 V
8 -19 V
7 -15 V
7 -13 V
7 -11 V
7 -9 V
7 -8 V
7 -7 V
7 -6 V
7 -4 V
7 -5 V
7 -3 V
7 -3 V
7 -3 V
7 -2 V
7 -2 V
7 -1 V
7 -1 V
7 -1 V
7 -1 V
7 0 V
7 -1 V
7 0 V
7 1 V
7 0 V
8 1 V
7 0 V
7 1 V
7 1 V
7 2 V
7 1 V
7 2 V
7 1 V
7 2 V
7 2 V
7 2 V
7 2 V
7 3 V
7 2 V
7 3 V
7 3 V
7 3 V
7 3 V
7 3 V
7 3 V
7 4 V
7 3 V
7 4 V
8 4 V
7 4 V
7 4 V
7 5 V
7 4 V
7 5 V
7 5 V
7 5 V
7 5 V
7 5 V
7 6 V
7 5 V
7 6 V
7 6 V
7 6 V
7 7 V
7 6 V
7 7 V
7 7 V
7 7 V
7 8 V
7 7 V
7 8 V
7 8 V
8 8 V
7 8 V
7 9 V
7 8 V
7 9 V
7 9 V
7 10 V
7 9 V
7 10 V
7 10 V
7 10 V
7 11 V
7 10 V
7 11 V
7 11 V
7 12 V
7 11 V
7 12 V
7 12 V
7 12 V
7 13 V
7 12 V
7 13 V
8 13 V
7 14 V
7 13 V
7 14 V
7 14 V
7 15 V
7 14 V
7 15 V
7 15 V
7 15 V
7 15 V
7 16 V
7 16 V
7 16 V
7 16 V
7 17 V
7 16 V
7 17 V
7 18 V
7 17 V
7 17 V
7 18 V
7 18 V
7 18 V
8 19 V
7 18 V
7 19 V
7 19 V
7 19 V
7 20 V
7 19 V
7 20 V
7 20 V
7 20 V
7 20 V
7 21 V
7 20 V
7 21 V
7 21 V
7 22 V
7 21 V
7 21 V
7 22 V
7 22 V
7 22 V
7 22 V
7 23 V
8 22 V
7 23 V
7 23 V
7 23 V
7 23 V
7 23 V
7 24 V
7 23 V
7 24 V
7 24 V
7 24 V
7 24 V
7 25 V
7 24 V
7 25 V
7 25 V
stroke
grestore
end
showpage
}
\put(2431,1686){\makebox(0,0)[r]{$Q^2=40$ GeV$^2$}}
\put(2431,1786){\makebox(0,0)[r]{$Q^2=10$ GeV$^2$}}
\put(2431,1886){\makebox(0,0)[r]{$Q^2= 4$ GeV$^2$}}
\put(2008,-149){\makebox(0,0){\Large $b$ (fm)}}
\put(100,1180){%
\special{ps: gsave currentpoint currentpoint translate
270 rotate neg exch neg exch translate}%
\makebox(0,0)[b]{\shortstack{\Large $\int dz | \Psi^{\gamma^*}_{T} |^2 / \int dz | \Psi^{\gamma^*}_{L} |^2 $}}%
\special{ps: currentpoint grestore moveto}%
}
\put(3417,151){\makebox(0,0){1}}
\put(3135,151){\makebox(0,0){0.9}}
\put(2854,151){\makebox(0,0){0.8}}
\put(2572,151){\makebox(0,0){0.7}}
\put(2290,151){\makebox(0,0){0.6}}
\put(2009,151){\makebox(0,0){0.5}}
\put(1727,151){\makebox(0,0){0.4}}
\put(1445,151){\makebox(0,0){0.3}}
\put(1163,151){\makebox(0,0){0.2}}
\put(882,151){\makebox(0,0){0.1}}
\put(600,151){\makebox(0,0){0}}
\put(540,2109){\makebox(0,0)[r]{25}}
\put(540,1737){\makebox(0,0)[r]{20}}
\put(540,1366){\makebox(0,0)[r]{15}}
\put(540,994){\makebox(0,0)[r]{10}}
\put(540,623){\makebox(0,0)[r]{5}}
\put(540,251){\makebox(0,0)[r]{0}}
\end{picture}

%% file: itb40.tex
% GNUPLOT: LaTeX picture with Postscript
\setlength{\unitlength}{0.1bp}
\special{!
%!PS-Adobe-2.0
%%Creator: gnuplot
%%DocumentFonts: Helvetica
%%BoundingBox: 50 50 770 554
%%Pages: (atend)
%%EndComments
/gnudict 40 dict def
gnudict begin
/Color false def
/Solid false def
/gnulinewidth 5.000 def
/vshift -33 def
/dl {10 mul} def
/hpt 31.5 def
/vpt 31.5 def
/M {moveto} bind def
/L {lineto} bind def
/R {rmoveto} bind def
/V {rlineto} bind def
/vpt2 vpt 2 mul def
/hpt2 hpt 2 mul def
/Lshow { currentpoint stroke M
  0 vshift R show } def
/Rshow { currentpoint stroke M
  dup stringwidth pop neg vshift R show } def
/Cshow { currentpoint stroke M
  dup stringwidth pop -2 div vshift R show } def
/DL { Color {setrgbcolor Solid {pop []} if 0 setdash }
 {pop pop pop Solid {pop []} if 0 setdash} ifelse } def
/BL { stroke gnulinewidth 2 mul setlinewidth } def
/AL { stroke gnulinewidth 2 div setlinewidth } def
/PL { stroke gnulinewidth setlinewidth } def
/LTb { BL [] 0 0 0 DL } def
/LTa { AL [1 dl 2 dl] 0 setdash 0 0 0 setrgbcolor } def
/LT0 { PL [] 0 1 0 DL } def
/LT1 { PL [4 dl 2 dl] 0 0 1 DL } def
/LT2 { PL [2 dl 3 dl] 1 0 0 DL } def
/LT3 { PL [1 dl 1.5 dl] 1 0 1 DL } def
/LT4 { PL [5 dl 2 dl 1 dl 2 dl] 0 1 1 DL } def
/LT5 { PL [4 dl 3 dl 1 dl 3 dl] 1 1 0 DL } def
/LT6 { PL [2 dl 2 dl 2 dl 4 dl] 0 0 0 DL } def
/LT7 { PL [2 dl 2 dl 2 dl 2 dl 2 dl 4 dl] 1 0.3 0 DL } def
/LT8 { PL [2 dl 2 dl 2 dl 2 dl 2 dl 2 dl 2 dl 4 dl] 0.5 0.5 0.5 DL } def
/P { stroke [] 0 setdash
  currentlinewidth 2 div sub M
  0 currentlinewidth V stroke } def
/D { stroke [] 0 setdash 2 copy vpt add M
  hpt neg vpt neg V hpt vpt neg V
  hpt vpt V hpt neg vpt V closepath stroke
  P } def
/A { stroke [] 0 setdash vpt sub M 0 vpt2 V
  currentpoint stroke M
  hpt neg vpt neg R hpt2 0 V stroke
  } def
/B { stroke [] 0 setdash 2 copy exch hpt sub exch vpt add M
  0 vpt2 neg V hpt2 0 V 0 vpt2 V
  hpt2 neg 0 V closepath stroke
  P } def
/C { stroke [] 0 setdash exch hpt sub exch vpt add M
  hpt2 vpt2 neg V currentpoint stroke M
  hpt2 neg 0 R hpt2 vpt2 V stroke } def
/T { stroke [] 0 setdash 2 copy vpt 1.12 mul add M
  hpt neg vpt -1.62 mul V
  hpt 2 mul 0 V
  hpt neg vpt 1.62 mul V closepath stroke
  P  } def
/S { 2 copy A C} def
end
}
\begin{picture}(3600,2160)(0,0)
\special{"
gnudict begin
gsave
50 50 translate
0.100 0.100 scale
0 setgray
/Helvetica findfont 100 scalefont setfont
newpath
-500.000000 -500.000000 translate
LTa
600 251 M
2817 0 V
600 251 M
0 1858 V
LTb
600 251 M
63 0 V
2754 0 R
-63 0 V
600 483 M
63 0 V
2754 0 R
-63 0 V
600 716 M
63 0 V
2754 0 R
-63 0 V
600 948 M
63 0 V
2754 0 R
-63 0 V
600 1180 M
63 0 V
2754 0 R
-63 0 V
600 1412 M
63 0 V
2754 0 R
-63 0 V
600 1645 M
63 0 V
2754 0 R
-63 0 V
600 1877 M
63 0 V
2754 0 R
-63 0 V
600 2109 M
63 0 V
2754 0 R
-63 0 V
600 251 M
0 63 V
0 1795 R
0 -63 V
882 251 M
0 63 V
0 1795 R
0 -63 V
1163 251 M
0 63 V
0 1795 R
0 -63 V
1445 251 M
0 63 V
0 1795 R
0 -63 V
1727 251 M
0 63 V
0 1795 R
0 -63 V
2009 251 M
0 63 V
0 1795 R
0 -63 V
2290 251 M
0 63 V
0 1795 R
0 -63 V
2572 251 M
0 63 V
0 1795 R
0 -63 V
2854 251 M
0 63 V
0 1795 R
0 -63 V
3135 251 M
0 63 V
0 1795 R
0 -63 V
3417 251 M
0 63 V
0 1795 R
0 -63 V
600 251 M
2817 0 V
0 1858 V
-2817 0 V
600 251 L
LT0
3114 1946 M
180 0 V
607 251 M
7 0 V
7 0 V
7 0 V
7 0 V
7 0 V
7 1 V
7 1 V
7 2 V
7 2 V
7 2 V
8 3 V
7 3 V
7 3 V
7 3 V
7 4 V
7 3 V
7 4 V
7 3 V
7 4 V
7 3 V
7 3 V
7 3 V
7 3 V
7 3 V
7 3 V
7 3 V
7 2 V
7 3 V
7 2 V
7 2 V
7 1 V
7 2 V
7 2 V
7 1 V
8 1 V
7 1 V
7 1 V
7 1 V
7 1 V
7 0 V
7 1 V
7 0 V
7 1 V
7 0 V
7 0 V
7 0 V
7 -1 V
7 0 V
7 0 V
7 0 V
7 0 V
7 0 V
7 -1 V
7 0 V
7 0 V
7 -1 V
7 0 V
8 -1 V
7 0 V
7 -1 V
7 -1 V
7 0 V
7 -1 V
7 -1 V
7 0 V
7 -1 V
7 -1 V
7 0 V
7 -1 V
7 -1 V
7 0 V
7 -1 V
7 -1 V
7 0 V
7 -1 V
7 0 V
7 -1 V
7 -1 V
7 0 V
7 -1 V
7 -1 V
8 0 V
7 -1 V
7 0 V
7 -1 V
7 -1 V
7 0 V
7 -1 V
7 0 V
7 -1 V
7 -1 V
7 0 V
7 -1 V
7 0 V
7 -1 V
7 0 V
7 -1 V
7 0 V
7 -1 V
7 -1 V
7 0 V
7 -1 V
7 0 V
7 -1 V
8 0 V
7 -1 V
7 0 V
7 0 V
7 -1 V
7 0 V
7 -1 V
7 0 V
7 -1 V
7 0 V
7 -1 V
7 0 V
7 0 V
7 -1 V
7 0 V
7 -1 V
7 0 V
7 0 V
7 -1 V
7 0 V
7 -1 V
7 0 V
7 0 V
7 -1 V
8 0 V
7 -1 V
7 0 V
7 0 V
7 -1 V
7 0 V
7 0 V
7 -1 V
7 0 V
7 0 V
7 -1 V
7 0 V
7 0 V
7 -1 V
7 0 V
7 0 V
7 0 V
7 -1 V
7 0 V
7 0 V
7 -1 V
7 0 V
7 0 V
8 -1 V
7 0 V
7 0 V
7 0 V
7 0 V
7 0 V
7 0 V
7 1 V
7 0 V
7 0 V
7 0 V
7 1 V
7 0 V
7 0 V
7 0 V
7 0 V
7 0 V
7 0 V
7 1 V
7 0 V
7 0 V
7 0 V
7 0 V
7 0 V
8 0 V
7 0 V
7 0 V
7 0 V
7 0 V
7 0 V
7 0 V
7 0 V
7 0 V
7 0 V
7 0 V
7 0 V
7 0 V
7 0 V
7 -1 V
7 0 V
7 0 V
7 0 V
7 0 V
7 0 V
7 0 V
7 0 V
7 0 V
8 0 V
7 0 V
7 -1 V
7 0 V
7 0 V
7 0 V
7 0 V
7 0 V
7 0 V
7 0 V
7 -1 V
7 0 V
7 0 V
7 0 V
7 0 V
7 0 V
7 0 V
7 -1 V
7 0 V
7 0 V
7 0 V
7 0 V
7 0 V
7 -1 V
8 0 V
7 0 V
7 0 V
7 0 V
7 0 V
7 0 V
7 -1 V
7 0 V
7 0 V
7 0 V
7 0 V
7 0 V
7 -1 V
7 0 V
7 0 V
7 0 V
7 0 V
7 0 V
7 -1 V
7 0 V
7 0 V
7 0 V
7 0 V
7 0 V
8 -1 V
7 0 V
7 0 V
7 0 V
7 0 V
7 0 V
7 -1 V
7 0 V
7 0 V
7 0 V
7 0 V
7 0 V
7 0 V
7 -1 V
7 0 V
7 0 V
7 0 V
7 -1 V
7 0 V
7 0 V
7 0 V
7 0 V
7 -1 V
8 0 V
7 0 V
7 0 V
7 0 V
7 0 V
7 -1 V
7 0 V
7 0 V
7 0 V
7 0 V
7 -1 V
7 0 V
7 0 V
7 0 V
7 0 V
7 0 V
7 -1 V
7 0 V
7 0 V
7 0 V
7 0 V
7 0 V
7 0 V
7 -1 V
8 0 V
7 0 V
7 0 V
7 0 V
7 0 V
7 0 V
7 -1 V
7 0 V
7 0 V
7 0 V
7 0 V
7 0 V
7 0 V
7 -1 V
7 0 V
7 0 V
7 0 V
7 0 V
7 0 V
7 0 V
7 0 V
7 -1 V
7 0 V
8 0 V
7 0 V
7 0 V
7 0 V
7 0 V
7 0 V
7 0 V
7 -1 V
7 0 V
7 0 V
7 0 V
7 0 V
7 0 V
7 0 V
7 0 V
7 0 V
7 0 V
7 -1 V
7 0 V
7 0 V
7 0 V
7 0 V
7 0 V
7 0 V
8 0 V
7 0 V
7 0 V
7 -1 V
7 0 V
7 0 V
7 0 V
7 0 V
7 0 V
7 0 V
7 0 V
7 0 V
7 0 V
7 0 V
7 0 V
7 -1 V
7 0 V
7 0 V
7 0 V
7 0 V
7 0 V
7 0 V
7 0 V
8 0 V
7 0 V
7 0 V
7 0 V
7 0 V
7 0 V
7 -1 V
7 0 V
7 0 V
7 0 V
7 0 V
7 0 V
7 0 V
7 0 V
7 0 V
7 0 V
7 0 V
7 0 V
7 0 V
7 0 V
7 0 V
7 -1 V
7 0 V
7 0 V
8 0 V
7 0 V
7 0 V
7 0 V
7 0 V
7 0 V
7 0 V
7 0 V
7 0 V
7 0 V
currentpoint stroke M
7 0 V
7 0 V
LT1
3114 1846 M
180 0 V
607 251 M
7 0 V
7 1 V
7 3 V
7 5 V
7 6 V
7 7 V
7 9 V
7 10 V
7 11 V
7 12 V
8 12 V
7 13 V
7 12 V
7 13 V
7 12 V
7 13 V
7 12 V
7 11 V
7 11 V
7 10 V
7 10 V
7 9 V
7 8 V
7 7 V
7 8 V
7 7 V
7 6 V
7 4 V
7 4 V
7 5 V
7 4 V
7 3 V
7 2 V
7 2 V
8 1 V
7 1 V
7 1 V
7 1 V
7 0 V
7 0 V
7 -1 V
7 -1 V
7 -1 V
7 -2 V
7 -1 V
7 -3 V
7 -2 V
7 -1 V
7 -2 V
7 -2 V
7 -2 V
7 -3 V
7 -2 V
7 -3 V
7 -2 V
7 -3 V
7 -2 V
8 -3 V
7 -3 V
7 -3 V
7 -3 V
7 -3 V
7 -3 V
7 -3 V
7 -3 V
7 -2 V
7 -3 V
7 -3 V
7 -3 V
7 -3 V
7 -3 V
7 -3 V
7 -2 V
7 -3 V
7 -2 V
7 -3 V
7 -2 V
7 -3 V
7 -2 V
7 -2 V
7 -3 V
8 -2 V
7 -2 V
7 -3 V
7 -2 V
7 -2 V
7 -2 V
7 -2 V
7 -2 V
7 -2 V
7 -2 V
7 -2 V
7 -2 V
7 -2 V
7 -2 V
7 -2 V
7 -2 V
7 -2 V
7 -1 V
7 -2 V
7 -2 V
7 -2 V
7 -1 V
7 -2 V
8 -2 V
7 -1 V
7 -2 V
7 -1 V
7 -2 V
7 -1 V
7 -2 V
7 -1 V
7 -2 V
7 -1 V
7 -1 V
7 -2 V
7 -1 V
7 -1 V
7 -2 V
7 -1 V
7 -1 V
7 -1 V
7 -2 V
7 -1 V
7 -1 V
7 -1 V
7 -1 V
7 -2 V
8 -1 V
7 -1 V
7 -1 V
7 -1 V
7 -1 V
7 -1 V
7 -1 V
7 -1 V
7 -1 V
7 -1 V
7 -1 V
7 -1 V
7 -1 V
7 -1 V
7 -1 V
7 -1 V
7 0 V
7 -1 V
7 -1 V
7 -1 V
7 -1 V
7 -1 V
7 -1 V
8 0 V
7 -1 V
7 -1 V
7 -1 V
7 0 V
7 -1 V
7 0 V
7 -1 V
7 0 V
7 -1 V
7 0 V
7 0 V
7 -1 V
7 0 V
7 -1 V
7 0 V
7 -1 V
7 0 V
7 0 V
7 -1 V
7 0 V
7 -1 V
7 0 V
7 -1 V
8 0 V
7 0 V
7 -1 V
7 0 V
7 -1 V
7 0 V
7 -1 V
7 0 V
7 0 V
7 -1 V
7 0 V
7 -1 V
7 0 V
7 0 V
7 -1 V
7 0 V
7 0 V
7 -1 V
7 0 V
7 -1 V
7 0 V
7 0 V
7 -1 V
8 0 V
7 0 V
7 -1 V
7 0 V
7 -1 V
7 0 V
7 0 V
7 -1 V
7 0 V
7 0 V
7 -1 V
7 0 V
7 0 V
7 -1 V
7 0 V
7 0 V
7 -1 V
7 0 V
7 0 V
7 -1 V
7 0 V
7 0 V
7 0 V
7 -1 V
8 0 V
7 0 V
7 -1 V
7 0 V
7 0 V
7 -1 V
7 0 V
7 0 V
7 0 V
7 -1 V
7 0 V
7 0 V
7 -1 V
7 0 V
7 0 V
7 0 V
7 -1 V
7 0 V
7 0 V
7 -1 V
7 0 V
7 0 V
7 0 V
7 -1 V
8 0 V
7 0 V
7 0 V
7 -1 V
7 0 V
7 0 V
7 0 V
7 -1 V
7 0 V
7 0 V
7 0 V
7 0 V
7 -1 V
7 0 V
7 0 V
7 -1 V
7 0 V
7 0 V
7 0 V
7 0 V
7 -1 V
7 0 V
7 0 V
8 0 V
7 -1 V
7 0 V
7 0 V
7 0 V
7 -1 V
7 0 V
7 0 V
7 0 V
7 0 V
7 -1 V
7 0 V
7 0 V
7 0 V
7 0 V
7 -1 V
7 0 V
7 0 V
7 0 V
7 0 V
7 -1 V
7 0 V
7 0 V
7 0 V
8 0 V
7 0 V
7 -1 V
7 0 V
7 0 V
7 0 V
7 0 V
7 0 V
7 -1 V
7 0 V
7 0 V
7 0 V
7 0 V
7 0 V
7 -1 V
7 0 V
7 0 V
7 0 V
7 0 V
7 0 V
7 0 V
7 -1 V
7 0 V
8 0 V
7 0 V
7 0 V
7 0 V
7 0 V
7 -1 V
7 0 V
7 0 V
7 0 V
7 0 V
7 0 V
7 0 V
7 0 V
7 -1 V
7 0 V
7 0 V
7 0 V
7 0 V
7 0 V
7 0 V
7 0 V
7 -1 V
7 0 V
7 0 V
8 0 V
7 0 V
7 0 V
7 0 V
7 0 V
7 0 V
7 0 V
7 -1 V
7 0 V
7 0 V
7 0 V
7 0 V
7 0 V
7 0 V
7 0 V
7 0 V
7 0 V
7 -1 V
7 0 V
7 0 V
7 0 V
7 0 V
7 0 V
8 0 V
7 0 V
7 0 V
7 0 V
7 0 V
7 -1 V
7 0 V
7 0 V
7 0 V
7 0 V
7 0 V
7 0 V
7 0 V
7 0 V
7 0 V
7 0 V
7 0 V
7 -1 V
7 0 V
7 0 V
7 0 V
7 0 V
7 0 V
7 0 V
8 0 V
7 0 V
7 0 V
7 0 V
7 0 V
7 0 V
7 0 V
7 -1 V
7 0 V
7 0 V
currentpoint stroke M
7 0 V
7 0 V
LT3
3114 1746 M
180 0 V
607 251 M
7 5 V
7 10 V
7 16 V
7 21 V
7 26 V
7 30 V
7 32 V
7 35 V
7 37 V
7 38 V
8 37 V
7 36 V
7 38 V
7 34 V
7 37 V
7 30 V
7 34 V
7 27 V
7 28 V
7 26 V
7 22 V
7 20 V
7 21 V
7 17 V
7 13 V
7 12 V
7 13 V
7 10 V
7 8 V
7 5 V
7 2 V
7 5 V
7 3 V
7 2 V
8 0 V
7 -1 V
7 -3 V
7 -4 V
7 -6 V
7 -4 V
7 -4 V
7 -5 V
7 -6 V
7 -6 V
7 -7 V
7 -7 V
7 -8 V
7 -8 V
7 -9 V
7 -8 V
7 -9 V
7 -9 V
7 -10 V
7 -9 V
7 -9 V
7 -8 V
7 -9 V
8 -8 V
7 -9 V
7 -8 V
7 -9 V
7 -9 V
7 -8 V
7 -9 V
7 -8 V
7 -8 V
7 -9 V
7 -8 V
7 -8 V
7 -8 V
7 -8 V
7 -8 V
7 -7 V
7 -8 V
7 -7 V
7 -7 V
7 -7 V
7 -8 V
7 -6 V
7 -7 V
7 -7 V
8 -6 V
7 -7 V
7 -6 V
7 -6 V
7 -6 V
7 -6 V
7 -6 V
7 -6 V
7 -6 V
7 -5 V
7 -6 V
7 -5 V
7 -5 V
7 -5 V
7 -5 V
7 -5 V
7 -5 V
7 -5 V
7 -4 V
7 -5 V
7 -4 V
7 -4 V
7 -5 V
8 -4 V
7 -5 V
7 -4 V
7 -5 V
7 -4 V
7 -4 V
7 -4 V
7 -5 V
7 -3 V
7 -4 V
7 -4 V
7 -4 V
7 -3 V
7 -4 V
7 -3 V
7 -4 V
7 -3 V
7 -3 V
7 -4 V
7 -3 V
7 -3 V
7 -3 V
7 -3 V
7 -3 V
8 -3 V
7 -2 V
7 -3 V
7 -3 V
7 -2 V
7 -3 V
7 -3 V
7 -2 V
7 -2 V
7 -3 V
7 -2 V
7 -2 V
7 -3 V
7 -2 V
7 -2 V
7 -2 V
7 -2 V
7 -2 V
7 -2 V
7 -2 V
7 -2 V
7 -2 V
7 -2 V
8 -2 V
7 -2 V
7 -1 V
7 -2 V
7 -2 V
7 -2 V
7 -1 V
7 -2 V
7 -1 V
7 -2 V
7 -1 V
7 -2 V
7 -1 V
7 -2 V
7 -1 V
7 -2 V
7 -1 V
7 -1 V
7 -2 V
7 -1 V
7 -1 V
7 -2 V
7 -1 V
7 -1 V
8 -1 V
7 -2 V
7 -1 V
7 -1 V
7 -1 V
7 -1 V
7 -1 V
7 -1 V
7 -1 V
7 -1 V
7 -1 V
7 -1 V
7 -1 V
7 -1 V
7 -1 V
7 -1 V
7 -1 V
7 -1 V
7 -1 V
7 -1 V
7 -1 V
7 -1 V
7 0 V
8 -1 V
7 -1 V
7 -1 V
7 -1 V
7 0 V
7 -1 V
7 -1 V
7 -1 V
7 0 V
7 -1 V
7 -1 V
7 -1 V
7 0 V
7 -1 V
7 -1 V
7 0 V
7 -1 V
7 -1 V
7 0 V
7 -1 V
7 0 V
7 -1 V
7 -1 V
7 0 V
8 -1 V
7 0 V
7 -1 V
7 0 V
7 -1 V
7 -1 V
7 0 V
7 -1 V
7 0 V
7 -1 V
7 0 V
7 -1 V
7 0 V
7 -1 V
7 0 V
7 -1 V
7 0 V
7 0 V
7 -1 V
7 0 V
7 -1 V
7 0 V
7 -1 V
7 0 V
8 0 V
7 -1 V
7 0 V
7 -1 V
7 0 V
7 0 V
7 -1 V
7 0 V
7 -1 V
7 0 V
7 0 V
7 -1 V
7 0 V
7 0 V
7 -1 V
7 0 V
7 0 V
7 -1 V
7 0 V
7 0 V
7 0 V
7 -1 V
7 0 V
8 0 V
7 -1 V
7 0 V
7 0 V
7 0 V
7 -1 V
7 0 V
7 0 V
7 0 V
7 -1 V
7 0 V
7 0 V
7 0 V
7 -1 V
7 0 V
7 0 V
7 0 V
7 -1 V
7 0 V
7 0 V
7 0 V
7 0 V
7 -1 V
7 0 V
8 0 V
7 0 V
7 0 V
7 -1 V
7 0 V
7 0 V
7 0 V
7 0 V
7 -1 V
7 0 V
7 0 V
7 0 V
7 0 V
7 -1 V
7 0 V
7 0 V
7 0 V
7 0 V
7 -1 V
7 0 V
7 0 V
7 0 V
7 0 V
8 0 V
7 -1 V
7 0 V
7 0 V
7 0 V
7 0 V
7 0 V
7 -1 V
7 0 V
7 0 V
7 0 V
7 0 V
7 0 V
7 0 V
7 -1 V
7 0 V
7 0 V
7 0 V
7 0 V
7 0 V
7 0 V
7 -1 V
7 0 V
7 0 V
8 0 V
7 0 V
7 0 V
7 0 V
7 -1 V
7 0 V
7 0 V
7 0 V
7 0 V
7 0 V
7 0 V
7 0 V
7 0 V
7 -1 V
7 0 V
7 0 V
7 0 V
7 0 V
7 0 V
7 0 V
7 0 V
7 -1 V
7 0 V
8 0 V
7 0 V
7 0 V
7 0 V
7 0 V
7 0 V
7 0 V
7 0 V
7 -1 V
7 0 V
7 0 V
7 0 V
7 0 V
7 0 V
7 0 V
7 0 V
7 0 V
7 0 V
7 0 V
7 -1 V
7 0 V
7 0 V
7 0 V
7 0 V
8 0 V
7 0 V
7 0 V
7 0 V
7 0 V
7 0 V
7 -1 V
7 0 V
7 0 V
7 0 V
currentpoint stroke M
7 0 V
7 0 V
3114 1646 M
180 0 V
607 256 M
7 23 V
7 42 V
7 58 V
7 72 V
7 81 V
7 91 V
7 97 V
7 99 V
7 100 V
7 98 V
8 98 V
7 99 V
7 88 V
7 92 V
7 79 V
7 82 V
7 64 V
7 74 V
7 60 V
7 49 V
7 55 V
7 44 V
7 31 V
7 35 V
7 32 V
7 23 V
7 14 V
7 8 V
7 15 V
7 10 V
7 4 V
7 -1 V
7 -6 V
7 -10 V
8 -9 V
7 -7 V
7 -10 V
7 -12 V
7 -16 V
7 -17 V
7 -19 V
7 -21 V
7 -23 V
7 -24 V
7 -25 V
7 -21 V
7 -21 V
7 -23 V
7 -22 V
7 -23 V
7 -23 V
7 -24 V
7 -23 V
7 -24 V
7 -24 V
7 -25 V
7 -24 V
8 -24 V
7 -24 V
7 -24 V
7 -23 V
7 -24 V
7 -23 V
7 -23 V
7 -22 V
7 -23 V
7 -22 V
7 -21 V
7 -22 V
7 -28 V
7 -27 V
7 -26 V
7 -26 V
7 -25 V
7 -24 V
7 -24 V
7 -22 V
7 -22 V
7 -22 V
7 -21 V
7 -20 V
8 -19 V
7 -19 V
7 -19 V
7 -18 V
7 -17 V
7 -17 V
7 -16 V
7 -16 V
7 -15 V
7 -15 V
7 -15 V
7 -14 V
7 -13 V
7 -14 V
7 -13 V
7 -12 V
7 -12 V
7 -12 V
7 -12 V
7 -11 V
7 -11 V
7 -10 V
7 -10 V
8 -10 V
7 -10 V
7 -9 V
7 -10 V
7 -9 V
7 -8 V
7 -9 V
7 -8 V
7 -8 V
7 -8 V
7 -7 V
7 -8 V
7 -7 V
7 -7 V
7 -7 V
7 -6 V
7 -7 V
7 -6 V
7 -6 V
7 -6 V
7 -6 V
7 -6 V
7 -5 V
7 -5 V
8 -6 V
7 -5 V
7 -5 V
7 -5 V
7 -5 V
7 -4 V
7 -5 V
7 -4 V
7 -5 V
7 -4 V
7 -4 V
7 -4 V
7 -4 V
7 -4 V
7 -4 V
7 -3 V
7 -4 V
7 -3 V
7 -4 V
7 -3 V
7 -3 V
7 -4 V
7 -3 V
8 -3 V
7 -3 V
7 -3 V
7 -3 V
7 -2 V
7 -3 V
7 -3 V
7 -3 V
7 -2 V
7 -3 V
7 -2 V
7 -2 V
7 -3 V
7 -2 V
7 -2 V
7 -3 V
7 -2 V
7 -2 V
7 -2 V
7 -2 V
7 -2 V
7 -2 V
7 -2 V
7 -2 V
8 -2 V
7 -2 V
7 -1 V
7 -2 V
7 -2 V
7 -2 V
7 -1 V
7 -2 V
7 -1 V
7 -2 V
7 -2 V
7 -1 V
7 -2 V
7 -1 V
7 -1 V
7 -2 V
7 -1 V
7 -1 V
7 -2 V
7 -1 V
7 -1 V
7 -2 V
7 -1 V
8 -1 V
7 -1 V
7 -1 V
7 -2 V
7 -1 V
7 -1 V
7 -1 V
7 -1 V
7 -1 V
7 -1 V
7 -1 V
7 -1 V
7 -1 V
7 -1 V
7 -1 V
7 -1 V
7 -1 V
7 -1 V
7 0 V
7 -1 V
7 -1 V
7 -1 V
7 -1 V
7 -1 V
8 0 V
7 -1 V
7 -1 V
7 -1 V
7 0 V
7 -1 V
7 -1 V
7 -1 V
7 0 V
7 -1 V
7 -1 V
7 0 V
7 -1 V
7 -1 V
7 0 V
7 -1 V
7 -1 V
7 0 V
7 -1 V
7 0 V
7 -1 V
7 -1 V
7 0 V
7 -1 V
8 0 V
7 -1 V
7 0 V
7 -1 V
7 0 V
7 -1 V
7 0 V
7 -1 V
7 0 V
7 -1 V
7 0 V
7 -1 V
7 0 V
7 0 V
7 -1 V
7 0 V
7 -1 V
7 0 V
7 0 V
7 -1 V
7 0 V
7 0 V
7 -1 V
8 0 V
7 0 V
7 -1 V
7 0 V
7 0 V
7 -1 V
7 0 V
7 0 V
7 -1 V
7 0 V
7 0 V
7 0 V
7 -1 V
7 0 V
7 0 V
7 -1 V
7 0 V
7 0 V
7 0 V
7 -1 V
7 0 V
7 0 V
7 -1 V
7 0 V
8 0 V
7 0 V
7 -1 V
7 0 V
7 0 V
7 0 V
7 -1 V
7 0 V
7 0 V
7 0 V
7 0 V
7 -1 V
7 0 V
7 0 V
7 0 V
7 -1 V
7 0 V
7 0 V
7 0 V
7 0 V
7 -1 V
7 0 V
7 0 V
8 0 V
7 0 V
7 -1 V
7 0 V
7 0 V
7 0 V
7 0 V
7 -1 V
7 0 V
7 0 V
7 0 V
7 0 V
7 -1 V
7 0 V
7 0 V
7 0 V
7 0 V
7 0 V
7 -1 V
7 0 V
7 0 V
7 0 V
7 0 V
7 0 V
8 -1 V
7 0 V
7 0 V
7 0 V
7 0 V
7 0 V
7 0 V
7 -1 V
7 0 V
7 0 V
7 0 V
7 0 V
7 0 V
7 0 V
7 -1 V
7 0 V
7 0 V
7 0 V
7 0 V
7 0 V
7 0 V
7 -1 V
7 0 V
8 0 V
7 0 V
7 0 V
7 0 V
7 0 V
7 0 V
7 -1 V
7 0 V
7 0 V
7 0 V
7 0 V
7 0 V
7 0 V
7 0 V
7 -1 V
7 0 V
7 0 V
7 0 V
7 0 V
7 0 V
7 0 V
7 0 V
7 0 V
7 0 V
8 -1 V
7 0 V
7 0 V
7 0 V
7 0 V
7 0 V
7 0 V
7 0 V
currentpoint stroke M
7 0 V
7 0 V
7 -1 V
7 0 V
stroke
grestore
end
showpage
}
\put(3054,1646){\makebox(0,0)[r]{$x=10^{-5}$}}
\put(3054,1746){\makebox(0,0)[r]{$x=10^{-4}$}}
\put(3054,1846){\makebox(0,0)[r]{$x=10^{-3}$}}
\put(3054,1946){\makebox(0,0)[r]{$Q^2 =40, x=10^{-2}$}}
\put(2008,-49){\makebox(0,0){\Large $b$ (fm)}}
\put(100,1180){%
\special{ps: gsave currentpoint currentpoint translate
270 rotate neg exch neg exch translate}%
\makebox(0,0)[b]{\shortstack{\Large  $I_{T} (b) $ }}%
\special{ps: currentpoint grestore moveto}%
}
\put(3417,151){\makebox(0,0){1}}
\put(3135,151){\makebox(0,0){0.9}}
\put(2854,151){\makebox(0,0){0.8}}
\put(2572,151){\makebox(0,0){0.7}}
\put(2290,151){\makebox(0,0){0.6}}
\put(2009,151){\makebox(0,0){0.5}}
\put(1727,151){\makebox(0,0){0.4}}
\put(1445,151){\makebox(0,0){0.3}}
\put(1163,151){\makebox(0,0){0.2}}
\put(882,151){\makebox(0,0){0.1}}
\put(600,151){\makebox(0,0){0}}
\put(540,2109){\makebox(0,0)[r]{0.08}}
\put(540,1877){\makebox(0,0)[r]{0.07}}
\put(540,1645){\makebox(0,0)[r]{0.06}}
\put(540,1412){\makebox(0,0)[r]{0.05}}
\put(540,1180){\makebox(0,0)[r]{0.04}}
\put(540,948){\makebox(0,0)[r]{0.03}}
\put(540,716){\makebox(0,0)[r]{0.02}}
\put(540,483){\makebox(0,0)[r]{0.01}}
\put(540,251){\makebox(0,0)[r]{0}}
\end{picture}

%% file: ilb40.tex
% GNUPLOT: LaTeX picture with Postscript
\setlength{\unitlength}{0.1bp}
\special{!
%!PS-Adobe-2.0
%%Creator: gnuplot
%%DocumentFonts: Helvetica
%%BoundingBox: 50 50 770 554
%%Pages: (atend)
%%EndComments
/gnudict 40 dict def
gnudict begin
/Color false def
/Solid false def
/gnulinewidth 5.000 def
/vshift -33 def
/dl {10 mul} def
/hpt 31.5 def
/vpt 31.5 def
/M {moveto} bind def
/L {lineto} bind def
/R {rmoveto} bind def
/V {rlineto} bind def
/vpt2 vpt 2 mul def
/hpt2 hpt 2 mul def
/Lshow { currentpoint stroke M
  0 vshift R show } def
/Rshow { currentpoint stroke M
  dup stringwidth pop neg vshift R show } def
/Cshow { currentpoint stroke M
  dup stringwidth pop -2 div vshift R show } def
/DL { Color {setrgbcolor Solid {pop []} if 0 setdash }
 {pop pop pop Solid {pop []} if 0 setdash} ifelse } def
/BL { stroke gnulinewidth 2 mul setlinewidth } def
/AL { stroke gnulinewidth 2 div setlinewidth } def
/PL { stroke gnulinewidth setlinewidth } def
/LTb { BL [] 0 0 0 DL } def
/LTa { AL [1 dl 2 dl] 0 setdash 0 0 0 setrgbcolor } def
/LT0 { PL [] 0 1 0 DL } def
/LT1 { PL [4 dl 2 dl] 0 0 1 DL } def
/LT2 { PL [2 dl 3 dl] 1 0 0 DL } def
/LT3 { PL [1 dl 1.5 dl] 1 0 1 DL } def
/LT4 { PL [5 dl 2 dl 1 dl 2 dl] 0 1 1 DL } def
/LT5 { PL [4 dl 3 dl 1 dl 3 dl] 1 1 0 DL } def
/LT6 { PL [2 dl 2 dl 2 dl 4 dl] 0 0 0 DL } def
/LT7 { PL [2 dl 2 dl 2 dl 2 dl 2 dl 4 dl] 1 0.3 0 DL } def
/LT8 { PL [2 dl 2 dl 2 dl 2 dl 2 dl 2 dl 2 dl 4 dl] 0.5 0.5 0.5 DL } def
/P { stroke [] 0 setdash
  currentlinewidth 2 div sub M
  0 currentlinewidth V stroke } def
/D { stroke [] 0 setdash 2 copy vpt add M
  hpt neg vpt neg V hpt vpt neg V
  hpt vpt V hpt neg vpt V closepath stroke
  P } def
/A { stroke [] 0 setdash vpt sub M 0 vpt2 V
  currentpoint stroke M
  hpt neg vpt neg R hpt2 0 V stroke
  } def
/B { stroke [] 0 setdash 2 copy exch hpt sub exch vpt add M
  0 vpt2 neg V hpt2 0 V 0 vpt2 V
  hpt2 neg 0 V closepath stroke
  P } def
/C { stroke [] 0 setdash exch hpt sub exch vpt add M
  hpt2 vpt2 neg V currentpoint stroke M
  hpt2 neg 0 R hpt2 vpt2 V stroke } def
/T { stroke [] 0 setdash 2 copy vpt 1.12 mul add M
  hpt neg vpt -1.62 mul V
  hpt 2 mul 0 V
  hpt neg vpt 1.62 mul V closepath stroke
  P  } def
/S { 2 copy A C} def
end
}
\begin{picture}(3600,2160)(0,0)
\special{"
gnudict begin
gsave
50 50 translate
0.100 0.100 scale
0 setgray
/Helvetica findfont 100 scalefont setfont
newpath
-500.000000 -500.000000 translate
LTa
600 251 M
2817 0 V
600 251 M
0 1858 V
LTb
600 251 M
63 0 V
2754 0 R
-63 0 V
600 561 M
63 0 V
2754 0 R
-63 0 V
600 870 M
63 0 V
2754 0 R
-63 0 V
600 1180 M
63 0 V
2754 0 R
-63 0 V
600 1490 M
63 0 V
2754 0 R
-63 0 V
600 1799 M
63 0 V
2754 0 R
-63 0 V
600 2109 M
63 0 V
2754 0 R
-63 0 V
600 251 M
0 63 V
0 1795 R
0 -63 V
882 251 M
0 63 V
0 1795 R
0 -63 V
1163 251 M
0 63 V
0 1795 R
0 -63 V
1445 251 M
0 63 V
0 1795 R
0 -63 V
1727 251 M
0 63 V
0 1795 R
0 -63 V
2009 251 M
0 63 V
0 1795 R
0 -63 V
2290 251 M
0 63 V
0 1795 R
0 -63 V
2572 251 M
0 63 V
0 1795 R
0 -63 V
2854 251 M
0 63 V
0 1795 R
0 -63 V
3135 251 M
0 63 V
0 1795 R
0 -63 V
3417 251 M
0 63 V
0 1795 R
0 -63 V
600 251 M
2817 0 V
0 1858 V
-2817 0 V
600 251 L
LT0
3114 1946 M
180 0 V
607 251 M
7 0 V
7 0 V
7 0 V
7 0 V
7 0 V
7 0 V
7 1 V
7 1 V
7 1 V
7 2 V
8 2 V
7 3 V
7 3 V
7 3 V
7 3 V
7 3 V
7 4 V
7 4 V
7 4 V
7 3 V
7 4 V
7 4 V
7 3 V
7 3 V
7 4 V
7 2 V
7 3 V
7 3 V
7 2 V
7 2 V
7 2 V
7 2 V
7 1 V
7 1 V
8 1 V
7 1 V
7 0 V
7 1 V
7 0 V
7 0 V
7 0 V
7 0 V
7 0 V
7 -1 V
7 -1 V
7 0 V
7 -1 V
7 -1 V
7 -1 V
7 -1 V
7 -1 V
7 -1 V
7 -2 V
7 -1 V
7 -1 V
7 -1 V
7 -2 V
8 -1 V
7 -1 V
7 -2 V
7 -1 V
7 -2 V
7 -1 V
7 -1 V
7 -2 V
7 -1 V
7 -1 V
7 -2 V
7 -1 V
7 -1 V
7 -1 V
7 -2 V
7 -1 V
7 -1 V
7 -1 V
7 -1 V
7 -1 V
7 -1 V
7 -1 V
7 -1 V
7 -1 V
8 -1 V
7 -1 V
7 -1 V
7 -1 V
7 -1 V
7 -1 V
7 0 V
7 -1 V
7 -1 V
7 -1 V
7 0 V
7 -1 V
7 -1 V
7 -1 V
7 0 V
7 -1 V
7 0 V
7 -1 V
7 0 V
7 -1 V
7 0 V
7 -1 V
7 0 V
8 -1 V
7 0 V
7 -1 V
7 0 V
7 -1 V
7 0 V
7 0 V
7 -1 V
7 0 V
7 0 V
7 -1 V
7 0 V
7 0 V
7 -1 V
7 0 V
7 0 V
7 -1 V
7 0 V
7 0 V
7 0 V
7 -1 V
7 0 V
7 0 V
7 0 V
8 -1 V
7 0 V
7 0 V
7 0 V
7 0 V
7 0 V
7 -1 V
7 0 V
7 0 V
7 0 V
7 0 V
7 0 V
7 -1 V
7 0 V
7 0 V
7 0 V
7 0 V
7 0 V
7 0 V
7 0 V
7 -1 V
7 0 V
7 0 V
8 0 V
7 0 V
7 0 V
7 0 V
7 0 V
7 0 V
7 0 V
7 0 V
7 0 V
7 0 V
7 0 V
7 0 V
7 0 V
7 0 V
7 0 V
7 0 V
7 0 V
7 -1 V
7 0 V
7 0 V
7 0 V
7 0 V
7 0 V
7 0 V
8 0 V
7 0 V
7 0 V
7 0 V
7 0 V
7 0 V
7 0 V
7 0 V
7 0 V
7 0 V
7 0 V
7 0 V
7 0 V
7 0 V
7 0 V
7 0 V
7 0 V
7 0 V
7 0 V
7 0 V
7 -1 V
7 0 V
7 0 V
8 0 V
7 0 V
7 0 V
7 0 V
7 0 V
7 0 V
7 0 V
7 0 V
7 0 V
7 0 V
7 0 V
7 0 V
7 0 V
7 0 V
7 0 V
7 0 V
7 0 V
7 0 V
7 0 V
7 0 V
7 0 V
7 0 V
7 0 V
7 0 V
8 0 V
7 0 V
7 0 V
7 0 V
7 0 V
7 0 V
7 0 V
7 -1 V
7 0 V
7 0 V
7 0 V
7 0 V
7 0 V
7 0 V
7 0 V
7 0 V
7 0 V
7 0 V
7 0 V
7 0 V
7 0 V
7 0 V
7 0 V
7 0 V
8 0 V
7 0 V
7 0 V
7 0 V
7 0 V
7 0 V
7 0 V
7 0 V
7 0 V
7 0 V
7 0 V
7 0 V
7 0 V
7 0 V
7 0 V
7 0 V
7 0 V
7 0 V
7 0 V
7 0 V
7 0 V
7 0 V
7 0 V
8 0 V
7 0 V
7 0 V
7 0 V
7 0 V
7 0 V
7 0 V
7 0 V
7 0 V
7 0 V
7 0 V
7 0 V
7 0 V
7 0 V
7 0 V
7 0 V
7 0 V
7 0 V
7 0 V
7 0 V
7 0 V
7 0 V
7 0 V
7 0 V
8 0 V
7 0 V
7 0 V
7 0 V
7 0 V
7 -1 V
7 0 V
7 0 V
7 0 V
7 0 V
7 0 V
7 0 V
7 0 V
7 0 V
7 0 V
7 0 V
7 0 V
7 0 V
7 0 V
7 0 V
7 0 V
7 0 V
7 0 V
8 0 V
7 0 V
7 0 V
7 0 V
7 0 V
7 0 V
7 0 V
7 0 V
7 0 V
7 0 V
7 0 V
7 0 V
7 0 V
7 0 V
7 0 V
7 0 V
7 0 V
7 0 V
7 0 V
7 0 V
7 0 V
7 0 V
7 0 V
7 0 V
8 0 V
7 0 V
7 0 V
7 0 V
7 0 V
7 0 V
7 0 V
7 0 V
7 0 V
7 0 V
7 0 V
7 0 V
7 0 V
7 0 V
7 0 V
7 0 V
7 0 V
7 0 V
7 0 V
7 0 V
7 0 V
7 0 V
7 0 V
8 0 V
7 0 V
7 0 V
7 0 V
7 0 V
7 0 V
7 0 V
7 0 V
7 0 V
7 0 V
7 0 V
7 0 V
7 0 V
7 0 V
7 0 V
7 0 V
7 0 V
7 0 V
7 0 V
7 0 V
7 0 V
7 0 V
7 0 V
7 0 V
8 0 V
7 0 V
7 0 V
7 0 V
7 0 V
7 0 V
7 0 V
7 0 V
7 0 V
7 0 V
currentpoint stroke M
7 0 V
7 0 V
LT1
3114 1846 M
180 0 V
607 251 M
7 0 V
7 0 V
7 1 V
7 2 V
7 2 V
7 5 V
7 5 V
7 7 V
7 9 V
7 9 V
8 11 V
7 12 V
7 12 V
7 13 V
7 13 V
7 14 V
7 14 V
7 13 V
7 13 V
7 12 V
7 12 V
7 11 V
7 10 V
7 9 V
7 10 V
7 8 V
7 7 V
7 5 V
7 5 V
7 5 V
7 4 V
7 3 V
7 2 V
7 1 V
8 1 V
7 -1 V
7 0 V
7 0 V
7 -1 V
7 -2 V
7 -3 V
7 -3 V
7 -3 V
7 -4 V
7 -4 V
7 -4 V
7 -5 V
7 -4 V
7 -4 V
7 -5 V
7 -4 V
7 -5 V
7 -5 V
7 -5 V
7 -4 V
7 -5 V
7 -5 V
8 -5 V
7 -5 V
7 -5 V
7 -5 V
7 -5 V
7 -5 V
7 -5 V
7 -4 V
7 -5 V
7 -4 V
7 -5 V
7 -4 V
7 -4 V
7 -4 V
7 -4 V
7 -4 V
7 -3 V
7 -4 V
7 -3 V
7 -4 V
7 -3 V
7 -3 V
7 -3 V
7 -3 V
8 -3 V
7 -3 V
7 -3 V
7 -2 V
7 -3 V
7 -2 V
7 -2 V
7 -3 V
7 -2 V
7 -2 V
7 -2 V
7 -2 V
7 -2 V
7 -2 V
7 -2 V
7 -1 V
7 -2 V
7 -2 V
7 -1 V
7 -2 V
7 -1 V
7 -1 V
7 -2 V
8 -1 V
7 -1 V
7 -1 V
7 -2 V
7 -1 V
7 -1 V
7 -1 V
7 -1 V
7 -1 V
7 -1 V
7 0 V
7 -1 V
7 -1 V
7 -1 V
7 -1 V
7 0 V
7 -1 V
7 -1 V
7 0 V
7 -1 V
7 -1 V
7 0 V
7 -1 V
7 0 V
8 -1 V
7 0 V
7 -1 V
7 0 V
7 -1 V
7 0 V
7 -1 V
7 0 V
7 -1 V
7 0 V
7 0 V
7 -1 V
7 0 V
7 0 V
7 -1 V
7 0 V
7 0 V
7 -1 V
7 0 V
7 0 V
7 0 V
7 -1 V
7 0 V
8 0 V
7 0 V
7 -1 V
7 0 V
7 0 V
7 0 V
7 0 V
7 -1 V
7 0 V
7 0 V
7 0 V
7 0 V
7 0 V
7 0 V
7 -1 V
7 0 V
7 0 V
7 0 V
7 0 V
7 0 V
7 0 V
7 0 V
7 -1 V
7 0 V
8 0 V
7 0 V
7 0 V
7 0 V
7 0 V
7 0 V
7 0 V
7 0 V
7 -1 V
7 0 V
7 0 V
7 0 V
7 0 V
7 0 V
7 0 V
7 0 V
7 0 V
7 0 V
7 0 V
7 0 V
7 0 V
7 -1 V
7 0 V
8 0 V
7 0 V
7 0 V
7 0 V
7 0 V
7 0 V
7 0 V
7 0 V
7 0 V
7 0 V
7 0 V
7 0 V
7 0 V
7 0 V
7 0 V
7 0 V
7 -1 V
7 0 V
7 0 V
7 0 V
7 0 V
7 0 V
7 0 V
7 0 V
8 0 V
7 0 V
7 0 V
7 0 V
7 0 V
7 0 V
7 0 V
7 0 V
7 0 V
7 0 V
7 0 V
7 0 V
7 0 V
7 0 V
7 0 V
7 0 V
7 0 V
7 0 V
7 0 V
7 0 V
7 0 V
7 0 V
7 -1 V
7 0 V
8 0 V
7 0 V
7 0 V
7 0 V
7 0 V
7 0 V
7 0 V
7 0 V
7 0 V
7 0 V
7 0 V
7 0 V
7 0 V
7 0 V
7 0 V
7 0 V
7 0 V
7 0 V
7 0 V
7 0 V
7 0 V
7 0 V
7 0 V
8 0 V
7 0 V
7 0 V
7 0 V
7 0 V
7 0 V
7 0 V
7 0 V
7 0 V
7 0 V
7 0 V
7 0 V
7 0 V
7 0 V
7 0 V
7 0 V
7 0 V
7 0 V
7 0 V
7 0 V
7 0 V
7 0 V
7 0 V
7 0 V
8 0 V
7 0 V
7 0 V
7 0 V
7 0 V
7 0 V
7 0 V
7 0 V
7 0 V
7 0 V
7 0 V
7 0 V
7 0 V
7 0 V
7 0 V
7 0 V
7 0 V
7 0 V
7 -1 V
7 0 V
7 0 V
7 0 V
7 0 V
8 0 V
7 0 V
7 0 V
7 0 V
7 0 V
7 0 V
7 0 V
7 0 V
7 0 V
7 0 V
7 0 V
7 0 V
7 0 V
7 0 V
7 0 V
7 0 V
7 0 V
7 0 V
7 0 V
7 0 V
7 0 V
7 0 V
7 0 V
7 0 V
8 0 V
7 0 V
7 0 V
7 0 V
7 0 V
7 0 V
7 0 V
7 0 V
7 0 V
7 0 V
7 0 V
7 0 V
7 0 V
7 0 V
7 0 V
7 0 V
7 0 V
7 0 V
7 0 V
7 0 V
7 0 V
7 0 V
7 0 V
8 0 V
7 0 V
7 0 V
7 0 V
7 0 V
7 0 V
7 0 V
7 0 V
7 0 V
7 0 V
7 0 V
7 0 V
7 0 V
7 0 V
7 0 V
7 0 V
7 0 V
7 0 V
7 0 V
7 0 V
7 0 V
7 0 V
7 0 V
7 0 V
8 0 V
7 0 V
7 0 V
7 0 V
7 0 V
7 0 V
7 0 V
7 0 V
7 0 V
7 0 V
currentpoint stroke M
7 0 V
7 0 V
LT3
3114 1746 M
180 0 V
607 251 M
7 0 V
7 3 V
7 4 V
7 9 V
7 13 V
7 18 V
7 21 V
7 26 V
7 30 V
7 33 V
8 36 V
7 36 V
7 40 V
7 38 V
7 40 V
7 36 V
7 40 V
7 35 V
7 34 V
7 34 V
7 29 V
7 27 V
7 27 V
7 23 V
7 18 V
7 17 V
7 16 V
7 14 V
7 9 V
7 7 V
7 3 V
7 4 V
7 3 V
7 0 V
8 -2 V
7 -4 V
7 -6 V
7 -8 V
7 -9 V
7 -9 V
7 -9 V
7 -10 V
7 -11 V
7 -12 V
7 -12 V
7 -14 V
7 -13 V
7 -14 V
7 -15 V
7 -15 V
7 -14 V
7 -15 V
7 -15 V
7 -15 V
7 -15 V
7 -13 V
7 -14 V
8 -14 V
7 -13 V
7 -14 V
7 -13 V
7 -13 V
7 -13 V
7 -12 V
7 -12 V
7 -12 V
7 -12 V
7 -12 V
7 -11 V
7 -10 V
7 -11 V
7 -10 V
7 -10 V
7 -9 V
7 -10 V
7 -9 V
7 -8 V
7 -9 V
7 -8 V
7 -8 V
7 -7 V
8 -7 V
7 -7 V
7 -7 V
7 -7 V
7 -6 V
7 -6 V
7 -6 V
7 -6 V
7 -5 V
7 -5 V
7 -5 V
7 -5 V
7 -5 V
7 -4 V
7 -5 V
7 -4 V
7 -4 V
7 -3 V
7 -4 V
7 -4 V
7 -3 V
7 -3 V
7 -3 V
8 -4 V
7 -3 V
7 -3 V
7 -2 V
7 -3 V
7 -3 V
7 -2 V
7 -2 V
7 -3 V
7 -2 V
7 -2 V
7 -2 V
7 -2 V
7 -2 V
7 -2 V
7 -1 V
7 -2 V
7 -1 V
7 -2 V
7 -1 V
7 -2 V
7 -1 V
7 -1 V
7 -2 V
8 -1 V
7 -1 V
7 -1 V
7 -1 V
7 -1 V
7 -1 V
7 -1 V
7 -1 V
7 -1 V
7 -1 V
7 0 V
7 -1 V
7 -1 V
7 -1 V
7 0 V
7 -1 V
7 -1 V
7 0 V
7 -1 V
7 0 V
7 -1 V
7 0 V
7 -1 V
8 0 V
7 -1 V
7 0 V
7 -1 V
7 0 V
7 -1 V
7 0 V
7 0 V
7 -1 V
7 0 V
7 -1 V
7 0 V
7 0 V
7 -1 V
7 0 V
7 0 V
7 0 V
7 -1 V
7 0 V
7 0 V
7 -1 V
7 0 V
7 0 V
7 0 V
8 -1 V
7 0 V
7 0 V
7 0 V
7 0 V
7 -1 V
7 0 V
7 0 V
7 0 V
7 0 V
7 0 V
7 -1 V
7 0 V
7 0 V
7 0 V
7 0 V
7 0 V
7 0 V
7 -1 V
7 0 V
7 0 V
7 0 V
7 0 V
8 0 V
7 0 V
7 0 V
7 0 V
7 -1 V
7 0 V
7 0 V
7 0 V
7 0 V
7 0 V
7 0 V
7 0 V
7 0 V
7 0 V
7 0 V
7 -1 V
7 0 V
7 0 V
7 0 V
7 0 V
7 0 V
7 0 V
7 0 V
7 0 V
8 0 V
7 0 V
7 0 V
7 0 V
7 0 V
7 0 V
7 0 V
7 -1 V
7 0 V
7 0 V
7 0 V
7 0 V
7 0 V
7 0 V
7 0 V
7 0 V
7 0 V
7 0 V
7 0 V
7 0 V
7 0 V
7 0 V
7 0 V
7 0 V
8 0 V
7 0 V
7 0 V
7 0 V
7 0 V
7 0 V
7 0 V
7 0 V
7 0 V
7 -1 V
7 0 V
7 0 V
7 0 V
7 0 V
7 0 V
7 0 V
7 0 V
7 0 V
7 0 V
7 0 V
7 0 V
7 0 V
7 0 V
8 0 V
7 0 V
7 0 V
7 0 V
7 0 V
7 0 V
7 0 V
7 0 V
7 0 V
7 0 V
7 0 V
7 0 V
7 0 V
7 0 V
7 0 V
7 0 V
7 0 V
7 0 V
7 0 V
7 0 V
7 0 V
7 0 V
7 0 V
7 0 V
8 0 V
7 0 V
7 0 V
7 0 V
7 0 V
7 0 V
7 0 V
7 0 V
7 0 V
7 0 V
7 0 V
7 0 V
7 0 V
7 0 V
7 0 V
7 0 V
7 0 V
7 0 V
7 0 V
7 0 V
7 0 V
7 0 V
7 0 V
8 0 V
7 0 V
7 0 V
7 0 V
7 0 V
7 0 V
7 0 V
7 -1 V
7 0 V
7 0 V
7 0 V
7 0 V
7 0 V
7 0 V
7 0 V
7 0 V
7 0 V
7 0 V
7 0 V
7 0 V
7 0 V
7 0 V
7 0 V
7 0 V
8 0 V
7 0 V
7 0 V
7 0 V
7 0 V
7 0 V
7 0 V
7 0 V
7 0 V
7 0 V
7 0 V
7 0 V
7 0 V
7 0 V
7 0 V
7 0 V
7 0 V
7 0 V
7 0 V
7 0 V
7 0 V
7 0 V
7 0 V
8 0 V
7 0 V
7 0 V
7 0 V
7 0 V
7 0 V
7 0 V
7 0 V
7 0 V
7 0 V
7 0 V
7 0 V
7 0 V
7 0 V
7 0 V
7 0 V
7 0 V
7 0 V
7 0 V
7 0 V
7 0 V
7 0 V
7 0 V
7 0 V
8 0 V
7 0 V
7 0 V
7 0 V
7 0 V
7 0 V
7 0 V
7 0 V
7 0 V
7 0 V
currentpoint stroke M
7 0 V
7 0 V
3114 1646 M
180 0 V
607 251 M
7 3 V
7 9 V
7 19 V
7 31 V
7 42 V
7 56 V
7 68 V
7 77 V
7 86 V
7 92 V
8 97 V
7 103 V
7 99 V
7 104 V
7 97 V
7 100 V
7 85 V
7 94 V
7 81 V
7 71 V
7 75 V
7 63 V
7 49 V
7 50 V
7 46 V
7 35 V
7 24 V
7 15 V
7 21 V
7 13 V
7 6 V
7 -1 V
7 -8 V
7 -14 V
8 -14 V
7 -14 V
7 -17 V
7 -22 V
7 -25 V
7 -28 V
7 -31 V
7 -33 V
7 -35 V
7 -36 V
7 -38 V
7 -35 V
7 -35 V
7 -36 V
7 -36 V
7 -37 V
7 -36 V
7 -37 V
7 -37 V
7 -36 V
7 -36 V
7 -36 V
7 -36 V
8 -35 V
7 -35 V
7 -33 V
7 -34 V
7 -32 V
7 -32 V
7 -31 V
7 -30 V
7 -29 V
7 -29 V
7 -27 V
7 -27 V
7 -31 V
7 -29 V
7 -28 V
7 -27 V
7 -25 V
7 -25 V
7 -23 V
7 -23 V
7 -21 V
7 -20 V
7 -20 V
7 -19 V
8 -17 V
7 -17 V
7 -17 V
7 -15 V
7 -15 V
7 -14 V
7 -13 V
7 -13 V
7 -12 V
7 -12 V
7 -11 V
7 -11 V
7 -10 V
7 -9 V
7 -9 V
7 -9 V
7 -9 V
7 -7 V
7 -8 V
7 -7 V
7 -7 V
7 -7 V
7 -6 V
8 -6 V
7 -6 V
7 -5 V
7 -5 V
7 -5 V
7 -5 V
7 -4 V
7 -5 V
7 -4 V
7 -4 V
7 -3 V
7 -4 V
7 -3 V
7 -3 V
7 -4 V
7 -3 V
7 -2 V
7 -3 V
7 -3 V
7 -2 V
7 -3 V
7 -2 V
7 -2 V
7 -2 V
8 -2 V
7 -2 V
7 -2 V
7 -1 V
7 -2 V
7 -2 V
7 -1 V
7 -2 V
7 -1 V
7 -2 V
7 -1 V
7 -1 V
7 -1 V
7 -1 V
7 -1 V
7 -2 V
7 -1 V
7 -1 V
7 0 V
7 -1 V
7 -1 V
7 -1 V
7 -1 V
8 -1 V
7 0 V
7 -1 V
7 -1 V
7 0 V
7 -1 V
7 -1 V
7 0 V
7 -1 V
7 0 V
7 -1 V
7 0 V
7 -1 V
7 0 V
7 -1 V
7 0 V
7 -1 V
7 0 V
7 0 V
7 -1 V
7 0 V
7 -1 V
7 0 V
7 0 V
8 -1 V
7 0 V
7 0 V
7 0 V
7 -1 V
7 0 V
7 0 V
7 -1 V
7 0 V
7 0 V
7 0 V
7 -1 V
7 0 V
7 0 V
7 0 V
7 0 V
7 -1 V
7 0 V
7 0 V
7 0 V
7 0 V
7 0 V
7 -1 V
8 0 V
7 0 V
7 0 V
7 0 V
7 0 V
7 0 V
7 -1 V
7 0 V
7 0 V
7 0 V
7 0 V
7 0 V
7 0 V
7 0 V
7 -1 V
7 0 V
7 0 V
7 0 V
7 0 V
7 0 V
7 0 V
7 0 V
7 0 V
7 0 V
8 0 V
7 0 V
7 -1 V
7 0 V
7 0 V
7 0 V
7 0 V
7 0 V
7 0 V
7 0 V
7 0 V
7 0 V
7 0 V
7 0 V
7 0 V
7 0 V
7 0 V
7 -1 V
7 0 V
7 0 V
7 0 V
7 0 V
7 0 V
7 0 V
8 0 V
7 0 V
7 0 V
7 0 V
7 0 V
7 0 V
7 0 V
7 0 V
7 0 V
7 0 V
7 0 V
7 0 V
7 0 V
7 0 V
7 0 V
7 0 V
7 0 V
7 0 V
7 0 V
7 0 V
7 -1 V
7 0 V
7 0 V
8 0 V
7 0 V
7 0 V
7 0 V
7 0 V
7 0 V
7 0 V
7 0 V
7 0 V
7 0 V
7 0 V
7 0 V
7 0 V
7 0 V
7 0 V
7 0 V
7 0 V
7 0 V
7 0 V
7 0 V
7 0 V
7 0 V
7 0 V
7 0 V
8 0 V
7 0 V
7 0 V
7 0 V
7 0 V
7 0 V
7 0 V
7 0 V
7 0 V
7 0 V
7 0 V
7 0 V
7 0 V
7 0 V
7 0 V
7 0 V
7 0 V
7 0 V
7 0 V
7 0 V
7 0 V
7 0 V
7 0 V
8 0 V
7 0 V
7 0 V
7 0 V
7 0 V
7 0 V
7 0 V
7 0 V
7 0 V
7 0 V
7 0 V
7 0 V
7 0 V
7 0 V
7 0 V
7 0 V
7 0 V
7 0 V
7 0 V
7 0 V
7 0 V
7 -1 V
7 0 V
7 0 V
8 0 V
7 0 V
7 0 V
7 0 V
7 0 V
7 0 V
7 0 V
7 0 V
7 0 V
7 0 V
7 0 V
7 0 V
7 0 V
7 0 V
7 0 V
7 0 V
7 0 V
7 0 V
7 0 V
7 0 V
7 0 V
7 0 V
7 0 V
8 0 V
7 0 V
7 0 V
7 0 V
7 0 V
7 0 V
7 0 V
7 0 V
7 0 V
7 0 V
7 0 V
7 0 V
7 0 V
7 0 V
7 0 V
7 0 V
7 0 V
7 0 V
7 0 V
7 0 V
7 0 V
7 0 V
7 0 V
7 0 V
8 0 V
7 0 V
7 0 V
7 0 V
7 0 V
7 0 V
7 0 V
7 0 V
currentpoint stroke M
7 0 V
7 0 V
7 0 V
7 0 V
stroke
grestore
end
showpage
}
\put(3054,1646){\makebox(0,0)[r]{$x=10^{-5}$}}
\put(3054,1746){\makebox(0,0)[r]{$x=10^{-4}$}}
\put(3054,1846){\makebox(0,0)[r]{$x=10^{-3}$}}
\put(3054,1946){\makebox(0,0)[r]{$Q^2 =40, x=10^{-2}$}}
\put(2008,51){\makebox(0,0){\Large $b$ (fm)}}
\put(100,1180){%
\special{ps: gsave currentpoint currentpoint translate
270 rotate neg exch neg exch translate}%
\makebox(0,0)[b]{\shortstack{\Large  $I_{L} (b) $ }}%
\special{ps: currentpoint grestore moveto}%
}
\put(3417,151){\makebox(0,0){1}}
\put(3135,151){\makebox(0,0){0.9}}
\put(2854,151){\makebox(0,0){0.8}}
\put(2572,151){\makebox(0,0){0.7}}
\put(2290,151){\makebox(0,0){0.6}}
\put(2009,151){\makebox(0,0){0.5}}
\put(1727,151){\makebox(0,0){0.4}}
\put(1445,151){\makebox(0,0){0.3}}
\put(1163,151){\makebox(0,0){0.2}}
\put(882,151){\makebox(0,0){0.1}}
\put(600,151){\makebox(0,0){0}}
\put(540,2109){\makebox(0,0)[r]{0.03}}
\put(540,1799){\makebox(0,0)[r]{0.025}}
\put(540,1490){\makebox(0,0)[r]{0.02}}
\put(540,1180){\makebox(0,0)[r]{0.015}}
\put(540,870){\makebox(0,0)[r]{0.01}}
\put(540,561){\makebox(0,0)[r]{0.005}}
\put(540,251){\makebox(0,0)[r]{0}}
\end{picture}

%% file: itb4.tex
% GNUPLOT: LaTeX picture with Postscript
\setlength{\unitlength}{0.1bp}
\special{!
%!PS-Adobe-2.0
%%Creator: gnuplot
%%DocumentFonts: Helvetica
%%BoundingBox: 50 50 770 554
%%Pages: (atend)
%%EndComments
/gnudict 40 dict def
gnudict begin
/Color false def
/Solid false def
/gnulinewidth 5.000 def
/vshift -33 def
/dl {10 mul} def
/hpt 31.5 def
/vpt 31.5 def
/M {moveto} bind def
/L {lineto} bind def
/R {rmoveto} bind def
/V {rlineto} bind def
/vpt2 vpt 2 mul def
/hpt2 hpt 2 mul def
/Lshow { currentpoint stroke M
  0 vshift R show } def
/Rshow { currentpoint stroke M
  dup stringwidth pop neg vshift R show } def
/Cshow { currentpoint stroke M
  dup stringwidth pop -2 div vshift R show } def
/DL { Color {setrgbcolor Solid {pop []} if 0 setdash }
 {pop pop pop Solid {pop []} if 0 setdash} ifelse } def
/BL { stroke gnulinewidth 2 mul setlinewidth } def
/AL { stroke gnulinewidth 2 div setlinewidth } def
/PL { stroke gnulinewidth setlinewidth } def
/LTb { BL [] 0 0 0 DL } def
/LTa { AL [1 dl 2 dl] 0 setdash 0 0 0 setrgbcolor } def
/LT0 { PL [] 0 1 0 DL } def
/LT1 { PL [4 dl 2 dl] 0 0 1 DL } def
/LT2 { PL [2 dl 3 dl] 1 0 0 DL } def
/LT3 { PL [1 dl 1.5 dl] 1 0 1 DL } def
/LT4 { PL [5 dl 2 dl 1 dl 2 dl] 0 1 1 DL } def
/LT5 { PL [4 dl 3 dl 1 dl 3 dl] 1 1 0 DL } def
/LT6 { PL [2 dl 2 dl 2 dl 4 dl] 0 0 0 DL } def
/LT7 { PL [2 dl 2 dl 2 dl 2 dl 2 dl 4 dl] 1 0.3 0 DL } def
/LT8 { PL [2 dl 2 dl 2 dl 2 dl 2 dl 2 dl 2 dl 4 dl] 0.5 0.5 0.5 DL } def
/P { stroke [] 0 setdash
  currentlinewidth 2 div sub M
  0 currentlinewidth V stroke } def
/D { stroke [] 0 setdash 2 copy vpt add M
  hpt neg vpt neg V hpt vpt neg V
  hpt vpt V hpt neg vpt V closepath stroke
  P } def
/A { stroke [] 0 setdash vpt sub M 0 vpt2 V
  currentpoint stroke M
  hpt neg vpt neg R hpt2 0 V stroke
  } def
/B { stroke [] 0 setdash 2 copy exch hpt sub exch vpt add M
  0 vpt2 neg V hpt2 0 V 0 vpt2 V
  hpt2 neg 0 V closepath stroke
  P } def
/C { stroke [] 0 setdash exch hpt sub exch vpt add M
  hpt2 vpt2 neg V currentpoint stroke M
  hpt2 neg 0 R hpt2 vpt2 V stroke } def
/T { stroke [] 0 setdash 2 copy vpt 1.12 mul add M
  hpt neg vpt -1.62 mul V
  hpt 2 mul 0 V
  hpt neg vpt 1.62 mul V closepath stroke
  P  } def
/S { 2 copy A C} def
end
}
\begin{picture}(3600,2160)(0,0)
\special{"
gnudict begin
gsave
50 50 translate
0.100 0.100 scale
0 setgray
/Helvetica findfont 100 scalefont setfont
newpath
-500.000000 -500.000000 translate
LTa
600 251 M
2817 0 V
600 251 M
0 1858 V
LTb
600 251 M
63 0 V
2754 0 R
-63 0 V
600 483 M
63 0 V
2754 0 R
-63 0 V
600 716 M
63 0 V
2754 0 R
-63 0 V
600 948 M
63 0 V
2754 0 R
-63 0 V
600 1180 M
63 0 V
2754 0 R
-63 0 V
600 1412 M
63 0 V
2754 0 R
-63 0 V
600 1645 M
63 0 V
2754 0 R
-63 0 V
600 1877 M
63 0 V
2754 0 R
-63 0 V
600 2109 M
63 0 V
2754 0 R
-63 0 V
600 251 M
0 63 V
0 1795 R
0 -63 V
882 251 M
0 63 V
0 1795 R
0 -63 V
1163 251 M
0 63 V
0 1795 R
0 -63 V
1445 251 M
0 63 V
0 1795 R
0 -63 V
1727 251 M
0 63 V
0 1795 R
0 -63 V
2009 251 M
0 63 V
0 1795 R
0 -63 V
2290 251 M
0 63 V
0 1795 R
0 -63 V
2572 251 M
0 63 V
0 1795 R
0 -63 V
2854 251 M
0 63 V
0 1795 R
0 -63 V
3135 251 M
0 63 V
0 1795 R
0 -63 V
3417 251 M
0 63 V
0 1795 R
0 -63 V
600 251 M
2817 0 V
0 1858 V
-2817 0 V
600 251 L
LT0
3114 1946 M
180 0 V
607 251 M
7 0 V
7 0 V
7 0 V
7 0 V
7 0 V
7 0 V
7 0 V
7 0 V
7 0 V
7 0 V
8 0 V
7 0 V
7 0 V
7 0 V
7 0 V
7 0 V
7 1 V
7 0 V
7 1 V
7 0 V
7 1 V
7 1 V
7 1 V
7 1 V
7 1 V
7 1 V
7 1 V
7 1 V
7 2 V
7 1 V
7 2 V
7 2 V
7 1 V
7 2 V
8 2 V
7 2 V
7 2 V
7 2 V
7 2 V
7 1 V
7 3 V
7 2 V
7 2 V
7 2 V
7 2 V
7 2 V
7 2 V
7 2 V
7 3 V
7 2 V
7 2 V
7 2 V
7 2 V
7 3 V
7 2 V
7 2 V
7 2 V
8 2 V
7 2 V
7 2 V
7 2 V
7 2 V
7 2 V
7 2 V
7 2 V
7 2 V
7 2 V
7 2 V
7 1 V
7 2 V
7 2 V
7 1 V
7 2 V
7 2 V
7 1 V
7 2 V
7 1 V
7 2 V
7 1 V
7 2 V
7 1 V
8 1 V
7 1 V
7 1 V
7 2 V
7 1 V
7 1 V
7 1 V
7 1 V
7 1 V
7 1 V
7 1 V
7 1 V
7 1 V
7 1 V
7 0 V
7 1 V
7 1 V
7 0 V
7 1 V
7 1 V
7 0 V
7 1 V
7 1 V
8 0 V
7 1 V
7 0 V
7 0 V
7 1 V
7 0 V
7 0 V
7 0 V
7 1 V
7 0 V
7 0 V
7 0 V
7 0 V
7 0 V
7 0 V
7 0 V
7 0 V
7 0 V
7 1 V
7 0 V
7 0 V
7 -1 V
7 0 V
7 0 V
8 0 V
7 0 V
7 0 V
7 0 V
7 -1 V
7 0 V
7 0 V
7 0 V
7 -1 V
7 0 V
7 0 V
7 -1 V
7 0 V
7 0 V
7 -1 V
7 0 V
7 -1 V
7 0 V
7 -1 V
7 0 V
7 -1 V
7 0 V
7 -1 V
8 0 V
7 0 V
7 -1 V
7 0 V
7 2 V
7 2 V
7 2 V
7 2 V
7 2 V
7 2 V
7 2 V
7 1 V
7 2 V
7 1 V
7 2 V
7 1 V
7 1 V
7 1 V
7 2 V
7 1 V
7 1 V
7 0 V
7 1 V
7 1 V
8 1 V
7 0 V
7 1 V
7 1 V
7 0 V
7 1 V
7 0 V
7 0 V
7 1 V
7 0 V
7 0 V
7 0 V
7 0 V
7 1 V
7 0 V
7 0 V
7 0 V
7 0 V
7 0 V
7 -1 V
7 0 V
7 0 V
7 0 V
8 0 V
7 0 V
7 -1 V
7 0 V
7 0 V
7 -1 V
7 0 V
7 0 V
7 -1 V
7 0 V
7 -1 V
7 0 V
7 0 V
7 -1 V
7 0 V
7 -1 V
7 0 V
7 -1 V
7 -1 V
7 0 V
7 -1 V
7 0 V
7 -1 V
7 0 V
8 -1 V
7 -1 V
7 0 V
7 -1 V
7 -1 V
7 0 V
7 -1 V
7 -1 V
7 0 V
7 -1 V
7 -1 V
7 0 V
7 -1 V
7 -1 V
7 -1 V
7 0 V
7 -1 V
7 -1 V
7 0 V
7 -1 V
7 -1 V
7 -1 V
7 0 V
7 -1 V
8 -1 V
7 -1 V
7 0 V
7 -1 V
7 -1 V
7 -1 V
7 0 V
7 -1 V
7 -1 V
7 -1 V
7 0 V
7 -1 V
7 -1 V
7 -1 V
7 -1 V
7 -1 V
7 -1 V
7 -1 V
7 -1 V
7 -1 V
7 -1 V
7 -1 V
7 -1 V
8 -1 V
7 -1 V
7 -1 V
7 -1 V
7 -1 V
7 -1 V
7 -1 V
7 -1 V
7 -1 V
7 0 V
7 -1 V
7 -1 V
7 -1 V
7 -1 V
7 -1 V
7 -1 V
7 0 V
7 -1 V
7 -1 V
7 -1 V
7 -1 V
7 0 V
7 -1 V
7 -1 V
8 -1 V
7 0 V
7 -1 V
7 -1 V
7 -1 V
7 0 V
7 -1 V
7 -1 V
7 0 V
7 -1 V
7 -1 V
7 0 V
7 -1 V
7 -1 V
7 0 V
7 -1 V
7 -1 V
7 0 V
7 -1 V
7 -1 V
7 0 V
7 -1 V
7 0 V
8 -1 V
7 -1 V
7 0 V
7 -1 V
7 0 V
7 -1 V
7 -1 V
7 0 V
7 -1 V
7 0 V
7 -1 V
7 0 V
7 -1 V
7 0 V
7 -1 V
7 0 V
7 -1 V
7 0 V
7 -1 V
7 0 V
7 -1 V
7 0 V
7 -1 V
7 0 V
8 -1 V
7 0 V
7 -1 V
7 0 V
7 -1 V
7 0 V
7 -1 V
7 0 V
7 0 V
7 -1 V
7 0 V
7 -1 V
7 0 V
7 -1 V
7 0 V
7 0 V
7 -1 V
7 0 V
7 -1 V
7 0 V
7 0 V
7 -1 V
7 0 V
8 0 V
7 -1 V
7 0 V
7 -1 V
7 0 V
7 0 V
7 -1 V
7 0 V
7 0 V
7 -1 V
7 0 V
7 0 V
7 -1 V
7 0 V
7 0 V
7 -1 V
7 0 V
7 0 V
7 -1 V
7 0 V
7 0 V
7 -1 V
7 0 V
7 0 V
8 -1 V
7 0 V
7 0 V
7 -1 V
7 0 V
7 0 V
7 0 V
7 -1 V
7 0 V
7 0 V
currentpoint stroke M
7 -1 V
7 0 V
LT1
3114 1846 M
180 0 V
607 251 M
7 0 V
7 0 V
7 0 V
7 0 V
7 0 V
7 1 V
7 0 V
7 1 V
7 2 V
7 1 V
8 2 V
7 3 V
7 2 V
7 3 V
7 3 V
7 4 V
7 4 V
7 4 V
7 4 V
7 5 V
7 4 V
7 5 V
7 5 V
7 6 V
7 5 V
7 6 V
7 5 V
7 6 V
7 6 V
7 6 V
7 6 V
7 6 V
7 7 V
7 6 V
8 5 V
7 7 V
7 6 V
7 6 V
7 6 V
7 7 V
7 6 V
7 6 V
7 6 V
7 6 V
7 6 V
7 6 V
7 6 V
7 6 V
7 5 V
7 5 V
7 6 V
7 5 V
7 6 V
7 5 V
7 5 V
7 5 V
7 4 V
8 4 V
7 5 V
7 5 V
7 5 V
7 4 V
7 4 V
7 4 V
7 3 V
7 3 V
7 4 V
7 3 V
7 4 V
7 4 V
7 3 V
7 3 V
7 3 V
7 2 V
7 2 V
7 2 V
7 2 V
7 2 V
7 2 V
7 3 V
7 2 V
8 2 V
7 2 V
7 1 V
7 2 V
7 1 V
7 1 V
7 1 V
7 1 V
7 0 V
7 1 V
7 0 V
7 0 V
7 1 V
7 1 V
7 0 V
7 1 V
7 0 V
7 0 V
7 0 V
7 0 V
7 0 V
7 0 V
7 0 V
8 -1 V
7 0 V
7 -1 V
7 -1 V
7 0 V
7 -1 V
7 -1 V
7 -1 V
7 -2 V
7 -1 V
7 -1 V
7 -1 V
7 -1 V
7 -1 V
7 -1 V
7 -1 V
7 -1 V
7 -1 V
7 -2 V
7 -1 V
7 -1 V
7 -2 V
7 -1 V
7 -2 V
8 -1 V
7 -2 V
7 -2 V
7 -1 V
7 -2 V
7 -2 V
7 -1 V
7 -2 V
7 -2 V
7 -2 V
7 -2 V
7 -2 V
7 -2 V
7 -2 V
7 -2 V
7 -2 V
7 -2 V
7 -2 V
7 -2 V
7 -2 V
7 -2 V
7 -3 V
7 -2 V
8 -2 V
7 -2 V
7 -2 V
7 -2 V
7 0 V
7 0 V
7 -1 V
7 0 V
7 -1 V
7 0 V
7 -1 V
7 -1 V
7 0 V
7 -1 V
7 -1 V
7 -1 V
7 -1 V
7 0 V
7 -1 V
7 -1 V
7 -1 V
7 -1 V
7 -1 V
7 -1 V
8 -1 V
7 -1 V
7 -1 V
7 -1 V
7 -1 V
7 -2 V
7 -1 V
7 -1 V
7 -1 V
7 -1 V
7 -1 V
7 -2 V
7 -1 V
7 -1 V
7 -1 V
7 -1 V
7 -2 V
7 -1 V
7 -1 V
7 -1 V
7 -2 V
7 -1 V
7 -1 V
8 -2 V
7 -1 V
7 -1 V
7 -2 V
7 -1 V
7 -1 V
7 -1 V
7 -2 V
7 -1 V
7 -1 V
7 -2 V
7 -1 V
7 -1 V
7 -2 V
7 -1 V
7 -1 V
7 -2 V
7 -1 V
7 -1 V
7 -1 V
7 -2 V
7 -1 V
7 -1 V
7 -2 V
8 -1 V
7 -1 V
7 -2 V
7 -1 V
7 -1 V
7 -1 V
7 -2 V
7 -1 V
7 -1 V
7 -1 V
7 -2 V
7 -1 V
7 -1 V
7 -1 V
7 -2 V
7 -1 V
7 -1 V
7 -1 V
7 -1 V
7 -2 V
7 -1 V
7 -1 V
7 -1 V
7 -1 V
8 -2 V
7 -1 V
7 -1 V
7 -1 V
7 -1 V
7 -1 V
7 -2 V
7 -1 V
7 -1 V
7 -1 V
7 -1 V
7 -1 V
7 -1 V
7 -2 V
7 -1 V
7 -1 V
7 -1 V
7 -2 V
7 -1 V
7 -1 V
7 -1 V
7 -1 V
7 -2 V
8 -1 V
7 -1 V
7 -1 V
7 -1 V
7 -1 V
7 -1 V
7 -2 V
7 -1 V
7 -1 V
7 -1 V
7 -1 V
7 -1 V
7 -1 V
7 -1 V
7 -1 V
7 -1 V
7 -1 V
7 -1 V
7 -1 V
7 -1 V
7 -1 V
7 -1 V
7 -1 V
7 0 V
8 -1 V
7 -1 V
7 -1 V
7 -1 V
7 -1 V
7 -1 V
7 -1 V
7 -1 V
7 0 V
7 -1 V
7 -1 V
7 -1 V
7 -1 V
7 0 V
7 -1 V
7 -1 V
7 -1 V
7 -1 V
7 0 V
7 -1 V
7 -1 V
7 0 V
7 -1 V
8 -1 V
7 -1 V
7 0 V
7 -1 V
7 -1 V
7 0 V
7 -1 V
7 -1 V
7 0 V
7 -1 V
7 -1 V
7 0 V
7 -1 V
7 -1 V
7 0 V
7 -1 V
7 -1 V
7 0 V
7 -1 V
7 0 V
7 -1 V
7 0 V
7 -1 V
7 -1 V
8 0 V
7 -1 V
7 0 V
7 -1 V
7 0 V
7 -1 V
7 -1 V
7 0 V
7 -1 V
7 0 V
7 -1 V
7 0 V
7 -1 V
7 0 V
7 -1 V
7 0 V
7 -1 V
7 0 V
7 -1 V
7 0 V
7 0 V
7 -1 V
7 0 V
8 -1 V
7 0 V
7 -1 V
7 0 V
7 -1 V
7 0 V
7 -1 V
7 0 V
7 0 V
7 -1 V
7 0 V
7 -1 V
7 0 V
7 0 V
7 -1 V
7 0 V
7 -1 V
7 0 V
7 0 V
7 -1 V
7 0 V
7 -1 V
7 0 V
7 0 V
8 -1 V
7 0 V
7 0 V
7 -1 V
7 0 V
7 0 V
7 -1 V
7 0 V
7 -1 V
7 0 V
currentpoint stroke M
7 0 V
7 -1 V
LT3
3114 1746 M
180 0 V
607 251 M
7 0 V
7 1 V
7 1 V
7 3 V
7 3 V
7 5 V
7 5 V
7 6 V
7 8 V
7 8 V
8 9 V
7 10 V
7 11 V
7 11 V
7 12 V
7 13 V
7 13 V
7 14 V
7 14 V
7 14 V
7 15 V
7 14 V
7 17 V
7 15 V
7 15 V
7 17 V
7 16 V
7 14 V
7 18 V
7 16 V
7 15 V
7 16 V
7 17 V
7 15 V
8 15 V
7 15 V
7 16 V
7 16 V
7 14 V
7 13 V
7 15 V
7 15 V
7 14 V
7 13 V
7 12 V
7 12 V
7 14 V
7 13 V
7 13 V
7 11 V
7 11 V
7 9 V
7 11 V
7 11 V
7 11 V
7 10 V
7 10 V
8 8 V
7 7 V
7 7 V
7 8 V
7 8 V
7 8 V
7 8 V
7 6 V
7 6 V
7 6 V
7 4 V
7 4 V
7 3 V
7 5 V
7 5 V
7 5 V
7 4 V
7 4 V
7 3 V
7 3 V
7 2 V
7 2 V
7 1 V
7 1 V
8 0 V
7 0 V
7 2 V
7 1 V
7 1 V
7 1 V
7 1 V
7 0 V
7 -1 V
7 0 V
7 -1 V
7 -1 V
7 -2 V
7 -1 V
7 -2 V
7 -3 V
7 -2 V
7 -3 V
7 -3 V
7 -3 V
7 -2 V
7 -2 V
7 -2 V
8 -3 V
7 -2 V
7 -3 V
7 -3 V
7 -3 V
7 -4 V
7 -3 V
7 -4 V
7 -4 V
7 -4 V
7 -4 V
7 -4 V
7 -5 V
7 -6 V
7 -6 V
7 -7 V
7 -6 V
7 -6 V
7 -7 V
7 -6 V
7 -6 V
7 -6 V
7 -6 V
7 -7 V
8 -6 V
7 -6 V
7 -6 V
7 -6 V
7 -6 V
7 -6 V
7 -6 V
7 -6 V
7 -6 V
7 -6 V
7 -5 V
7 -6 V
7 -6 V
7 -6 V
7 -5 V
7 -6 V
7 -5 V
7 -6 V
7 -5 V
7 -6 V
7 -5 V
7 -6 V
7 -5 V
8 -5 V
7 -6 V
7 -5 V
7 -5 V
7 -5 V
7 -5 V
7 -5 V
7 -5 V
7 -5 V
7 -5 V
7 -5 V
7 -5 V
7 -4 V
7 -5 V
7 -5 V
7 -4 V
7 -5 V
7 -4 V
7 -5 V
7 -4 V
7 -5 V
7 -4 V
7 -4 V
7 -5 V
8 -4 V
7 -4 V
7 -4 V
7 -4 V
7 -4 V
7 -4 V
7 -4 V
7 -4 V
7 -4 V
7 -4 V
7 -4 V
7 -4 V
7 -3 V
7 -4 V
7 -4 V
7 -3 V
7 -4 V
7 -3 V
7 -4 V
7 -3 V
7 -4 V
7 -3 V
7 -3 V
8 -4 V
7 -3 V
7 -3 V
7 -3 V
7 -4 V
7 -3 V
7 -3 V
7 -3 V
7 -3 V
7 -3 V
7 -3 V
7 -3 V
7 -3 V
7 -2 V
7 -3 V
7 -3 V
7 -3 V
7 -3 V
7 -2 V
7 -3 V
7 -3 V
7 -2 V
7 -3 V
7 -2 V
8 -3 V
7 -2 V
7 -3 V
7 -2 V
7 -3 V
7 -2 V
7 -2 V
7 -3 V
7 -2 V
7 -2 V
7 -2 V
7 -3 V
7 -2 V
7 -2 V
7 -2 V
7 -2 V
7 -2 V
7 -2 V
7 -2 V
7 -3 V
7 -2 V
7 -1 V
7 -2 V
7 -2 V
8 -2 V
7 -2 V
7 -2 V
7 -2 V
7 -2 V
7 -1 V
7 -2 V
7 -2 V
7 -2 V
7 -2 V
7 -1 V
7 -2 V
7 -2 V
7 -1 V
7 -2 V
7 -1 V
7 -2 V
7 -1 V
7 -2 V
7 -1 V
7 -2 V
7 -1 V
7 -1 V
8 -2 V
7 -1 V
7 -2 V
7 -1 V
7 -1 V
7 -2 V
7 -1 V
7 -1 V
7 -2 V
7 -1 V
7 -1 V
7 -1 V
7 -2 V
7 -1 V
7 -1 V
7 -1 V
7 -1 V
7 -2 V
7 -1 V
7 -1 V
7 -1 V
7 -1 V
7 -1 V
7 -1 V
8 -1 V
7 -2 V
7 -1 V
7 -1 V
7 -1 V
7 -1 V
7 -1 V
7 -1 V
7 -1 V
7 -1 V
7 -1 V
7 -1 V
7 -1 V
7 -1 V
7 -1 V
7 -1 V
7 0 V
7 -1 V
7 -1 V
7 -1 V
7 -1 V
7 -1 V
7 -1 V
8 -1 V
7 -1 V
7 0 V
7 -1 V
7 -1 V
7 -1 V
7 -1 V
7 -1 V
7 0 V
7 -1 V
7 -1 V
7 -1 V
7 0 V
7 -1 V
7 -1 V
7 -1 V
7 0 V
7 -1 V
7 -1 V
7 -1 V
7 0 V
7 -1 V
7 -1 V
7 0 V
8 -1 V
7 -1 V
7 0 V
7 -1 V
7 -1 V
7 0 V
7 -1 V
7 -1 V
7 0 V
7 -1 V
7 -1 V
7 0 V
7 -1 V
7 0 V
7 -1 V
7 -1 V
7 0 V
7 -1 V
7 0 V
7 -1 V
7 0 V
7 -1 V
7 -1 V
8 0 V
7 -1 V
7 0 V
7 -1 V
7 0 V
7 -1 V
7 0 V
7 -1 V
7 0 V
7 -1 V
7 0 V
7 -1 V
7 0 V
7 -1 V
7 0 V
7 -1 V
7 0 V
7 -1 V
7 0 V
7 -1 V
7 0 V
7 -1 V
7 0 V
7 0 V
8 -1 V
7 0 V
7 -1 V
7 0 V
7 -1 V
7 0 V
7 0 V
7 -1 V
7 0 V
7 -1 V
currentpoint stroke M
7 0 V
7 -1 V
3114 1646 M
180 0 V
607 251 M
7 2 V
7 6 V
7 8 V
7 12 V
7 15 V
7 17 V
7 20 V
7 23 V
7 24 V
7 27 V
8 29 V
7 31 V
7 31 V
7 33 V
7 35 V
7 33 V
7 38 V
7 35 V
7 39 V
7 37 V
7 38 V
7 40 V
7 35 V
7 42 V
7 38 V
7 35 V
7 42 V
7 38 V
7 34 V
7 39 V
7 39 V
7 35 V
7 31 V
7 39 V
8 37 V
7 33 V
7 30 V
7 32 V
7 35 V
7 32 V
7 29 V
7 26 V
7 27 V
7 31 V
7 29 V
7 26 V
7 23 V
7 20 V
7 22 V
7 26 V
7 24 V
7 23 V
7 19 V
7 18 V
7 14 V
7 15 V
7 19 V
8 19 V
7 16 V
7 15 V
7 12 V
7 11 V
7 9 V
7 7 V
7 8 V
7 12 V
7 10 V
7 9 V
7 8 V
7 6 V
7 6 V
7 4 V
7 2 V
7 2 V
7 0 V
7 1 V
7 4 V
7 3 V
7 2 V
7 2 V
7 0 V
8 0 V
7 -1 V
7 -13 V
7 -18 V
7 -19 V
7 -18 V
7 -18 V
7 -18 V
7 -18 V
7 -18 V
7 -18 V
7 -19 V
7 -18 V
7 -17 V
7 -18 V
7 -18 V
7 -18 V
7 -17 V
7 -18 V
7 -17 V
7 -17 V
7 -18 V
7 -17 V
8 -17 V
7 -16 V
7 -17 V
7 -17 V
7 -16 V
7 -16 V
7 -16 V
7 -16 V
7 -16 V
7 -16 V
7 -16 V
7 -15 V
7 -15 V
7 -15 V
7 -15 V
7 -15 V
7 -15 V
7 -14 V
7 -15 V
7 -14 V
7 -14 V
7 -14 V
7 -13 V
7 -14 V
8 -13 V
7 -13 V
7 -14 V
7 -12 V
7 -13 V
7 -13 V
7 -12 V
7 -13 V
7 -12 V
7 -12 V
7 -12 V
7 -11 V
7 -12 V
7 -11 V
7 -12 V
7 -11 V
7 -11 V
7 -10 V
7 -11 V
7 -11 V
7 -10 V
7 -10 V
7 -10 V
8 -10 V
7 -10 V
7 -10 V
7 -10 V
7 -9 V
7 -9 V
7 -10 V
7 -9 V
7 -9 V
7 -8 V
7 -9 V
7 -9 V
7 -8 V
7 -8 V
7 -9 V
7 -8 V
7 -8 V
7 -8 V
7 -7 V
7 -8 V
7 -8 V
7 -7 V
7 -8 V
7 -7 V
8 -7 V
7 -7 V
7 -7 V
7 -7 V
7 -6 V
7 -7 V
7 -7 V
7 -6 V
7 -7 V
7 -6 V
7 -6 V
7 -6 V
7 -6 V
7 -6 V
7 -6 V
7 -6 V
7 -5 V
7 -6 V
7 -5 V
7 -6 V
7 -5 V
7 -6 V
7 -5 V
8 -5 V
7 -5 V
7 -5 V
7 -5 V
7 -5 V
7 -5 V
7 -4 V
7 -5 V
7 -5 V
7 -4 V
7 -5 V
7 -4 V
7 -4 V
7 -4 V
7 -5 V
7 -4 V
7 -4 V
7 -4 V
7 -4 V
7 -4 V
7 -4 V
7 -4 V
7 -3 V
7 -4 V
8 -4 V
7 -3 V
7 -4 V
7 -3 V
7 -4 V
7 -3 V
7 -4 V
7 -3 V
7 -3 V
7 -3 V
7 -4 V
7 -3 V
7 -3 V
7 -3 V
7 -3 V
7 -3 V
7 -3 V
7 -3 V
7 -3 V
7 -2 V
7 -3 V
7 -3 V
7 -3 V
7 -2 V
8 -3 V
7 -3 V
7 -2 V
7 -3 V
7 -2 V
7 -3 V
7 -2 V
7 -2 V
7 -3 V
7 -2 V
7 -2 V
7 -3 V
7 -2 V
7 -2 V
7 -2 V
7 -1 V
7 -2 V
7 -2 V
7 -2 V
7 -2 V
7 -1 V
7 -2 V
7 -2 V
8 -1 V
7 -2 V
7 -2 V
7 -1 V
7 -2 V
7 -2 V
7 -1 V
7 -2 V
7 -1 V
7 -2 V
7 -1 V
7 -2 V
7 -1 V
7 -2 V
7 -1 V
7 -2 V
7 -1 V
7 -2 V
7 -1 V
7 -1 V
7 -2 V
7 -1 V
7 -1 V
7 -2 V
8 -1 V
7 -1 V
7 -2 V
7 -1 V
7 -1 V
7 -1 V
7 -2 V
7 -1 V
7 -1 V
7 -1 V
7 -1 V
7 -2 V
7 -1 V
7 -1 V
7 -1 V
7 -1 V
7 -1 V
7 -1 V
7 -2 V
7 -1 V
7 -1 V
7 -1 V
7 -1 V
8 -1 V
7 -1 V
7 -1 V
7 -1 V
7 -1 V
7 -1 V
7 -1 V
7 -1 V
7 -1 V
7 -1 V
7 -1 V
7 -1 V
7 0 V
7 -1 V
7 -1 V
7 -1 V
7 -1 V
7 -1 V
7 -1 V
7 -1 V
7 -1 V
7 0 V
7 -1 V
7 -1 V
8 -1 V
7 -1 V
7 -1 V
7 0 V
7 -1 V
7 -1 V
7 -1 V
7 0 V
7 -1 V
7 -1 V
7 -1 V
7 0 V
7 -1 V
7 -1 V
7 -1 V
7 0 V
7 -1 V
7 -1 V
7 0 V
7 -1 V
7 -1 V
7 0 V
7 -1 V
8 -1 V
7 0 V
7 -1 V
7 -1 V
7 0 V
7 -1 V
7 -1 V
7 0 V
7 -1 V
7 0 V
7 -1 V
7 -1 V
7 0 V
7 -1 V
7 0 V
7 -1 V
7 -1 V
7 0 V
7 -1 V
7 0 V
7 -1 V
7 0 V
7 -1 V
7 0 V
8 -1 V
7 0 V
7 -1 V
7 0 V
7 -1 V
7 -1 V
7 0 V
7 -1 V
currentpoint stroke M
7 0 V
7 -1 V
7 0 V
7 0 V
stroke
grestore
end
showpage
}
\put(3054,1646){\makebox(0,0)[r]{$x=10^{-5}$}}
\put(3054,1746){\makebox(0,0)[r]{$x=10^{-4}$}}
\put(3054,1846){\makebox(0,0)[r]{$x=10^{-3}$}}
\put(3054,1946){\makebox(0,0)[r]{$Q^2 =4, x=10^{-2}$}}
\put(2008,-49){\makebox(0,0){\Large $b$ (fm)}}
\put(100,1180){%
\special{ps: gsave currentpoint currentpoint translate
270 rotate neg exch neg exch translate}%
\makebox(0,0)[b]{\shortstack{\Large  $I_{T} (b) $ }}%
\special{ps: currentpoint grestore moveto}%
}
\put(3417,151){\makebox(0,0){1}}
\put(3135,151){\makebox(0,0){0.9}}
\put(2854,151){\makebox(0,0){0.8}}
\put(2572,151){\makebox(0,0){0.7}}
\put(2290,151){\makebox(0,0){0.6}}
\put(2009,151){\makebox(0,0){0.5}}
\put(1727,151){\makebox(0,0){0.4}}
\put(1445,151){\makebox(0,0){0.3}}
\put(1163,151){\makebox(0,0){0.2}}
\put(882,151){\makebox(0,0){0.1}}
\put(600,151){\makebox(0,0){0}}
\put(540,2109){\makebox(0,0)[r]{0.16}}
\put(540,1877){\makebox(0,0)[r]{0.14}}
\put(540,1645){\makebox(0,0)[r]{0.12}}
\put(540,1412){\makebox(0,0)[r]{0.1}}
\put(540,1180){\makebox(0,0)[r]{0.08}}
\put(540,948){\makebox(0,0)[r]{0.06}}
\put(540,716){\makebox(0,0)[r]{0.04}}
\put(540,483){\makebox(0,0)[r]{0.02}}
\put(540,251){\makebox(0,0)[r]{0}}
\end{picture}

%% file: ilb4.tex
% GNUPLOT: LaTeX picture with Postscript
\setlength{\unitlength}{0.1bp}
\special{!
%!PS-Adobe-2.0
%%Creator: gnuplot
%%DocumentFonts: Helvetica
%%BoundingBox: 50 50 770 554
%%Pages: (atend)
%%EndComments
/gnudict 40 dict def
gnudict begin
/Color false def
/Solid false def
/gnulinewidth 5.000 def
/vshift -33 def
/dl {10 mul} def
/hpt 31.5 def
/vpt 31.5 def
/M {moveto} bind def
/L {lineto} bind def
/R {rmoveto} bind def
/V {rlineto} bind def
/vpt2 vpt 2 mul def
/hpt2 hpt 2 mul def
/Lshow { currentpoint stroke M
  0 vshift R show } def
/Rshow { currentpoint stroke M
  dup stringwidth pop neg vshift R show } def
/Cshow { currentpoint stroke M
  dup stringwidth pop -2 div vshift R show } def
/DL { Color {setrgbcolor Solid {pop []} if 0 setdash }
 {pop pop pop Solid {pop []} if 0 setdash} ifelse } def
/BL { stroke gnulinewidth 2 mul setlinewidth } def
/AL { stroke gnulinewidth 2 div setlinewidth } def
/PL { stroke gnulinewidth setlinewidth } def
/LTb { BL [] 0 0 0 DL } def
/LTa { AL [1 dl 2 dl] 0 setdash 0 0 0 setrgbcolor } def
/LT0 { PL [] 0 1 0 DL } def
/LT1 { PL [4 dl 2 dl] 0 0 1 DL } def
/LT2 { PL [2 dl 3 dl] 1 0 0 DL } def
/LT3 { PL [1 dl 1.5 dl] 1 0 1 DL } def
/LT4 { PL [5 dl 2 dl 1 dl 2 dl] 0 1 1 DL } def
/LT5 { PL [4 dl 3 dl 1 dl 3 dl] 1 1 0 DL } def
/LT6 { PL [2 dl 2 dl 2 dl 4 dl] 0 0 0 DL } def
/LT7 { PL [2 dl 2 dl 2 dl 2 dl 2 dl 4 dl] 1 0.3 0 DL } def
/LT8 { PL [2 dl 2 dl 2 dl 2 dl 2 dl 2 dl 2 dl 4 dl] 0.5 0.5 0.5 DL } def
/P { stroke [] 0 setdash
  currentlinewidth 2 div sub M
  0 currentlinewidth V stroke } def
/D { stroke [] 0 setdash 2 copy vpt add M
  hpt neg vpt neg V hpt vpt neg V
  hpt vpt V hpt neg vpt V closepath stroke
  P } def
/A { stroke [] 0 setdash vpt sub M 0 vpt2 V
  currentpoint stroke M
  hpt neg vpt neg R hpt2 0 V stroke
  } def
/B { stroke [] 0 setdash 2 copy exch hpt sub exch vpt add M
  0 vpt2 neg V hpt2 0 V 0 vpt2 V
  hpt2 neg 0 V closepath stroke
  P } def
/C { stroke [] 0 setdash exch hpt sub exch vpt add M
  hpt2 vpt2 neg V currentpoint stroke M
  hpt2 neg 0 R hpt2 vpt2 V stroke } def
/T { stroke [] 0 setdash 2 copy vpt 1.12 mul add M
  hpt neg vpt -1.62 mul V
  hpt 2 mul 0 V
  hpt neg vpt 1.62 mul V closepath stroke
  P  } def
/S { 2 copy A C} def
end
}
\begin{picture}(3600,2160)(0,0)
\special{"
gnudict begin
gsave
50 50 translate
0.100 0.100 scale
0 setgray
/Helvetica findfont 100 scalefont setfont
newpath
-500.000000 -500.000000 translate
LTa
600 251 M
2817 0 V
600 251 M
0 1858 V
LTb
600 251 M
63 0 V
2754 0 R
-63 0 V
600 561 M
63 0 V
2754 0 R
-63 0 V
600 870 M
63 0 V
2754 0 R
-63 0 V
600 1180 M
63 0 V
2754 0 R
-63 0 V
600 1490 M
63 0 V
2754 0 R
-63 0 V
600 1799 M
63 0 V
2754 0 R
-63 0 V
600 2109 M
63 0 V
2754 0 R
-63 0 V
600 251 M
0 63 V
0 1795 R
0 -63 V
882 251 M
0 63 V
0 1795 R
0 -63 V
1163 251 M
0 63 V
0 1795 R
0 -63 V
1445 251 M
0 63 V
0 1795 R
0 -63 V
1727 251 M
0 63 V
0 1795 R
0 -63 V
2009 251 M
0 63 V
0 1795 R
0 -63 V
2290 251 M
0 63 V
0 1795 R
0 -63 V
2572 251 M
0 63 V
0 1795 R
0 -63 V
2854 251 M
0 63 V
0 1795 R
0 -63 V
3135 251 M
0 63 V
0 1795 R
0 -63 V
3417 251 M
0 63 V
0 1795 R
0 -63 V
600 251 M
2817 0 V
0 1858 V
-2817 0 V
600 251 L
LT0
3114 1946 M
180 0 V
607 251 M
7 0 V
7 0 V
7 0 V
7 0 V
7 0 V
7 0 V
7 0 V
7 0 V
7 0 V
7 0 V
8 0 V
7 0 V
7 0 V
7 0 V
7 0 V
7 0 V
7 0 V
7 0 V
7 1 V
7 0 V
7 0 V
7 1 V
7 0 V
7 1 V
7 0 V
7 1 V
7 1 V
7 1 V
7 1 V
7 1 V
7 1 V
7 1 V
7 1 V
7 2 V
8 1 V
7 2 V
7 1 V
7 2 V
7 2 V
7 1 V
7 2 V
7 2 V
7 2 V
7 2 V
7 2 V
7 2 V
7 2 V
7 2 V
7 2 V
7 3 V
7 2 V
7 2 V
7 2 V
7 3 V
7 2 V
7 2 V
7 2 V
8 3 V
7 2 V
7 2 V
7 3 V
7 2 V
7 2 V
7 2 V
7 2 V
7 2 V
7 3 V
7 2 V
7 2 V
7 2 V
7 2 V
7 2 V
7 1 V
7 2 V
7 2 V
7 2 V
7 2 V
7 2 V
7 1 V
7 2 V
7 2 V
8 1 V
7 1 V
7 2 V
7 1 V
7 2 V
7 1 V
7 1 V
7 2 V
7 1 V
7 1 V
7 1 V
7 1 V
7 1 V
7 1 V
7 0 V
7 1 V
7 1 V
7 1 V
7 0 V
7 1 V
7 1 V
7 0 V
7 1 V
8 0 V
7 1 V
7 0 V
7 0 V
7 0 V
7 1 V
7 0 V
7 0 V
7 0 V
7 0 V
7 0 V
7 0 V
7 0 V
7 0 V
7 -1 V
7 0 V
7 0 V
7 0 V
7 0 V
7 0 V
7 -1 V
7 0 V
7 0 V
7 -1 V
8 0 V
7 0 V
7 -1 V
7 0 V
7 -1 V
7 0 V
7 -1 V
7 -1 V
7 0 V
7 -1 V
7 -1 V
7 0 V
7 -1 V
7 -1 V
7 0 V
7 -1 V
7 -1 V
7 -1 V
7 -1 V
7 0 V
7 -1 V
7 -1 V
7 -1 V
8 -1 V
7 -1 V
7 0 V
7 -1 V
7 1 V
7 2 V
7 1 V
7 2 V
7 1 V
7 1 V
7 1 V
7 1 V
7 1 V
7 0 V
7 1 V
7 1 V
7 0 V
7 1 V
7 0 V
7 0 V
7 1 V
7 0 V
7 0 V
7 0 V
8 0 V
7 0 V
7 0 V
7 0 V
7 -1 V
7 0 V
7 0 V
7 0 V
7 -1 V
7 0 V
7 -1 V
7 0 V
7 -1 V
7 0 V
7 -1 V
7 0 V
7 -1 V
7 0 V
7 -1 V
7 -1 V
7 -1 V
7 0 V
7 -1 V
8 -1 V
7 -1 V
7 0 V
7 -1 V
7 -1 V
7 -1 V
7 -1 V
7 -1 V
7 -1 V
7 0 V
7 -1 V
7 -1 V
7 -1 V
7 -1 V
7 -1 V
7 -1 V
7 -1 V
7 -1 V
7 -1 V
7 -1 V
7 -1 V
7 -1 V
7 -1 V
7 0 V
8 -1 V
7 -1 V
7 -1 V
7 -1 V
7 -1 V
7 -1 V
7 -1 V
7 -1 V
7 -1 V
7 -1 V
7 -1 V
7 -1 V
7 -1 V
7 -1 V
7 -1 V
7 -1 V
7 0 V
7 -1 V
7 -1 V
7 -1 V
7 -1 V
7 -1 V
7 -1 V
7 -1 V
8 0 V
7 -1 V
7 -1 V
7 -1 V
7 -1 V
7 -1 V
7 -1 V
7 0 V
7 -1 V
7 -1 V
7 -1 V
7 -1 V
7 0 V
7 -1 V
7 -1 V
7 -1 V
7 -1 V
7 -1 V
7 -1 V
7 -1 V
7 0 V
7 -1 V
7 -1 V
8 -1 V
7 -1 V
7 0 V
7 -1 V
7 -1 V
7 -1 V
7 0 V
7 -1 V
7 -1 V
7 -1 V
7 0 V
7 -1 V
7 -1 V
7 0 V
7 -1 V
7 -1 V
7 0 V
7 -1 V
7 0 V
7 -1 V
7 -1 V
7 0 V
7 -1 V
7 0 V
8 -1 V
7 0 V
7 -1 V
7 0 V
7 -1 V
7 0 V
7 -1 V
7 0 V
7 -1 V
7 0 V
7 -1 V
7 0 V
7 -1 V
7 0 V
7 -1 V
7 0 V
7 0 V
7 -1 V
7 0 V
7 -1 V
7 0 V
7 0 V
7 -1 V
8 0 V
7 -1 V
7 0 V
7 0 V
7 -1 V
7 0 V
7 0 V
7 -1 V
7 0 V
7 0 V
7 -1 V
7 0 V
7 0 V
7 -1 V
7 0 V
7 0 V
7 -1 V
7 0 V
7 0 V
7 0 V
7 -1 V
7 0 V
7 0 V
7 0 V
8 -1 V
7 0 V
7 0 V
7 0 V
7 -1 V
7 0 V
7 0 V
7 0 V
7 -1 V
7 0 V
7 0 V
7 0 V
7 0 V
7 -1 V
7 0 V
7 0 V
7 0 V
7 -1 V
7 0 V
7 0 V
7 0 V
7 0 V
7 0 V
8 -1 V
7 0 V
7 0 V
7 0 V
7 0 V
7 -1 V
7 0 V
7 0 V
7 0 V
7 0 V
7 0 V
7 0 V
7 -1 V
7 0 V
7 0 V
7 0 V
7 0 V
7 0 V
7 0 V
7 -1 V
7 0 V
7 0 V
7 0 V
7 0 V
8 0 V
7 0 V
7 0 V
7 -1 V
7 0 V
7 0 V
7 0 V
7 0 V
7 0 V
7 0 V
currentpoint stroke M
7 0 V
7 -1 V
LT1
3114 1846 M
180 0 V
607 251 M
7 0 V
7 0 V
7 0 V
7 0 V
7 0 V
7 0 V
7 0 V
7 0 V
7 1 V
7 0 V
8 1 V
7 0 V
7 1 V
7 1 V
7 2 V
7 1 V
7 2 V
7 2 V
7 2 V
7 3 V
7 3 V
7 3 V
7 3 V
7 3 V
7 4 V
7 4 V
7 4 V
7 4 V
7 5 V
7 5 V
7 4 V
7 6 V
7 5 V
7 5 V
8 6 V
7 6 V
7 6 V
7 5 V
7 6 V
7 6 V
7 7 V
7 6 V
7 6 V
7 6 V
7 7 V
7 6 V
7 7 V
7 6 V
7 6 V
7 6 V
7 6 V
7 7 V
7 6 V
7 7 V
7 5 V
7 6 V
7 5 V
8 6 V
7 6 V
7 6 V
7 6 V
7 5 V
7 5 V
7 5 V
7 5 V
7 4 V
7 4 V
7 5 V
7 5 V
7 5 V
7 4 V
7 4 V
7 4 V
7 3 V
7 3 V
7 3 V
7 3 V
7 3 V
7 3 V
7 3 V
7 3 V
8 3 V
7 2 V
7 3 V
7 2 V
7 2 V
7 1 V
7 2 V
7 1 V
7 1 V
7 1 V
7 0 V
7 1 V
7 1 V
7 1 V
7 1 V
7 0 V
7 1 V
7 0 V
7 0 V
7 0 V
7 0 V
7 0 V
7 0 V
8 -1 V
7 -1 V
7 0 V
7 -1 V
7 -1 V
7 -2 V
7 -1 V
7 -1 V
7 -2 V
7 -1 V
7 -2 V
7 -1 V
7 -2 V
7 -1 V
7 -2 V
7 -1 V
7 -2 V
7 -1 V
7 -2 V
7 -2 V
7 -2 V
7 -2 V
7 -2 V
7 -2 V
8 -2 V
7 -3 V
7 -2 V
7 -2 V
7 -3 V
7 -2 V
7 -3 V
7 -2 V
7 -3 V
7 -2 V
7 -3 V
7 -2 V
7 -3 V
7 -3 V
7 -2 V
7 -3 V
7 -3 V
7 -3 V
7 -2 V
7 -3 V
7 -3 V
7 -3 V
7 -3 V
8 -2 V
7 -3 V
7 -2 V
7 -3 V
7 -1 V
7 -2 V
7 -1 V
7 -1 V
7 -2 V
7 -1 V
7 -1 V
7 -2 V
7 -1 V
7 -2 V
7 -2 V
7 -1 V
7 -2 V
7 -1 V
7 -2 V
7 -2 V
7 -2 V
7 -1 V
7 -2 V
7 -2 V
8 -2 V
7 -1 V
7 -2 V
7 -2 V
7 -2 V
7 -2 V
7 -1 V
7 -2 V
7 -2 V
7 -2 V
7 -2 V
7 -2 V
7 -1 V
7 -2 V
7 -2 V
7 -2 V
7 -2 V
7 -2 V
7 -1 V
7 -2 V
7 -2 V
7 -2 V
7 -2 V
8 -1 V
7 -2 V
7 -2 V
7 -2 V
7 -2 V
7 -1 V
7 -2 V
7 -2 V
7 -2 V
7 -1 V
7 -2 V
7 -2 V
7 -1 V
7 -2 V
7 -2 V
7 -1 V
7 -2 V
7 -2 V
7 -1 V
7 -2 V
7 -2 V
7 -1 V
7 -2 V
7 -1 V
8 -2 V
7 -1 V
7 -2 V
7 -1 V
7 -2 V
7 -1 V
7 -2 V
7 -1 V
7 -2 V
7 -1 V
7 -1 V
7 -2 V
7 -1 V
7 -2 V
7 -1 V
7 -1 V
7 -2 V
7 -1 V
7 -1 V
7 -1 V
7 -2 V
7 -1 V
7 -1 V
7 -1 V
8 -2 V
7 -1 V
7 -1 V
7 -1 V
7 -1 V
7 -1 V
7 -1 V
7 -2 V
7 -1 V
7 -1 V
7 -1 V
7 -1 V
7 -1 V
7 -1 V
7 -1 V
7 -1 V
7 -1 V
7 -1 V
7 -1 V
7 -1 V
7 -1 V
7 -1 V
7 -1 V
8 -1 V
7 -1 V
7 -1 V
7 -1 V
7 -1 V
7 -1 V
7 -1 V
7 -1 V
7 -1 V
7 0 V
7 -1 V
7 -1 V
7 -1 V
7 -1 V
7 0 V
7 -1 V
7 -1 V
7 -1 V
7 0 V
7 -1 V
7 -1 V
7 0 V
7 -1 V
7 -1 V
8 0 V
7 -1 V
7 -1 V
7 0 V
7 -1 V
7 -1 V
7 0 V
7 -1 V
7 0 V
7 -1 V
7 0 V
7 -1 V
7 -1 V
7 0 V
7 -1 V
7 0 V
7 -1 V
7 0 V
7 -1 V
7 0 V
7 -1 V
7 0 V
7 -1 V
8 0 V
7 0 V
7 -1 V
7 0 V
7 -1 V
7 0 V
7 -1 V
7 0 V
7 0 V
7 -1 V
7 0 V
7 -1 V
7 0 V
7 0 V
7 -1 V
7 0 V
7 0 V
7 -1 V
7 0 V
7 0 V
7 -1 V
7 0 V
7 0 V
7 -1 V
8 0 V
7 0 V
7 -1 V
7 0 V
7 0 V
7 0 V
7 -1 V
7 0 V
7 0 V
7 -1 V
7 0 V
7 0 V
7 0 V
7 -1 V
7 0 V
7 0 V
7 0 V
7 -1 V
7 0 V
7 0 V
7 0 V
7 -1 V
7 0 V
8 0 V
7 0 V
7 0 V
7 -1 V
7 0 V
7 0 V
7 0 V
7 0 V
7 -1 V
7 0 V
7 0 V
7 0 V
7 0 V
7 0 V
7 -1 V
7 0 V
7 0 V
7 0 V
7 0 V
7 -1 V
7 0 V
7 0 V
7 0 V
7 0 V
8 0 V
7 0 V
7 -1 V
7 0 V
7 0 V
7 0 V
7 0 V
7 0 V
7 0 V
7 -1 V
currentpoint stroke M
7 0 V
7 0 V
LT3
3114 1746 M
180 0 V
607 251 M
7 0 V
7 0 V
7 0 V
7 0 V
7 1 V
7 0 V
7 2 V
7 1 V
7 2 V
7 2 V
8 3 V
7 4 V
7 4 V
7 5 V
7 5 V
7 6 V
7 7 V
7 7 V
7 8 V
7 9 V
7 10 V
7 9 V
7 11 V
7 11 V
7 11 V
7 13 V
7 12 V
7 12 V
7 15 V
7 13 V
7 14 V
7 14 V
7 16 V
7 14 V
8 15 V
7 15 V
7 16 V
7 16 V
7 15 V
7 14 V
7 16 V
7 17 V
7 16 V
7 15 V
7 14 V
7 15 V
7 16 V
7 16 V
7 15 V
7 14 V
7 14 V
7 12 V
7 14 V
7 15 V
7 14 V
7 13 V
7 13 V
8 12 V
7 11 V
7 10 V
7 11 V
7 12 V
7 11 V
7 11 V
7 10 V
7 9 V
7 9 V
7 7 V
7 7 V
7 7 V
7 7 V
7 8 V
7 8 V
7 7 V
7 6 V
7 6 V
7 5 V
7 4 V
7 4 V
7 4 V
7 2 V
8 3 V
7 1 V
7 3 V
7 4 V
7 2 V
7 2 V
7 2 V
7 1 V
7 1 V
7 1 V
7 0 V
7 -1 V
7 -1 V
7 -1 V
7 -1 V
7 -2 V
7 -3 V
7 -2 V
7 -3 V
7 -3 V
7 -2 V
7 -2 V
7 -2 V
8 -3 V
7 -3 V
7 -3 V
7 -3 V
7 -4 V
7 -4 V
7 -4 V
7 -4 V
7 -5 V
7 -5 V
7 -4 V
7 -6 V
7 -5 V
7 -8 V
7 -7 V
7 -7 V
7 -7 V
7 -8 V
7 -7 V
7 -7 V
7 -8 V
7 -7 V
7 -7 V
7 -8 V
8 -7 V
7 -7 V
7 -7 V
7 -8 V
7 -7 V
7 -7 V
7 -7 V
7 -7 V
7 -8 V
7 -7 V
7 -7 V
7 -7 V
7 -7 V
7 -7 V
7 -6 V
7 -7 V
7 -7 V
7 -7 V
7 -6 V
7 -7 V
7 -7 V
7 -6 V
7 -7 V
8 -6 V
7 -6 V
7 -7 V
7 -6 V
7 -6 V
7 -6 V
7 -6 V
7 -6 V
7 -6 V
7 -6 V
7 -6 V
7 -6 V
7 -6 V
7 -5 V
7 -6 V
7 -5 V
7 -6 V
7 -5 V
7 -6 V
7 -5 V
7 -5 V
7 -5 V
7 -5 V
7 -5 V
8 -5 V
7 -5 V
7 -5 V
7 -5 V
7 -5 V
7 -4 V
7 -5 V
7 -4 V
7 -5 V
7 -4 V
7 -5 V
7 -4 V
7 -4 V
7 -5 V
7 -4 V
7 -4 V
7 -4 V
7 -4 V
7 -4 V
7 -4 V
7 -3 V
7 -4 V
7 -4 V
8 -3 V
7 -4 V
7 -4 V
7 -3 V
7 -4 V
7 -3 V
7 -3 V
7 -4 V
7 -3 V
7 -3 V
7 -3 V
7 -3 V
7 -3 V
7 -3 V
7 -3 V
7 -3 V
7 -3 V
7 -3 V
7 -3 V
7 -3 V
7 -2 V
7 -3 V
7 -3 V
7 -2 V
8 -3 V
7 -2 V
7 -3 V
7 -2 V
7 -3 V
7 -2 V
7 -2 V
7 -2 V
7 -3 V
7 -2 V
7 -2 V
7 -2 V
7 -2 V
7 -2 V
7 -2 V
7 -2 V
7 -2 V
7 -2 V
7 -2 V
7 -2 V
7 -2 V
7 -2 V
7 -2 V
7 -1 V
8 -2 V
7 -2 V
7 -2 V
7 -1 V
7 -2 V
7 -1 V
7 -2 V
7 -2 V
7 -1 V
7 -2 V
7 -1 V
7 -1 V
7 -2 V
7 -1 V
7 -2 V
7 -1 V
7 -1 V
7 -1 V
7 -2 V
7 -1 V
7 -1 V
7 -1 V
7 -2 V
8 -1 V
7 -1 V
7 -1 V
7 -1 V
7 -1 V
7 -1 V
7 -1 V
7 -1 V
7 -1 V
7 -1 V
7 -1 V
7 -1 V
7 -1 V
7 -1 V
7 -1 V
7 -1 V
7 -1 V
7 -1 V
7 -1 V
7 -1 V
7 0 V
7 -1 V
7 -1 V
7 -1 V
8 -1 V
7 0 V
7 -1 V
7 -1 V
7 -1 V
7 0 V
7 -1 V
7 -1 V
7 -1 V
7 0 V
7 -1 V
7 -1 V
7 0 V
7 -1 V
7 0 V
7 -1 V
7 -1 V
7 0 V
7 -1 V
7 0 V
7 -1 V
7 -1 V
7 0 V
8 -1 V
7 0 V
7 -1 V
7 0 V
7 -1 V
7 0 V
7 -1 V
7 0 V
7 -1 V
7 0 V
7 -1 V
7 0 V
7 -1 V
7 0 V
7 -1 V
7 0 V
7 0 V
7 -1 V
7 0 V
7 -1 V
7 0 V
7 0 V
7 -1 V
7 0 V
8 -1 V
7 0 V
7 0 V
7 -1 V
7 0 V
7 0 V
7 -1 V
7 0 V
7 0 V
7 -1 V
7 0 V
7 0 V
7 -1 V
7 0 V
7 0 V
7 -1 V
7 0 V
7 0 V
7 0 V
7 -1 V
7 0 V
7 0 V
7 0 V
8 -1 V
7 0 V
7 0 V
7 0 V
7 -1 V
7 0 V
7 0 V
7 0 V
7 -1 V
7 0 V
7 0 V
7 0 V
7 -1 V
7 0 V
7 0 V
7 0 V
7 0 V
7 -1 V
7 0 V
7 0 V
7 0 V
7 0 V
7 -1 V
7 0 V
8 0 V
7 0 V
7 0 V
7 0 V
7 -1 V
7 0 V
7 0 V
7 0 V
7 0 V
7 0 V
currentpoint stroke M
7 -1 V
7 0 V
3114 1646 M
180 0 V
607 251 M
7 0 V
7 0 V
7 1 V
7 1 V
7 2 V
7 3 V
7 4 V
7 6 V
7 6 V
7 9 V
8 10 V
7 11 V
7 13 V
7 15 V
7 17 V
7 17 V
7 20 V
7 21 V
7 23 V
7 24 V
7 25 V
7 28 V
7 26 V
7 31 V
7 30 V
7 29 V
7 34 V
7 33 V
7 31 V
7 35 V
7 36 V
7 35 V
7 32 V
7 38 V
8 37 V
7 36 V
7 33 V
7 36 V
7 38 V
7 37 V
7 34 V
7 33 V
7 33 V
7 37 V
7 36 V
7 33 V
7 31 V
7 30 V
7 30 V
7 33 V
7 33 V
7 30 V
7 29 V
7 26 V
7 25 V
7 23 V
7 28 V
8 27 V
7 25 V
7 23 V
7 22 V
7 19 V
7 18 V
7 15 V
7 17 V
7 19 V
7 18 V
7 16 V
7 16 V
7 14 V
7 12 V
7 11 V
7 9 V
7 8 V
7 6 V
7 7 V
7 9 V
7 9 V
7 7 V
7 6 V
7 5 V
8 4 V
7 3 V
7 -9 V
7 -14 V
7 -15 V
7 -14 V
7 -16 V
7 -15 V
7 -16 V
7 -16 V
7 -16 V
7 -16 V
7 -17 V
7 -16 V
7 -17 V
7 -17 V
7 -17 V
7 -17 V
7 -17 V
7 -17 V
7 -17 V
7 -17 V
7 -18 V
8 -17 V
7 -17 V
7 -17 V
7 -18 V
7 -17 V
7 -17 V
7 -17 V
7 -17 V
7 -17 V
7 -17 V
7 -17 V
7 -16 V
7 -17 V
7 -16 V
7 -17 V
7 -16 V
7 -16 V
7 -16 V
7 -16 V
7 -16 V
7 -16 V
7 -15 V
7 -16 V
7 -15 V
8 -15 V
7 -15 V
7 -15 V
7 -14 V
7 -15 V
7 -14 V
7 -14 V
7 -14 V
7 -14 V
7 -14 V
7 -14 V
7 -13 V
7 -13 V
7 -13 V
7 -13 V
7 -13 V
7 -12 V
7 -13 V
7 -12 V
7 -12 V
7 -12 V
7 -12 V
7 -11 V
8 -12 V
7 -11 V
7 -11 V
7 -11 V
7 -11 V
7 -11 V
7 -10 V
7 -10 V
7 -11 V
7 -10 V
7 -9 V
7 -10 V
7 -10 V
7 -9 V
7 -10 V
7 -9 V
7 -9 V
7 -9 V
7 -8 V
7 -9 V
7 -8 V
7 -9 V
7 -8 V
7 -8 V
8 -8 V
7 -8 V
7 -7 V
7 -8 V
7 -7 V
7 -8 V
7 -7 V
7 -7 V
7 -7 V
7 -7 V
7 -6 V
7 -7 V
7 -7 V
7 -6 V
7 -6 V
7 -6 V
7 -6 V
7 -6 V
7 -6 V
7 -6 V
7 -6 V
7 -5 V
7 -6 V
8 -5 V
7 -5 V
7 -6 V
7 -5 V
7 -5 V
7 -5 V
7 -5 V
7 -4 V
7 -5 V
7 -5 V
7 -4 V
7 -5 V
7 -4 V
7 -4 V
7 -4 V
7 -4 V
7 -5 V
7 -4 V
7 -3 V
7 -4 V
7 -4 V
7 -4 V
7 -3 V
7 -4 V
8 -4 V
7 -3 V
7 -3 V
7 -4 V
7 -3 V
7 -3 V
7 -3 V
7 -3 V
7 -4 V
7 -3 V
7 -2 V
7 -3 V
7 -3 V
7 -3 V
7 -3 V
7 -2 V
7 -3 V
7 -3 V
7 -2 V
7 -3 V
7 -2 V
7 -3 V
7 -2 V
7 -2 V
8 -2 V
7 -3 V
7 -2 V
7 -2 V
7 -2 V
7 -2 V
7 -2 V
7 -2 V
7 -2 V
7 -2 V
7 -2 V
7 -2 V
7 -2 V
7 -1 V
7 -2 V
7 -2 V
7 -1 V
7 -2 V
7 -1 V
7 -2 V
7 -1 V
7 -2 V
7 -1 V
8 -1 V
7 -2 V
7 -1 V
7 -2 V
7 -1 V
7 -1 V
7 -1 V
7 -2 V
7 -1 V
7 -1 V
7 -1 V
7 -1 V
7 -2 V
7 -1 V
7 -1 V
7 -1 V
7 -1 V
7 -1 V
7 -1 V
7 -1 V
7 -1 V
7 -1 V
7 -1 V
7 -1 V
8 -1 V
7 -1 V
7 -1 V
7 -1 V
7 -1 V
7 -1 V
7 0 V
7 -1 V
7 -1 V
7 -1 V
7 -1 V
7 -1 V
7 0 V
7 -1 V
7 -1 V
7 -1 V
7 0 V
7 -1 V
7 -1 V
7 0 V
7 -1 V
7 -1 V
7 -1 V
8 0 V
7 -1 V
7 0 V
7 -1 V
7 -1 V
7 0 V
7 -1 V
7 0 V
7 -1 V
7 -1 V
7 0 V
7 -1 V
7 0 V
7 -1 V
7 0 V
7 -1 V
7 0 V
7 -1 V
7 0 V
7 -1 V
7 0 V
7 -1 V
7 0 V
7 -1 V
8 0 V
7 -1 V
7 0 V
7 0 V
7 -1 V
7 0 V
7 -1 V
7 0 V
7 -1 V
7 0 V
7 0 V
7 -1 V
7 0 V
7 0 V
7 -1 V
7 0 V
7 0 V
7 -1 V
7 0 V
7 -1 V
7 0 V
7 0 V
7 0 V
8 -1 V
7 0 V
7 0 V
7 -1 V
7 0 V
7 0 V
7 -1 V
7 0 V
7 0 V
7 0 V
7 -1 V
7 0 V
7 0 V
7 0 V
7 -1 V
7 0 V
7 0 V
7 0 V
7 -1 V
7 0 V
7 0 V
7 0 V
7 -1 V
7 0 V
8 0 V
7 0 V
7 0 V
7 -1 V
7 0 V
7 0 V
7 0 V
7 -1 V
currentpoint stroke M
7 0 V
7 0 V
7 0 V
7 0 V
stroke
grestore
end
showpage
}
\put(3054,1646){\makebox(0,0)[r]{$x=10^{-5}$}}
\put(3054,1746){\makebox(0,0)[r]{$x=10^{-4}$}}
\put(3054,1846){\makebox(0,0)[r]{$x=10^{-3}$}}
\put(3054,1946){\makebox(0,0)[r]{$Q^2 =4, x=10^{-2}$}}
\put(2008,51){\makebox(0,0){\Large $b$ (fm)}}
\put(100,1180){%
\special{ps: gsave currentpoint currentpoint translate
270 rotate neg exch neg exch translate}%
\makebox(0,0)[b]{\shortstack{\Large  $I_{L} (b) $ }}%
\special{ps: currentpoint grestore moveto}%
}
\put(3417,151){\makebox(0,0){1}}
\put(3135,151){\makebox(0,0){0.9}}
\put(2854,151){\makebox(0,0){0.8}}
\put(2572,151){\makebox(0,0){0.7}}
\put(2290,151){\makebox(0,0){0.6}}
\put(2009,151){\makebox(0,0){0.5}}
\put(1727,151){\makebox(0,0){0.4}}
\put(1445,151){\makebox(0,0){0.3}}
\put(1163,151){\makebox(0,0){0.2}}
\put(882,151){\makebox(0,0){0.1}}
\put(600,151){\makebox(0,0){0}}
\put(540,2109){\makebox(0,0)[r]{0.06}}
\put(540,1799){\makebox(0,0)[r]{0.05}}
\put(540,1490){\makebox(0,0)[r]{0.04}}
\put(540,1180){\makebox(0,0)[r]{0.03}}
\put(540,870){\makebox(0,0)[r]{0.02}}
\put(540,561){\makebox(0,0)[r]{0.01}}
\put(540,251){\makebox(0,0)[r]{0}}
\end{picture}